\documentclass[twocolumn]{aastex6}

\begin{document}
\title{Star Formation In Nearby Clouds (SFiNCs): X-ray And Infrared Source Catalogs And Membership}

\author{Konstantin V.\ Getman and Patrick S.\ Broos}
\affil{Department of Astronomy \& Astrophysics, 525 Davey Laboratory, Pennsylvania State University, University Park PA 16802}
\author{Michael A.\ Kuhn}
\affil{Instituto de Fisica y Astronomia, Universidad de Valparaiso, Gran Bretana 1111, Playa Ancha, Valparaiso, Chile; Millennium Institute of Astrophysics, MAS, Chile}
\affil{Millenium Institute of Astrophysics, Av. Vicuna Mackenna 4860, 782-0436 Macul, Santiago, Chile}
\author{Eric D.\ Feigelson and Alexander J.\ W.\ Richert and Yosuke Ota}
\affil{Department of Astronomy \& Astrophysics, 525 Davey Laboratory, Pennsylvania State University, University Park PA 16802}
\author{Matthew R.\ Bate}
\affil{Department of Physics and Astronomy, University of Exeter, Stocker Road, Exeter, Devon EX4 4SB, UK}
\and
\author{Gordon P.\ Garmire}
\affil{Huntingdon Institute for X-ray Astronomy, LLC, 10677 Franks Road, Huntingdon, PA 16652, USA}

\begin{abstract}
The Star Formation in Nearby Clouds (SFiNCs) project is aimed at providing detailed study of the young stellar populations and star cluster formation in nearby 22 star forming regions (SFRs) for comparison with our earlier MYStIX survey of richer, more distant clusters. As a foundation for the SFiNCs science studies, here, homogeneous data analyses of the {\it Chandra} X-ray and {\it Spitzer} mid-infrared archival SFiNCs data are described, and the resulting catalogs of over 15300 X-ray and over 1630000 mid-infrared point sources are presented. On the basis of their X-ray/infrared properties and spatial distributions, nearly 8500 point sources have been identified as probable young stellar members of the SFiNCs regions. Compared to the existing X-ray/mid-infrared publications, the SFiNCs member list increases the census of YSO members by 6--200\% for individual SFRs and by 40\% for the merged sample of all 22 SFiNCs SFRs.
\end{abstract}

\keywords{infrared: stars; stars: early-type; open clusters and associations: individual (NGC 7822, IRAS 00013+6817, NGC 1333, IC 348, LkH$\alpha$ 101, NGC 2068-2071, Orion Nebula, OMC 2-3, Mon R2, GGD 12-15, RCW 120, Serpens Main, Serpens South, IRAS 20050+2720, Sh 2-106, IC 5146, NGC 7160, LDN 1251B, Cep OB3b, Cep A, Cep C); open clusters and associations: general; stars: formation; stars:pre-main sequence; X-rays: stars}

\section{Intriduction} \label{intro_section}

\subsection{Motivation For SFiNCs} \label{sfincs_motivation_subsection}

Most stars today, like our Sun, are field stars.  Yet examination of star formation in molecular clouds shows that most form in compact bound clusters with $10^2-10^4$ members \citep{Lada2003} or distributed, unbound stellar associations \citep{Kruijssen2012}.  Most young stellar structures rapidly disperse, often when the gravitational potential decreases due to the dispersal of the molecular cloud material \citep{Tutukov1978, Hills1980} or due to the tidal interactions with other giant molecular clouds \citep{Kruijssen2012b}. Although these basic concepts seem physically reasonable, we have a poor empirical characterization, and hence uncertain astrophysical understanding of the detailed processes of cluster formation and early evolution.  Do clusters form `top-down', rapidly in a dense molecular cloud core \citep{Clarke2000}?  Or, since clouds are turbulent, do clusters form `bottom-up' by merging subclusters produced in small kinematically-distinct molecular structures \citep{Bonnell2003,McMillan2007,Bate2009}?  Do clusters principally form in elongated molecular structures such as cold Infrared Dark Clouds \citep{Krumholz2007} and filaments found to be pervasive in giant molecular clouds by the Herschel satellite \citep{Andre2010}? Do massive stars form early in the life cycle of a stellar cluster or are they the last to form, halting further star formation \citep{Zinnecker2007}? How important is the role of the feedback by young stars (such as protostellar outflows) or, for the more massive clusters with hot OB stars, by ultraviolet radiation and stellar winds on star cluster formation processes \citep{Krumholz2014}? 

One of the central reasons for slow progress in resolving these questions is the lack of homogeneous and reliable census of young stellar object (YSO) members for a wide range of star forming environments.  Early studies focused on stars with H$\alpha$ emission (which arises in accreting columns) and/or photometric variability. Such variability is due to rotational modulation of cool magnetic and hot accretion spots and variable obscuration from circumstellar dust \citep[e.g.,][]{Herbig1988, Herbst1994}.  More recent studies of young stellar populations identify the disk-bearing stellar subpopulations that are easily found through their photometric infrared (IR) excess arising from blackbody emission of the protoplanetary disk. Most studies focus on single star forming regions (SFRs). However, due to the heterogeneity of the analysis procedures and resulting YSO data sets in such studies, it is not trivial to compare stellar populations of SFRs with different distances and absorptions.

Recent progress, by our group and others, has been made by combining the selection of young stars through the synergy of X-ray and infra-red (IR) surveys. For instance,  2MASS, UKIDSS near-IR (NIR) and $Spitzer$ mid-IR (MIR) surveys can cover large areas, readily identifying young Class~I protostars embedded in cloud cores as well as older accreting Class~II (or `classical T Tauri') stars \citep[e.g.,][]{Gutermuth2009, Megeath2012}.  These IR surveys are effectively restricted to stars with IR excesses from dusty protoplanetary disks; disk-free young stars appear in the IR photometric catalogs but are indistinguishable from foreground and background Galactic field stars.

X-ray surveys have different selection criteria based on magnetic reconnection flaring near the stellar surface; they efficiently capture disk-free (Class III or `weak-lined T Tauri' stars) stars as well as a good fraction of Class I and II stars \citep[e.g.,][]{Feigelson99, Feigelson2013}.  Since X-ray emission from old Galactic stars is reduced by factors of $10^{2-3}$ below that of pre-main sequence (PMS) stars \citep{Preibisch2005}, field star contamination in the X-ray images of SFRs is significantly smaller than that of IR and optical images. Quasars, principal contaminants in nearby SFRs (\S \ref{source_contamination_section}), are easily removed by the faintness and red colors of their IR counterparts. Class III cluster members can be individually identified \citep[e.g.,][]{Getman2011, Broos2013}. An additional advantage of the X-ray surveys of young stellar clusters is that the X-ray Luminosity Function (XLF) of PMS stars is empirically found to be nearly universal and is closely linked to the stellar initial mass function (IMF) \citep[e.g.,][]{Getman2012, Kuhn2015a}, so that X-ray samples obtained at a known sensitivity are roughly complete to a specific mass; albeit, the scatter in the X-ray luminosity versus mass relationship differs in various studies \citep{Preibisch2005b,Getman2006,Telleschi2007}.

We are now engaged in an effort called MYStIX,  Massive Young Star-Forming Complex Study in Infrared and X-ray \citep[][\url{http://astro.psu.edu/mystix}]{Feigelson2013}. It combines a reanalysis of the {\it Chandra} data archive with new reductions of UKIRT+2MASS NIR and $Spitzer$ MIR surveys to identify young stars in a wide range of evolutionary stages, from protostars to disk-free pre-main sequence stars, in 20 massive SFRs at distances from 0.4 to 3.6 kpc. Stars with published spectra indicative of OB spectral types are added to the sample.  Each MYStIX region was chosen to have a rich OB-dominated cluster. By combining X-ray selected and infrared-excess selected stars, and using sophisticated statistical methods to reduce field star and quasar contamination, MYStIX obtained a uniquely rich sample of $>30,000$ young stars in the 20 massive SFRs. MYStIX is briefly reviewed in \S \ref{more_mystix_subsection}.

The SFiNCs project extends the MYStIX effort to an archive study of 22 generally nearer and smaller SFRs where the stellar clusters are often dominated by a single massive star --- typically a late-O, or early-B --- rather than by numerous O stars as in the MYStIX fields. The SFiNCs science goals are closely tied to the diverse MYStIX science program. Both projects are committed to comparative study of a reasonably large sample of SFRs with stellar populations derived from X-ray and IR surveys using uniform methodologies. 

The scientific goals of SFiNCs could be simply stated: to perform analyses similar to those of MYStIX in order to examine whether the behaviors of clustered star formation are similar $-$ or different $-$ in smaller (SFiNCs) and giant (MYStIX) molecular clouds. The two projects together will establish, in a uniform fashion, empirical properties and correlations among properties for $\sim 200$ subclusters, each with $10-3000$ detected stars, in SFRs on scales of $0.1-30$ pc. Do the SFiNCs subclusters occupy the same loci in parameter space as MYStIX subclusters?  Do they show similar spatio-temporal gradients?  Are certain stellar cluster properties different due to reduced turbulence in smaller molecular clouds, or are population characteristics absent in SFiNCs SFRs due to the absence of OB star feedback that play important roles in MYStIX regions? It is possible, for example, that smaller molecular clouds have less turbulence and thus produce small clusters in single events with simple structure, rather than clusters with complicated substructure suggesting subcluster mergers, as seen in many MYStIX SFRs.  It is possible that smaller clusters are formed with different initial central star densities and expand at different times or rates.

\subsection{MYStIX As A Foundation For SFiNCs} \label{more_mystix_subsection}

Since the SFiNCs and MYStIX programs are closely tied, it is important to review the status of the MYStIX project. The MYStIX studies will serve as the foundation for various SFiNCs efforts. 

MYStIX  has recently emerged with eight technical/catalog papers, seven published science papers, and two science papers that are nearing completion. The technical papers, reviewed in \citet{Feigelson2013}, describe the following innovative methods designed for crowded and nebular star forming regions that are often found lying near the Galactic Plane: {\it Chandra} X-ray, {\it Spitzer} MIR, UKIRT NIR data reduction \citep{Kuhn2013a, Townsley2014, Kuhn2013b, King2013}; X-ray/IR source matching \citep{Naylor2013}; and X-ray/IR membership classifications \citep{Povich2013,Broos2013}. 

In the current work, the same MYStIX-based X-ray and MIR data analysis methods are used for the re-analyses of the archived SFiNCs {\it Chandra} and {\it Spitzer} data (\S \ref{data_section}). However, SFiNCs can simplify the MYStIX analysis work-flow in two ways. First, since SFiNCs stars are on average brighter (due to the proximity of the SFRs) and suffer less Galactic field star contamination (due to higher Galactic latitudes), infrared counterpart identification can be achieved using traditional proximity methods. That is, the closest IR star to a {\it Chandra} source is a reliable counterpart in most cases (\S \ref{source_ir_matching_section}). The more complicated Bayesian probabilistic counterpart identification method developed for MYStIX \citep{Naylor2013} is not needed. Second, again because field star contamination is greatly reduced, MYStIX's complex naive Bayes source classification method \citep{Broos2013} is not needed here; and a simpler approach based on a decision tree classification is used instead (\S \ref{yso_selection_section}).

The current major MYStIX science results include: identification of over 140 MYStIX stellar subclusters and demonstration of their diverse morphologies from simple ellipsoids to elongated, clumpy substructures \citep{Kuhn2014a}; development of an X-ray/IR age stellar chronometer and demonstration of spatio-age gradients on scales of $\sim 1 - 30$~pc \citep{Getman2014a}; discovery of core-halo age gradients within two rich nearby clusters on scales $\leq 1$~pc \citep{Getman2014b}; demonstration of a universal XLF and discovery of wide ranges of the surface stellar density distribution in young stellar clusters \citep{Kuhn2015a}; demonstration of correlations among subcluster properties providing empirical signs of dynamical evolution and cluster expansion \citep{Kuhn2015b}; and no evidence for protoplanetary destruction by OB stars in the MYStIX clusters \citep{Richert2015}. Complementary to the MYStIX sample of $>30,000$ probable cluster members \citep{Broos2013}, \citet{Romine2016} provide the catalog of $>1000$  MYStIX candidate Class~I protostars. The science studies on stellar mass segregation (Kuhn et al. in prep.) and the catalog of probable new OB stars (Povich et al. in prep.) are underway.

The MYStIX methods and data will be effectively used in the following future planned SFiNCs science efforts. Identification of SFiNCs subclusters, derivation of their apparent properties, and comparison of these properties to those of the MYStIX subclusters; the effort is in many respects reminiscent of the MYStIX work of \citet{Kuhn2014a}. Examination of age gradients within the SFiNCs SFRs and subclusters; this is analogous to the MYStIX studies of \citet{Getman2014a,Getman2014b}. Derivation of intrinsic properties of the SFiNCs subclusters, their comparison with MYStIX, and examination of implications for dynamical evolution and cluster expansion, similar to \citet{Kuhn2015a,Kuhn2015b}. Revision of the longevities of protoplanetary disks using homogeneous rich datasets of ages and disk fractions for the MYStIX$+$SFiNCs subclusters with a proper account for stellar mass completeness and relative sensitivity to different clusters in IR and X-ray bands. Comparison of the multivariate MYStIX$+$SFiNCs cluster properties to astrophysical models of cluster formation and early dynamical evolution in search of empirical constraints on a variety of predictions made by theoretical models.

\subsection{Outline Of This Paper} \label{outline_subsection}

As a foundation for the SFiNCs science studies, here, homogeneous data and YSO membership analyses are described, and the resulting catalogs of X-ray/IR point sources and YSO members of the SFiNCs SFRs are presented. The SFiNCs sample is introduced in \S \ref{sample_section}. The paper further describes {\it Chandra}-ACIS X-ray and {\it Spitzer}-IRAC MIR observations and their data reduction, source detection and characterization (\S \ref{data_section}), cross matching among the 2MASS NIR, IRAC, and ACIS source catalogs (\S \ref{source_ir_matching_section}). These are followed by disk classification and YSO membership (\S \ref{yso_selection_section}), spatial distribution of the YSOs (\S \ref{spcm_spatial_section}), IR/optical diagrams and global properties of the YSOs (\S\ref{oir_diagrams_section}, \S \ref{spcm_global_section}), comparison between bright and faint X-ray YSOs (\S \ref{xray_bright_vs_faint_section}), and comparison with the previous member lists from the literature (\S \ref{sfincs_vs_published_section}). Extensive tables of the SFiNCs X-ray/IR sources and YSO members and their properties, as well as a visual atlas with various YSO's characteristics (Appendix \S \ref{sec_appendix_source_atlas}) are provided. Other SFiNCs papers will discuss various science issues emerging from these data.

\section{SFiNCs Sample Selection} \label{sample_section}

The {\it Chandra X-ray Observatory} mission has observed several dozen SFRs in the nearby Galaxy. A large portion of massive SFRs with typical distances in the range $0.4$ to $3.6$~kpc was treated in MYStIX  \citep[Table 1 in][]{Feigelson2013}. Here we have selected 22 generally smaller SFRs with the following criteria: nearby $0.2 < d \la 1$~kpc; young (cluster ages $ \la 10$~Myr, when estimated); populations typically dominated by B type stars; archived {\it Chandra} observations sensitive (but not complete) down to $\log (L_{X}) \la 29.5$~erg/s corresponding to PMS stars with $\la 0.3$~M$_\odot$; archived $Spitzer$-IRAC data are available. The properties of the 22 SFiNCs targets are given in Table \ref{tbl_sfincs_sfrs}. Note that we do not set a criterion based on obscuration because the X-ray emission of many (although not all) young stars is often hard enough to penetrate $A_V > 10-20$~mag of obscuration, which is comparable to that of the NIR 2MASS survey. Half of our targets are in the \citet{Lada2003} catalog of nearby embedded star clusters. Unlike the MYStIX SFRs, all but two SFiNCs SFRs lie away from the Galactic Plane. Note that as for MYStIX \citep[Table 1 in][]{Feigelson2013}, here we omit reporting heterogeneous SFiNCs age estimates published in the literature. A homogeneous set of median ages for individual SFiNCs clusters, using the cluster and age methods of \citet{Kuhn2014a} and \citet{Getman2014a}, will be reported in a future SFiNCs publication.

The nearest ($d < 0.2$~kpc) SFRs are omitted from SFiNCs because on the sky they often subtend areas much greater than the $Chandra$ detector. Excellent {\it Chandra} exposures are available for the two cluster regions in the Perseus cloud, IC 348 and NGC 1333 at $\sim 300$~pc. Two single exposures are available for the Serpens core cloud and Serpens South cluster; both are parts of the Serpens-Aquila Rift region at $\sim 400$~pc. We proceed to the Orion clouds at a distance of $\sim 400$~pc. Single {\it Chandra} exposures are available for the regions adjacent to the Orion Nebula Cluster (ONC) in the Orion A cloud (ONC Flank S, ONC Flank N, and OMC 2-3). Multiple {\it Chandra} exposures/pointings exist for the NGC 2068-2071 complex in the Orion B cloud. Two famous rich clusters in the Orion region, ONC and Flame Nebula, are in the MYStIX sample. Two parts of the Monoceros R2 cloud (Mon R2 and GGD 12-15) at a distance of $\sim 800$~pc are covered by single {\it Chandra} exposures. Three parts of the large Cepheus cloud (Cep OB3b, Cep A, and Cep C) at a distance of $\sim 700$~pc and two parts of the large Cepheus Loop H\,{\sc{ii}} bubble (NGC 7822 and IRAS 00013+6817) at a distance of $\sim 900$~pc are covered, as well as a variety of isolated SFRs.

\section{$\it Chandra$-ACIS And {\it Spitzer}-IRAC Observations And Data Reduction} \label{data_section}

\subsection{X-ray Data} \label{xray_data_section}

For many SFiNCs SFRs, the {\it Chandra Source Catalog} (CSC; current release version 1.1\footnote{Current release version, v1.1, of the {\it Chandra} Source Catalog is available at \url{http://cxc.cfa.harvard.edu/csc1/}. The production of the release v2.0 is in progress, \url{http://cxc.cfa.harvard.edu/csc2/}.}) provides catalogs of X-ray point sources and their X-ray properties. However, the CSC v1.1 is limited to relatively bright X-ray sources and does not adequately treat multiple ObsID mosaics. Our experience with MYStIX \citep{Feigelson2013, Kuhn2013a} suggested that our SFiNCs X-ray catalogs would have $2-3$ times the population as the CSC v1.1. We do not use published X-ray source catalogs either, because they were produced with heterogeneous methods that are often less sensitive to point sources than the MYStIX-based methods \citep[see \S 4 in][]{Kuhn2013a}, and they often do not provide the absorption and X-ray luminosity measurements for faint X-ray PMS stars needed for our age and population analyses. Instead, we opted to re-analyze the archive {\it Chandra} data using the MYStIX-based methods.

Sixty five X-ray observations for the 22 SFiNCs SFRs, made with the imaging array on the Advanced CCD Imaging Spectrometer \citep[ACIS][]{Weisskopf2002,Garmire2003}, were pulled from the {\it Chandra} archive\footnote{\url{http://cxc.cfa.harvard.edu/cda/}}. The details on these observations are provided in Table \ref{tbl_chandra_obslog}. For half of the SFiNCs regions, multiple observations often with multiple pointings were collected. The left panels in Figure \ref{fig_img_fov} present the low-resolution adaptively smoothed images of these mosaicked {\it Chandra}-ACIS exposures. The broadening of the X-ray point sources at the halos of the SFiNCs fields is due to the considerable degradation of the {\it Chandra} telescope point-spread function (PSF) far off-axis. A typical net exposure time for a single SFiNCs pointing ranges between 50 and 100~ks. All but one (ObsID 6401) observations were taken in the imaging mode with the imaging array ACIS-I, an array of four abutted $1024 \times 1024$ pixel front-side illuminated charge-coupled devices (CCDs) covering about $17\arcmin \times 17 \arcmin$ on the sky; the aimpoints for these observations are located on the I3 chip. In some of these observations, the S2 and/or S3 chips from the spectroscopic ACIS-S array were also operational. One observation (ObsID 6401 for IC~5146) was taken in the imaging mode with the aimpont located on the S3 chip; in this case the S2 and all ACIS-I chips were turned on as well.  
 
Our X-ray data reduction follows the procedures described in detail by the MYStIX studies of \citet{Kuhn2013a,Townsley2014} and earlier studies of \citet{Broos2010,Broos2011}. The Level 1 processed event lists provided by the pipeline processing at the {\it Chandra} X-ray Center were calibrated and cleaned using mostly standard methods and tools. Briefly, using the tool {\it acis\_process\_events} from the CIAO version 4.6 the latest calibration information (CALDB 4.6.2) on time-dependent gain and our custom bad pixel mask \citep[\S 3 in][]{Broos2010} are applied; background event candidates are identified. Using the {\it acis\_detect\_afterglow} tool, additional afterglow events not detected with the standard Chandra X-ray Center (CXC) pipeline are flagged. The event list is cleaned by ``grade'' (only ASCA grades 0, 2, 3, 4, 6 are accepted), ``status'', ``good-time interval'', and energy filters. The slight PSF broadening from the CXC software position randomizations is removed. Instrumental background events were identified and removed using an aggressive algorithm when searching for sources and using an algorithm with few false positives when extracting sources \citep[\S 3 in][]{Broos2010}. Event positions were adjusted to better align with the 2MASS catalog \citep{Skrutskie2006}. 

Detection of candidate point sources is performed using two methods, the wavelet transform method \citep{Freeman2002} and the maximum likelihood image deconvolution with local PSFs \citep[see \S 4.2 in][]{Broos2010}; the latter is better suited for resolving closely spaced sources. Source photon extraction and characterization from multi-ObsID ACIS data based on local PSFs, and updated position estimates were obtained for candidate sources using the {\it ACIS Extract} \citep[{\it AE};][]{Broos2010,Broos2012} software package\footnote{The {\it ACIS Extract} software package and User's Guide are available at \url{http://www2.astro.psu.edu/xray/docs/TARA/ae\_users\_guide.html}.}. Through numerous iterations over spatially crowded source candidates, {\it AE} produces optimal source and background extraction regions; and based on Poisson statistics, {\it AE} calculates the probability that a source candidate is a background fluctuation ($P_B$). As in MYStIX, here the SFiNCs {\it Chandra} catalog retains all X-ray point sources for which $P_B < 1$\%; this criterion sometimes results in on-axis X-ray sources with as few as 3 net counts.

\subsection{X-ray Source Catalog} \label{acis_catalog_section}

Our final {\it Chandra}-ACIS catalog for the 22 SFiNCs SFRs comprises 15364 X-ray sources (Table \ref{tbl_acis_src_properties}). The {\it AE} package provides a variety of source characteristics, including celestial position, off-axis angle, net and background counts, fraction of the PSF enclosed within the extraction region, source significance and probability for a source being a background fluctuation assuming Poisson statistics, a variability indicator extracted from the one-sided Kolmogorov-Smirnov test, median energy after background subtraction, and occasional anomalies related to chip gap, filed edge positions, photon pile-up, location on a readout streak, and a possible contamination by afterglow events.

It has to be stressed here that our deliberately aggressive strategy of pushing the X-ray source detection down to 3 net counts on-axis, similar to that of the MYStIX \citep{Kuhn2013a,Townsley2014} and earlier CCCP \citep[Chandra Carina Complex Project]{Townsley2011,Broos2011} projects, produces most sensitive X-ray source catalogs. For instance, the comparison between the SFiNCs and previously published X-ray catalogs, which are available for 13 SFiNCs SFRs (Appendix \ref{sec_appendix_flux_comparison} and Table \ref{tbl_new_sfincs_vs_pub}), indicates that the number of X-ray sources in SFiNCs is typically by a factor of $>3$ higher than that in the earlier studies. The X-ray color-magnitude diagrams presented further in \S \ref{xray_bright_vs_faint_section} suggest that the vast majority of the newly discovered faint X-ray sources have hard X-ray spectra (median energy above 2-3~keV), characteristic of AGN or highly absorbed YSOs. Older background stars in the high Galactic latitude SFiNCs SFRs would generally contribute only a small fraction to these weak, extremely hard X-ray sources \citep[Figure 2 in][]{Broos2013}. But the MYStIX, CCCP, and SFiNCs catalogs are undoubtedly subject to contamination by faint spurious X-ray sources. This choice is obvious --- by producing most sensitive X-ray catalogs we aim at identifying larger numbers of faint X-ray sources with IR counterparts, many of which would be new low-mass and/or highly absorbed YSO members of star forming regions. Our choice of a threshold of $P_B <1$\% for detection has been justified in previous studies of the Carina Nebula star forming complex, in which the number of X-ray sources, both with and without IR counterparts, increased smoothly (without jumps) as the threshold was increased \citep[see Figure 9 in][]{Broos2011}. Similar trends are seen in MYStIX \citep{Kuhn2013a}. Due to the complexity of our X-ray methods (a fusion of two source detection methods, iterative background optimization, and extraction from multi-ObsID ACIS data) a simulation aimed at estimating the fraction of spurious X-ray sources is deemed to be infeasible \citep[\S 6.2 in][]{Broos2011}. Nevertheless, the comparison of the numbers of the simulated Galactic field and extragalactic X-ray contaminants with the numbers of the observed X-ray non-members for the MYStIX fields suggest that the fraction of the spurious X-ray sources in our X-ray catalogs is likely below a few-$10$ percent \citep[see Table 8 in][]{Broos2013}.

Another caveat pertain to our methods of detecting and identifying X-ray sources and to the resulting SFiNCs as well as MYStIX and CCCP X-ray catalogs. The very weak X-ray sources often tend to concentrate towards the central parts of {\it Chandra}-ACIS fields producing a ring-shaped spatial distribution. This, so called ``egg-crate'' effect, is due to the variation in detection completeness with off-axis angle \citep{Broos2011}. Our simulations of the MYStIX extragalactic X-ray contaminants successfully reproduced such a distribution \citep[see Figure 6f in][]{Broos2013}. 

The X-ray fluxes and absorbing column densities of the SFiNCs X-ray sources are generated using the non-parametric method {\it XPHOT} \citep{Getman2010}. The concept of the method is similar to the long-standing use of color-magnitude diagrams in optical and infrared astronomy, with X-ray median energy replacing color index and X-ray source counts replacing magnitude. Empirical X-ray spectral templates derived from bright sources from the {\it Chandra} Orion Ultradeep Project \citep[COUP;][]{Getman2005} are further used to translate apparent photometric properties of weak PMS stars into their intrinsic properties. The advantage of the {\it XPHOT} method over a traditional parametric spectral modeling is that it is more accurate for very faint sources and provides both statistical and systematic (due to uncertainty in X-ray model) errors on derived intrinsic fluxes and absorptions. $XPHOT$ allows a recovery of the soft ($\la 1$~keV) X-ray plasma component, which is often missed in the X-ray data of weak and/or highly absorbed sources when using traditional methods of parametric model fitting. The comparison of the $XPHOT$ fluxes and column densities with the previously published X-ray fluxes and densities (Appendix \ref{sec_appendix_flux_comparison}) supports these notions: the number of faint X-ray sources with available flux estimates in SFiNCs is higher than that in the previous catalogs; and the SFiNCs $XPHOT$ fluxes are systematically higher than the earlier published X-ray fluxes.

The incident photon fluxes produced by {\it AE} and the apparent/intrinsic X-ray fluxes and absorbing column densities produced by {\it XPHOT} are given in Table \ref{tbl_acis_src_fluxes}. Since {\it XPHOT} assumes the X-ray spectral shapes of young, low-mass stars, the estimates of the column densities and intrinsic X-ray fluxes (Columns 10-18 in Table \ref{tbl_acis_src_fluxes}) could be inaccurate for high-mass stellar members of the SFiNCs SFRs as well as for non-members, such as Galactic field stars and extragalactic objects, whereas the incident and apparent X-ray fluxes (Columns 4-9 in Table \ref{tbl_acis_src_fluxes}) are valid for any class of X-ray sources.

X-ray luminosity in the total X-ray band ($L_{tc}$) can be derived from the $F_{tc}$ quantity (Column 16 in Table \ref{tbl_acis_src_fluxes}) by multiplying it by $4 \pi d^2$ where $d$ is the object's distance, in cm, assuming that the object is a YSO member of a SFiNCs region. For the sets of the SFiNCs cluster members (defined further in \S \ref{yso_selection_section}) with available $L_{tc}$ measurements, their X-ray luminosity functions are presented in Figure \ref{fig_xlf_bright}; the XLFs are sorted by the SFR distances.  The X-ray sensitivity to a PMS population is a function of the X-ray exposure time, the SFR distance, and the X-ray absoprtion of the PMS population. Figure \ref{fig_xlf_bright} shows that for the nearest SFiNCs regions NGC~1333 and IC~348 their XLFs peak at $\log(L_{tc}) \sim 29-29.5$~erg~s$^{-1}$ whereas for the most distant regions RCW~120 and Sh~2-106 the peaks are at $\log(L_{tc}) \sim 30.3-30.8$~erg~s$^{-1}$. 

For the PMS members of young (age $\la$ 5-10~Myr) stellar clusters, their X-ray luminosities can be translated to stellar masses \citep{Getman2014a} using the empirical X-ray luminosity/mass relation calibrated to well-studied Taurus PMS stars \citep{Telleschi2007}. We thus expect that our SFiNCs X-ray source catalog includes PMS sub-samples that are roughly complete down to $\sim 0.3$~M$_{\odot}$ and $\sim 2$~M$_{\odot}$ for the nearest and the most distant SFiNCs SFRs, respectively.

\subsection{{\it Spitzer}-IRAC Data} \label{mir_data_section}

$Spitzer$ IRAC infrared photometry (often together with $JHK$ measurements) is used to establish the presence of protoplanetary disks around cluster members.  All but two SFiNCs regions have published catalogs of $Spitzer$-selected disk-bearing stellar populations (Table \ref{tbl_sfincs_sfrs}); typical regions have $100-200$ published infrared-excess young objects. For MYStIX, we found published catalogs were often inadequate due to crowding and strong PAH-band nebulosity so that a new analysis was needed.  For SFiNCs, both crowding and nebulosity are greatly reduced and we believe that published catalogs of disky YSOs have sufficient high quality to provide the basis for constructing our catalogs. The catalogs for 12 of the 22 SFiNCs regions have been reduced in a consistent way by \citet{Gutermuth2009}, and most of the other regions have catalogs derived using similar consistent procedures by \citet{Megeath2012} and \citet{Allen2012}. Nevertheless, most of these published {\it Spitzer} catalogs are limited to disky YSOs. 

To obtain MIR photometry for X-ray objects and to identify and measure MIR photometry for additional non-{\it Chandra} disky stars that were missed in previous studies of the SFiNCs regions (typically faint YSOs), we have reduced the archived {\it Spitzer}-IRAC data by homogeneously applying the MYStIX-based {\it Spitzer}-IRAC data reduction methods of \citet{Kuhn2013b} to the 423 Astronomical Object Request (AORs) datasets for the 22 SFiNCs SFRs (Table \ref{tbl_irac_obslog}). 

These observations were taken with the IRAC (Infrared Array Camera) detector \citep{Fazio2004}, which operates simultaneously on four wavelengths in two pairs of channels (3.6 and 5.8; 4.5 and 8.0~$\mu$m), providing $5.2\arcmin \times 5.2\arcmin$ images with spatial resolution of FWHM = $1.6\arcsec$ to $1.9\arcsec$ from 3.6 to 8~$\mu$m. Most of these observations were taken in the high-dynamic-range mode, to provide unsaturated photometry for both brighter and fainter sources. Over 70\% of these were taken in the Post-Cryo mode during the warm mission of the {\it Spitzer} observatory with only two shortest wavelengths in operation. NGC~7822 is the only SFiNCs region lacking data in the two longest IRAC wavelenghts, 5.8 and 8.0~$\mu$m. 

Typical total integration times per pixel for long frames in the channel 3.6, for all combined observations per SFiNCs SFR, are $\ga 50$~s~pix$^{-1}$. For nine SFiNCs SFRs (NGC~7822, IC~348, ONC Flank S, Serpens Main, IRAS~20050$+$2720, Sh~2-106, NGC~7160, Cep~OB3b, and Cep~A) the integration times are close to or over 100~s~pix$^{-1}$. Only one SFR, RCW~120, is limited to a short, GLIMPSE exposure of $2 \times 1.2$~s~pix$^{-1}$. For all but seven SFiNCs regions their {\it Chandra}-ACIS-I fields have full IRAC coverage; for the seven regions (IRAS~00013$+$6817, LkH$\alpha$~101, ONC~Flank~S, ONC~Flank~N, IRAS~20050$+$2720, LDN~1251B, and Cep~C) the IRAC observations cover over $80-90$\% of the {\it Chandra}-ACIS-I fields (Figure \ref{fig_img_fov}).

Our {\it Spitzer}-IRAC data reduction follows the procedures described in detail by the MYStIX study of \citet{Kuhn2013b}. Briefly, Basic Calibrated Data (BCD) products from the {\it Spitzer} Science Center's IRAC pipeline were automatically treated with the WCSmosaic IDL package developed by R. Gutermuth from the IRAC instrumental team. Starting with BCD data products, the package mosaics individual exposures while treating bright source artifacts, cosmic ray rejection, distortion correction, subpixel offsetting, and background matching \citep{Gutermuth2008}. We selected a plate scale of 0.86$\arcsec$ for the reduced IRAC mosaics, which is the native scale divided by $\sqrt{2}$. Source detection was performed on mosaicked images using the IRAF task STARFIND. 

Aperture photometry of IRAC sources was obtained using the IRAF task PHOT. The photometry was calculated in circular apertures with radius of 2, 3, 4  and 14 pixels ($1.7\arcsec$, $2.6\arcsec$, $3.5\arcsec$, $12.1\arcsec$). For all the SFiNCs IRAC sources 1 pixel wide background was adopted. There is no improvement in photometry if a 4 pixel wide background is used instead \citep{Getman2012}. Using the IRAC PSF for the [3.6] band re-sampled to a plate scale of 0.86$\arcsec$, pairs of sources are simulated with wide ranges of source separations, orientations, and flux ratios to derive the flux contribution from a nearby source within 2, 3, and 4 pixel apertures as a function of separation angle and flux ratio \citep{Kuhn2013b}. For the real sources, their photometry is derived using 2, 3, and 4 pixel source apertures, and we report photometry from the largest aperture for which the expected contamination from a nearby source is less than 5\%; larger apertures were favored for uncrowded sources and smaller apertures favored for crowded sources. Extractions from the 14 pixel ($\sim 10$ native pixel) apertures for the relatively isolated, bright, and unsaturated SFiNCs sources were employed to estimate the aperture correction values. We adopt the same zero-point IRAC magnitudes for the different cases of apertures as those used in MYStIX \citep[\S3.3 in][]{Kuhn2013b}.

\subsection{IRAC Source Catalog} \label{irac_catalog_section}

As in MYStIX, here the SFiNCs IRAC source catalog retains all point sources with the photometric signal-to-noise ratio $> 5$ in both [3.6] and [4.5] channels. This catalog covers the 22 SFiNCs SFRs and their vicinities on the sky, and comprises 1638654 IRAC sources with available photometric measurements for 100\%, 100\%, 29\%, and 23\% of these sources in the 3.6, 4.5, 5.8, and 8.0~$\mu$m bands, respectively. Over 90\% of these sources are from the extraction of the wide mosaics in/around the following six SFiNCs SFRs: Mon~R2, RCW~120, Serpens South, Sh~2-106, Cep~OB3b, and Cep~A. Table \ref{tbl_irac_sources} lists IRAC source' positions, IRAC band magnitudes and their uncertainties, and aperture size flag. The magnitude uncertainties include the statistical uncertainty, uncertainty in the calibration of the IRAC detector, and uncertainty in the aperture correction. The cases for which the contamination from a nearby source exceeds $10$\% in the smallest 2-pixel aperture are flagged as ``-1'' (last column in Table \ref{tbl_irac_sources}). Similar to MYStIX, the fractions of the SFiNCs IRAC sources with different aperture sizes are: 13\% with 4-pixel apertures, 10\% with 3-pixel apertures, and 77\% with 2-pixel apertures (44\% of the sources have flags indicating crowding). 

From the entire SFiNCs IRAC catalog we consider a sub-sample of the IRAC sources ($\sim 22$\%) that covers only the {\it Chandra} ACIS fields and their {\it immediate} vicinities (marked as red polygons in Figure \ref{fig_img_fov}) and thus harbors a significant fraction of SFiNCs YSO members; we call this the ``cut-out'' sample. Figure \ref{fig_mlf} shows the histograms of the [3.6]-band magnitude for the ``cut-out'' sample. The sensitivity of the sample depends on a number of factors: IRAC exposure time, presence/absence of a diffuse nebular background, source extinction and distance to an object of interest. For most SFiNCs SFRs, the histograms peak near $m_{3.6} \sim 17$~mag, which would translate to a mass of a lightly obscured, diskless SFiNCs YSO, located in a region with low nebulosity,  of $<0.1$~M$_{\odot}$. Thus, the {\it Spitzer}-based sample of disk-bearing PMS stars goes down to lower masses than the {\it Chandra}-based sample of PMS stars (M$\ga 0.3$~M$_{\odot}$, \S \ref{acis_catalog_section}).

The comparison between the SFiNCs IRAC catalog and the previously published {\it Spitzer} catalogs is presented in Appendix \ref{sec_appendix_flux_comparison}. Since most of the published catalogs are limited to diskbearing YSO objects, the number of IRAC sources in SFiNCs is typically by a factor of $>60$ higher than that in the earlier studies (Table \ref{tbl_new_sfincs_vs_pub_ir}). As in MYStIX \citep{Kuhn2013b}, the SFiNCs MIR photometry is in a good agreement with that of the previous MIR studies (Figure \ref{fig_comparison_of_mags}).

\subsection{X-ray/Infrared Matching} \label{source_ir_matching_section}

Due to the deterioration of the {\it Chandra} telescope PSF with the off-axis angle, the SFiNCs X-ray sources lying on the ACIS-S chips for all but one SFiNCs regions (IC~5146) were further omitted from our identification and membership analyses. For IC~5146, the X-ray data from the ObsID 6401, with the aimpoint on S3 chip, were retained for further analyses. 

Source position cross-correlations between the SFiNCs {\it Chandra} X-ray source catalog (\S \ref{acis_catalog_section}) and an IR catalog, either the ``cut-out'' IRAC (\S \ref{irac_catalog_section}) or 2MASS \citep{Skrutskie2006}, were made using the following steps. 

First, the trivial matching of close source pairs within the constant $2\arcsec$ radius was used to identify candidate matches. The $2\arcsec$ tolerance size was chosen based on the X-ray-IR source positional offsets obtained in the COUP project \citep[Figure 9 in][]{Getman2005}. 

Second, the more sophisticated matching with the {IDL} tool {\it match\_xy} \citep[\S 8 in][]{Broos2010} was applied to all the candidate matches from the previous step. The $match\_xy$ package takes into consideration the positional statistical uncertainties of individual sources, which is particularly relevant for {\it Chandra} X-ray source catalogs since the {\it Chandra} PSF significantly degrades off-axis\footnote{The X-ray source position uncertainty is a function of both the number of extracted counts and the off-axis angle. The SFiNCs X-ray random position uncertainties vary from $0.1\arcsec$ in the core to $>1\arcsec$ in the halo parts of the SFiNCs fields; whereas the 2MASS and IRAC SFiNCs source position uncertainties remain fixed across the fields at the $\la 0.1\arcsec$ and $\la 0.1-0.3\arcsec$ levels for the relatively bright ($J < 16$~mag) 2MASS and IRAC sources, respectively.}. The {\it match\_xy} tool was applied only to closest pairs, discarding multiple matches, and was run to accept matches, for which the source separations are less than 2.3 times the combined uncertainty on positions. 

Third, we performed a careful visual inspection of all source pairs with the X-ray-IR separations of $<2\arcsec$ (candidate matches from the first step) that were rejected by the {\it match\_xy} procedure during the second step. The results of this inspection suggest that for many of the rejected matches (typically 10\% out of all possible X-ray-IR matches per a SFiNCs region) their unusually large separations arise from systematic effects, such as inaccurate measurements of X-ray/IR source positions due to the presence of multiple resolved or un-resolved sources. The {\it match\_xy} tool does not account for such effects. As an example, expended views for a few of such X-ray-IR pairs are shown in Figure \ref{fig_acis_ir_matches}. Typical cases include: a cataloged double X-ray source and a single registered IR source visually recognized as a single (first panel from above); a cataloged double X-ray source and a double IR source with a single registered companion (second panel); a single registered and visually recognized X-ray source and a double IR source with a single registered companion (third panel); both X-ray and IR sources viewed and registered as single sources (forth panel); a single X-ray source near a registered IR double (fifth panel). More complex cases involving triple and multiple visual systems exist. In all of these cases, the X-ray and IR source extraction regions still highly overlap and their X-ray-IR properties are often consistent with the X-ray-IR trends seen for the source pairs accepted by {\it match\_xy}, such as the trends of the X-ray flux vs. the $J$-band magnitude and the X-ray median energy vs. the $J-H$ color. For the SFiNCs membership study, we consider to retain such pairs as legitimate X-ray-IR matches. In cases where there are double/multiple cataloged X-ray sources near a single cataloged IR source, notes on multiplicity are added for all such X-ray sources that are probable YSO members of the SFiNCs SFRs (see the YSO membership tables below), but only the X-ray source closest to the IR source is assigned as a formal counterpart to the IR source.

\subsection{Sources Of Contamination In SFiNCs} \label{source_contamination_section}

X-ray surveys of star-forming regions are subject to contamination by extragalactic active galactic nuclei (AGNs) and Galactic foreground and background stars \citep{Getman2011,Broos2013}. Unlike the MYStIX SFRs, the SFiNCs regions lie away from the Galactic plane where the field star contamination is greatly reduced. While detailed and sophisticated simulations of the X-ray contaminants for the SFiNCs fields are not feasible due to the lack of computing and manpower resources, we can evaluate the levels of the SFiNCs X-ray contaminants based on the results of the recent MYStIX simulations by \citet{Broos2013} for the two nearby MYStIX SFRs, Flame Nebula and RCW~36. Both have typical SFiNCs distances and {\it Chandra} observation exposure times \citep{Feigelson2013}: Flame Nebula ($d = 414$~pc) and RCW~36 ($d = 700$~pc) were captured in single {\it Chandra}-ACIS-I images, each with $70-80$~ks exposure time. As most of the SFiNCs SFRs, Flame Nebula is a high-Galactic latitude region ($b = -16.4\arcdeg$), whereas RCW~36 lies close to the Galactic plane ($b=+1.4\arcdeg$). As the majority of the SFiNCs SFRs, both Flame and RCW~36 are relatively young regions featuring prominent molecular cloud structures.  

Considering the results of the contamination simulations for this two SFRs \citep[Table 8 in][]{Broos2013}, one may reasonably guess the typical numbers of the X-ray contaminants within the SFiNCs {\it  Chandra} fields (per $17\arcmin \times 17\arcmin$ field): a dozen X-ray foreground stars, from a few to a dozen background stars, and roughly a hundred AGNs. The foreground stars are expected to have soft X-ray spectra ($E_{median} < 1$~keV); the background stars would have X-ray median energies similar to the bulk of PMS stars ($1 < E_{median} < 2.5$~keV); while the AGNs would have high median energies above $2-3$~keV comparable mainly to deeply embedded YSOs and protostellar objects \citep[Figure 2 in ][]{Broos2013}. A quarter of these foreground stars, more than half of the background stars, and all the AGNs are expected to be undetected by 2MASS \citep[Figure 3 in][]{Broos2013}.

Although the overwhelming contamination from unrelated Galactic field stars prohibits IR surveys from providing complete censuses of YSO populations, these surveys are known to be very effective in isolating YSOs with IR excesses. The selection of disky YSOs is subject to further contamination by star-forming galaxies and obscured AGNs, and nebular knot emission \citep{Gutermuth2009}. For SFRs close to the Galactic midplane, additional major sources of contamination include dusty asymptotic giant branch (AGB) stars and YSOs from unrelated SFRs \citep{Povich2013}. For most of the SFiNCs SFRs located away from the Galactic midplane, the contamination of {\it Spitzer}-selected disky YSO samples by AGB stars is expected to be small, no more than a few percent \citep[e.g.][and references therein]{Dunham2015}; and the contamination by unrelated YSOs to be negligible or absent.

\section{YSO Membership Of The SFiNCs SFRs} \label{membership_section}

\subsection{YSO Selection Procedure} \label{yso_selection_section}

Since the Galactic field star contamination is greatly reduced in SFiNCs, the MYStIX's complex probabilistic X-ray and IR source classifications \citep{Povich2013,Broos2013} are not needed here. Instead, the SFiNCs YSO classification can be achieved using simpler IR and X-ray classification approaches given in \citet{Gutermuth2009} and \citet{Getman2012}. The major ten steps of our membership analysis are presented below.

First, we start by applying the YSO classification scheme of \citet{Gutermuth2009} to the SFiNCs IRAC ``cut-out'' catalog. The Phase 1 of the scheme identifies and removes star-forming galaxies and broad-line AGNs, as well as knots of shocked emission; the disky YSO candidates with available photometric measurements in all 4 IRAC bands are selected. The Phase 2 analysis is further applied to identify additional disky YSO candidates with IR photometry available only in the $JHK_s$[3.6][4.5] bands. 

Second, following \citet{Getman2012}, we construct the observed infrared spectral energy distributions (SEDs) in 2MASS$+$IRAC IR bands for all the SFiNCs IRAC disky YSO candidates as well as for all the SFiNCs X-ray sources with available IR photometric measurements, and compare them to the (de)reddened median SED templates of the IC~348 PMS stellar photosphers given by \citet{Lada2006}. Figure \ref{fig_seds} shows examples of such SEDs for several SFiNCs X-ray YSOs in the OMC~2-3 SFR. According to this SED-based analysis, the four sources shown in the upper panels are diskless YSOs, the next two sources (third row) are disky Class~II YSOs, and the last two (bottom row) are disky Class~I protostars. The two X-ray Class~II YSOs (third row) are not listed in the previous disky YSO catalog of \citet{Megeath2012}. While generally in agreement with the color-color approach of \citet{Gutermuth2009} for the IRAC-selected YSOs, the SED-based method is found to be extremely useful in selecting disky (flagged in the SFiNCs membership catalog as ``DSK'') and diskless (``NOD'') X-ray YSO candidates.

Third, the X-ray sample is culled of X-ray selected YSO disky candidates with very faint MIR counterparts ([3.6]$ \ga 15.5$~mag and [4.5]$ \ga 14.5$~mag) whose spatial distribution is inconsistent with that of other YSO IR-X-ray candidates (if clustered) and/or molecular cloud cores seen in Herschel-$SPIRE$ images. Most of those sources are considered to be X-ray AGN candidates \citep{Getman2012}. Their positions on the X-ray color-magnitude diagrams are consistent with those of extragalactic background sources (black points on the X-ray color-magnitude diagrams in \S \ref{xray_bright_vs_faint_section}).

\begin{figure*}
\centering
\includegraphics[angle=0.,width=7.0in]{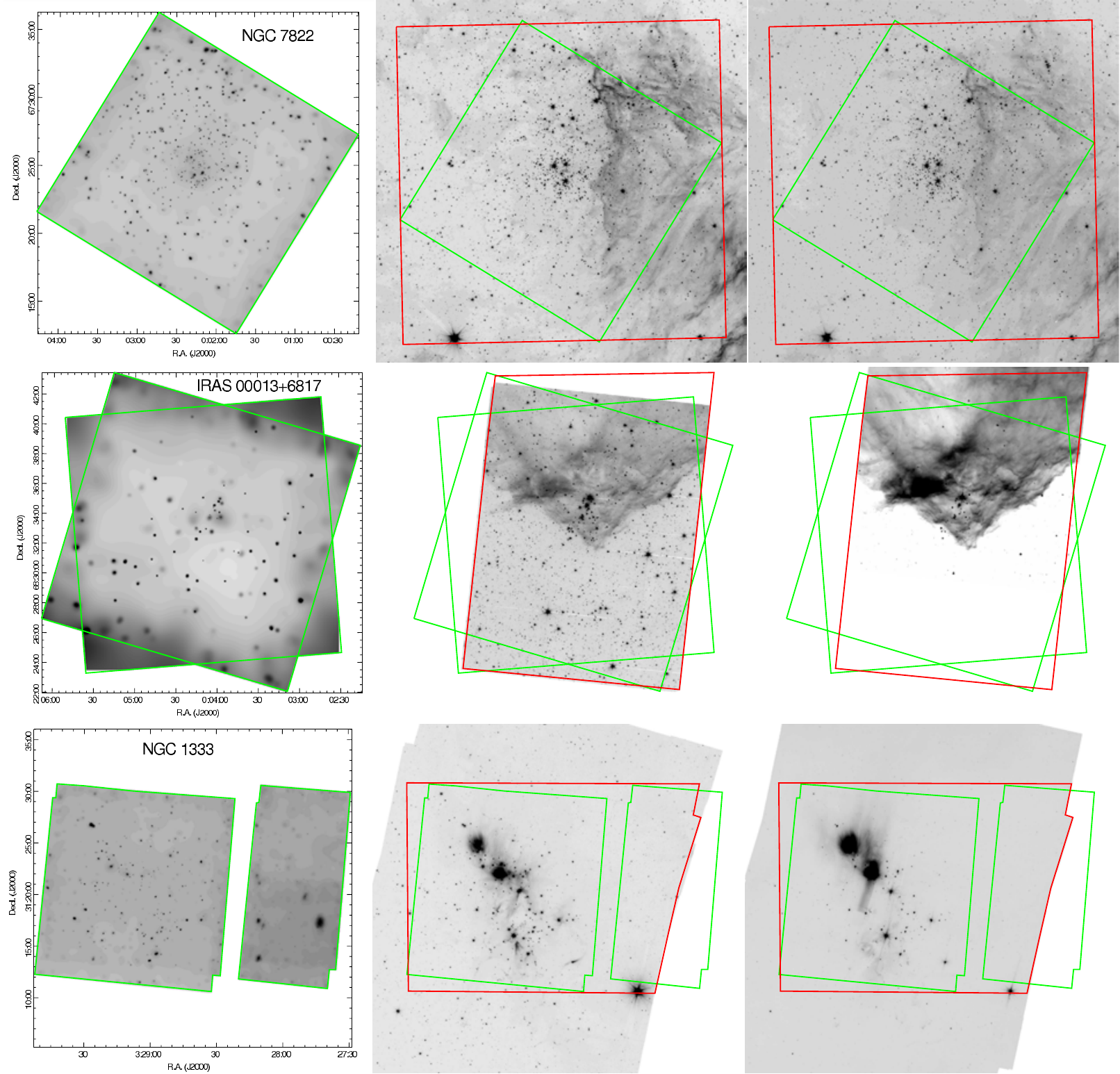}
\caption{Low-resolution images of the SFiNCs SFRs: adaptively smoothed {\it Chandra}-ACIS image in the total ($0.5-8$)~keV band (left), {\it Spitzer}-IRAC in the 3.6~$\mu$m band (middle), and {\it Spitzer}-IRAC 8.0~$\mu$m band (right). For NGC~7822, no observations were taken in the IRAC 8.0~$\mu$m band; instead, the {\it Spitzer}-IRAC image in the 4.5~$\mu$m band is shown. {\it Chandra}-ACIS field of view is outlined in green; for the reference on the angular size of the field, recall that the size of a single square {\it Chandra}-ACIS-I field is $\sim 17\arcmin \times 17\arcmin$. The field of view for the ``cut-out'' IRAC catalog (\S \ref{irac_catalog_section}) is outlined in red. \label{fig_img_fov}}
\end{figure*}

The following steps locate additional SFiNCs YSO candidates that we flag as possible members (``PMB''). Forth, the X-ray sources without IRAC (and often without 2MASS) counterparts that lie in the centers of YSO clusters/groups and are subject to significant MIR diffuse nebular background, especially for LkH$\alpha$~101, Mon~R2, RCW~120, Sh~2-106, and Cep~A (Figure \ref{fig_img_fov}), are flagged as ``PMB''. In LkH$\alpha$~101, the high diffuse background is due to the contamination from the point spread function wings and trails of the bright EM$^{\star}$~LkH$\alpha$~101 star; in the remaining four regions the high background nebular emission is likely due to heated dust in the cluster cores. Figure \ref{fig_pmb_map} shows that most of these ``PMB'' sources are relatively bright X-ray sources (outlined in cyan) that are clearly visible by eye on the X-ray images but are often missed from the MIR images since the MIR point source sensitivity is dramatically reduced by the high nebular background. Their positions on the X-ray color-magnitude diagrams (presented below in \S \ref{xray_bright_vs_faint_section}) are consistent with those of highly absorbed YSOs, AGN, or background field stars. However, since these are located at the very centers of dense stellar clusters and molecular cores (presented below as Figure \ref{fig_spcm_maps}), which subtend $<1$\% of the $17\arcmin \times 17\arcmin$ ACIS-I field, the probability of being an AGN or a background star is tiny (\S \ref{source_contamination_section}). X-ray YSO clusters/groups without IR counterparts have been previously seen in other SFRs. For instance, in MYStIX SFRs, such as NGC~3576, M~17, NGC~6357, and RCW~38 \citep[\S 2.3 and Figure 1 in][]{Broos2013} and in the Carina Complex \citep[\S 5.2.2 in][]{Townsley2011}. Fifth, the X-ray sources with 2MASS counterparts that are located outside the field of view of the SFiNCs IRAC ``cut-out'' catalog for the seven SFiNCs SFRs (IRAS~00013$+$6817, LkH$\alpha$~101, ONC~Flank~S, ONC~Flank~N, IRAS~20050$+$2720, LDN~1251B, and Cep~C in Figure \ref{fig_img_fov}) are flagged as ``PMB''. The positions of these sources on the X-ray and NIR color-color and color-magnitude diagrams are consistent with those of other SFiNCs YSO members (the diagrams are presented below in \S \ref{oir_diagrams_section}, \S \ref{xray_bright_vs_faint_section}, and \S \ref{sec_appendix_source_atlas}). Sixth, based on the cross-correlation between the source positions of the SFiNCs X-ray/IR catalogs and the catalogs of massive OB-type stars from SIMBAD and \citet{Skiff2009}, all known OB-type stars in the SFiNCs fields are identified; the non-disky IR and/or undetected in X-rays OB stars are added to the SFiNCs membership catalog and flagged as ``PMB''. Seventh, the X-ray sources that are part of close visual double/multiple X-ray systems with their companions previously identified as YSO candidates, are also flagged as ``PMB''.

Eight, based on the criteria, $E_{median} < 1$~keV and $J-H < 0.5$~mag, representing negligible interstellar absorption \citep{Getman2012}, we select foreground X-ray candidates, typically a few to several per SFiNCs field. However, these criteria are not sufficient for choosing definite foreground stars. The soft X-ray spectra and low NIR colors often pertain to intermediate and high-mass young stars. Many of these foreground candidates are found to be associated with the B-type probable members of the SFiNCs regions. For instance, 3 out of 4 foreground candidates in ONC Flank N are known B-type stars, including one of the ionizing stars of the region, B1V star c~Ori. While many of the SFiNCs foreground candidates are diskless stars, 3 out of 4 of those in ONC Flank S possess IR excesses. At this stage, we keep these objects in the SFiNCs membership database, with the suffix ``FRG'' appended to the name of their YSO class. During our final stage of the membership analysis (see below), the SFiNCs member sample is culled of a substantial number of these and additional foreground objects.

Ninth, our results on the YSO membership in SFiNCs are compared to those from the previous {\it Spitzer} studies of the SFiNCs SFRs (Column 9 in Table \ref{tbl_sfincs_sfrs}). Three major different types of inconsistencies can be noted. 1) With the exception of the Serpens South SFR, roughly several percent of sources in the compared catalogs are relatively faint MIR sources with their measurement uncertainties in either of the two catalogs (SFiNCs vs. literature) being slightly larger than those imposed in the Phase 1/2 classification scheme of \citet{Gutermuth2009}. Those that are too bright to be classified as background galaxies and pass our SED-based classification, were added to the SFiNCs membership catalog as additional disky YSOs. In the case of Serpens South, roughly 50\% of additional disky YSOs from \citet{Povich2013} that passed our SED-based analysis were added to the SFiNCs membership catalog. 2) A few percent of sources in the compared catalogs have inconsistent classifications --- transition disks in the published studies but diskless (based on our SED-based analysis) in the SFiNCs membership catalog. This is likely due to the use of additional longer wavelength MIPS data in the previous studies. For compatibility with the MYStIX data, we prefer to retain the diskless class for these sources; those that have X-ray counterparts further appear in our SFiNCs membership catalog as ``NOD'' members. 3) A few percent of sources that are identified as protostars in the previous studies but are not present in our SFiNCs IRAC catalog due to the imposed constraint on the photometric signal-to-noise ratio of $> 5$ in both [3.6] and [4.5] channels for our catalog (\S \ref{irac_catalog_section}). Such protostars are not included in our SFiNCs membership catalog.

Tenth, at this final stage, for every SFiNCs YSO member candidate a source atlas is created, similar to the one provided in the Appendix~\ref{sec_appendix_source_atlas}. For each source, the atlas conveniently collects various source's properties into a two-page digest, including the source's spatial position, X-ray/IR photometric quantities, source's IR spectral energy distribution, source's locations on X-ray/IR color-magnitude and color-color diagrams, information on the presence/absence of a counterpart from previously published YSO catalogs, etc. For each source, we perform a visual inspection of the source atlas combined with the information on source's X-ray light-curve (not shown in the atlas), on source's spectral type from the SIMBAD database (when available), and on source's parallax measurements (when available) from the Gaia-Tycho catalog \citep{GaiaCollaboration2016}. This source examination allowed us to identify and remove over 500 ambiguous member candidates. These include a hundred foreground candidates (judging mainly from their parallax measurements and positions on the color-color and color-magnitude diagrams), a hundred weak IR non-X-ray sources (mainly based on their ambiguous SED shapes), a hundred ``PMB'' sources with very weak X-ray counterparts, and two hundred weak X-ray sources, for which their X-ray median energy is inconsistent with their NIR colors. The final SFiNCs YSO member list comprises 8492 sources.

\subsection{Catalog Of SFiNCs Probable Cluster Members} \label{spcm_catalog_section}

Tables \ref{tbl_spcm_irprops} and \ref{tbl_spcm_otherprops} provide the list of 8492 SFINCs probable cluster members (SPCMs) and their main IR and X-ray properties. In this list, 66\%, 30\%, and 4\% were classified as disky (``DSK'' in column 11 of Table~\ref{tbl_spcm_otherprops}), diskless (``NOD''), and ``PMB'' members, respectively. Their reported properties include source names and positions, 2MASS NIR and {\it Spitzer}-IRAC MIR photometry, {\it Chandra}-ACIS-I X-ray net counts, median energies, incident fluxes, column densities, and intrinsic luminosities, apparent IRAC SED slopes, visual source extinctions, and stellar ages. The IRAC IR and X-ray properties of the SFiNCs SPCMs were excerpted from Tables \ref{tbl_irac_sources}, \ref{tbl_acis_src_properties}, and \ref{tbl_acis_src_fluxes}.  Also provided are: the positional flag indicating whether the SPCM source lies within the {\it Chandra}-ACIS field of view (FOV); the flag indicating an association with an OB-type star and the related notes on the basic OB properties; as well as the flag and the related notes indicating complicated X-ray-IR stellar identifications often involving double or multiple sources.

The source extinctions (Table~\ref{tbl_spcm_otherprops}, column 12) were estimated for over half of the SFiNCs SPCMs with certain IR properties, including reliable NIR photometric measurements, using the NIR color-color method described in \citet{Getman2014a}. The stellar ages (Table~\ref{tbl_spcm_otherprops}, column 13) were estimated for 22\% of the SPCMs, whose NIR and X-ray properties satisfy certain criteria discussed in \citet{Getman2014a}. This age estimator, called $AgeJX$, is based on an empirical X-ray luminosity - mass relation calibrated to well-studied Taurus PMS stars \citep{Telleschi2007} and to theoretical evolutionary tracks calculated by \citet{Siess2000}. While individual $AgeJX$ values are very noisy, median ages for stellar clusters, such as the MYStIX clusters, are reasonably precise \citep{Getman2014a}.

\subsection{Spatial Distribution Of SPCMs} \label{spcm_spatial_section}

Figure \ref{fig_spcm_maps} shows the spatial distributions of different classes of SPCMs superimposed on the far-infrared (FIR) images of the SFiNCs fields taken by $AKARI$-FIS at 160~$\mu$m (for NGC 7822, IRAS 00013+6817, IRAS 20050+2720, NGC 7160, CepOB3b) and $Herschel$-SPIRE at 500~$\mu$m (for the rest of the SFiNCs SFRs). The images trace the locations of the SFiNCs molecular clouds. Upon the inspection of this figure a few noteworthy items emerge.

First, it is important to stress here that the frequent apparent prevalence of disky populations in SFiNCs is due to the combination of two effects: an astrophysical from the youth of many SFiNCs regions; and an observational as a sample selection bias from the higher sensitivity of the {\it Spitzer}-IRAC observations (primary disky selector) to low-mass and/or extremely absorbed YSOs compared to the {\it Chandra}-ACIS observations (primary diskless selector). In the future SFiNCs studies, care must be taken to account for the unseen YSO populations using the XLF/IMF analyses, as it was done in MYStIX \citep{Kuhn2015a}.

Second, for many relatively young SFRs (e.g., IRAS~00013+6817,  LkH$\alpha$~101, ONC Flank fields, OMC~2-3, Mon~R2, GGD~12-15, Serpens Main, IRAS~20050$+$2720, LDN~1251B, Cep~A, and Cep~C), the disky SPCMs often lie projected close to the molecular structures whereas the diskless SPCMs are often more widely distributed around the clouds suggesting processes of continuous star formation or possibly several episodes of a rapid star formation followed by kinematic drifting over several million years \citep{Feigelson1996}. Some of the distributed YSOs are likely young stars that have been dynamically ejected from star-forming clouds \citep{Bate2003}. 

Third, the efficacy of the X-ray selection can be immediately noticed in cases of relatively older SFRs: for instance, the apparent number of diskless YSOs becomes comparable or higher than that of the disky YSOs in the revealed rich sub-clusters of the IC~348, ONC Flank N, and Cep~OB3b SFRs, and --- undoubtedly predominant in the oldest SFiNCs SFR NGC~7160. The X-ray selection also plays the major role in discovering stellar clusters in areas with strong IR nebular background: for instance, the central parts of the rich stellar sub-clusters in Mon~R2 and Sh~2-106 remain undetected in purely IR studies (compare Figure \ref{fig_spcm_maps} of this paper with Figure~1 of \citet{Gutermuth2009}).

\subsection{IR and Optical Color-Magnitude And Color-Color Diagrams} \label{oir_diagrams_section}

The positions of the SPCM sources in the NIR color-magnitude and color-color diagrams (Figures \ref{fig_j_vs_jh} and \ref{fig_jh_vs_hk}) are consistent with the expected loci of PMS stars, i.e., to the right (on the color-magnitude) and generally above and to the right (on the color-color diagrams) from the unabsorbed 3 and 10~Myr (cyan and magenta) PMS isochrones. Many SPCM populations are subject to a wide range of source extinction; for instance, $A_V$ changes from 0 to $>20$~mag in the NGC~1333 and NGC~2068-2071 SFRs. Some SFRs are subject to a uniform foreground absorption of $A_V \ga 2$~mag, such as Be~59, LkH$\alpha$~101, and Mon~R2. For some SFRs, such as NGC~7160 and the ONC Flank sub-regions, their SPCM populations are mainly unobscured. For the oldest SFiNCs SFR, NGC~7160, the color-magnitude positions of a relatively large fraction of diskless SPCMs are clustered around the 10~Myr PMS isohrone; this is consistent with the average age of $10-13$~Myr previously estimated for the sample of bright optical stars \citep{Sicilia-Aguilar2004,Bell2013}. For the most distant SFRs (RCW~120 and Sh~2-106) the SPCM datasets are sensitive down to only $\sim 0.8-1$~M$_{\odot}$, whereas for the nearest SFRs (NGC~1333 and IC~348) the SPCM datasets comprise many low-mass YSOs with masses $\la 0.1$~M$_{\odot}$.

Our YSO selection method (\S \ref{yso_selection_section}), although benefiting from the synergy of multiple data components (such as, X-ray and/or NIR and/or MIR photometry, and spatial distribution), is flexible enough to allow reasonably reliable YSO classification for some cases of missing data, such as non-X-ray disky YSOs, non-IR PMB YSOs, and X-ray$+$MIR YSOs with uncertain NIR data. With regards to the latter, it is important to note here that the positions of the SPCMs with uncertain NIR magnitudes (orange circles) on the NIR diagrams should be considered with caution because many of them remain undetected in the 2MASS $J$- and/or $H$-bands (typically due to high absorption and/or nebulosity) and their $J,H$-band magnitudes are often reported by 2MASS as upper limits. For instance, several disky SPCMs in OMC~2-3 (\#\# 105, 141, 420, 423, 430, 439, and 460) have abnormal positions on the NIR color-magnitude diagram (Figures \ref{fig_j_vs_jh}), to the left of the 10~Myr PMS isohrones. Such positions can not be trusted due to the uncertain NIR photometry. Nevertheless, the detailed inspection of SPCM Atlas (Appendix \ref{sec_appendix_source_atlas}) and other auxiliary data provide strong evidences for YSO nature of these sources, such as the spatial location against the OMC molecular filaments and cores, the X-ray variability and high X-ray absorption, the IR SED shapes (in the $K_s$ and IRAC bands) reminiscent of Class~II/I YSOs, and the previous YSO identification by \citet{Megeath2012}.

A few relatively bright ($M_J \sim 0$~mag) and red ($J-H > 2$~mag) outliers can be noticed on the NIR color-magnitude diagram (Figures \ref{fig_j_vs_jh}); these are the SPCM sources \#136 in LkH$\alpha$~101, \#105 in Sh~2-106, and \#103 in Serpens Main. While the first two are clearly associated with the main ionizing sources of the LkH$\alpha$~101 and Sh~2-106 SFRs (see SPCM Atlas in Appendix \ref{sec_appendix_source_atlas}), the latter is a non-X-ray source previously identified by \citet[][their source \#71]{Winston2009} as a Class~II YSO of spectral type M9. Its unusually high IR brightness for an M9 PMS star, red colors, disky SED, and spatial location at the outskirts of the main cluster (see SPCM Atlas in Appendix \ref{sec_appendix_source_atlas}) suggest that the source could be considered as a candidate AGB star unrelated to the region. A small contamination of the SPCM sample by AGBs is expected (\S \ref{source_contamination_section}).

The positions of SPCMs in the MIR color-magnitude and color-color diagrams (Figures \ref{fig_ch1_vs_ch12}, \ref{fig_ksch1_vs_ch12}, \ref{fig_ch12_vs_ch23}, and \ref{fig_ch12_vs_ch34}) generally agree with our independent IR SED classification of objects as disk-bearing (red and pink) and diskless (green) based on comparison with IC 348 stars (second stage of our membership analysis in \S \ref{yso_selection_section}). On the MIR color-magnitude diagram (Figures \ref{fig_ch1_vs_ch12}), the diskless SPCMs generally follow well the locus of the diskless IC~348 stars (cyan). On the MIR color-color diagrams, the majority of SPCMs with their disk classification are consistent with the simple color criteria that diskless stars have $K_{s} - [3.6] \la 0.6$~mag, $[3.6] - [4.5] \la 0.2$~mag, $[4.5] - [5.8] \la 0.2$~mag, and $[5.8] - [8.0] \la 0.2$~mag. Disky sources with discrepant locations on such diagrams may represent cases of transition disks; for instance, the SPCM sources \#\# 109, 231, and 258 in the well studied IC~348 SFR (see SPCM Atlas in Appendix \ref{sec_appendix_source_atlas}). 

Serpens South, likely the youngest SFiNCs SFR, harbors an exceptionally rich population of disky SPCMs that lies projected against an IR dark molecular cloud. A dozen disky SPCMs in Serpens South have anomalously blue colors, such as $[4.5]-[5.8]<0$~mag (Figure \ref{fig_ch12_vs_ch23}). These can be divided into two groups. The first group is composed of likely outflow candidate YSOs with elevated [4.5] emission from molecular shocks \citep{Povich2013}. These  Serpens South SPCMs (\#\# 158, 169, 255, 256, 379, 389, and 409) generally lie projected against the dense parts of the molecular filaments (see SPCM Atlas in Appendix \ref{sec_appendix_source_atlas}). The second group constitutes transition disk YSOs with an [8.0]-excess (\#\# 199, 281, 430, and 603) that are typically located away from the dense filaments in the areas with low background nebular emission in the [8.0]-band. Another set of Serpens South disky SPCMs (such as \#\# 115, 119, 196, 251, 258, 295, 318, 343, and 413) has abnormally blue $[5.8]-[8.0]$ colors (see SPCMs with $[5.8]-[8.0] \la 0$~mag in Figure \ref{fig_ch12_vs_ch34}). These lie projected against the dense parts of the cloud and have SED shapes characterized by a flux rise in shorter following by a flux decline in longer IRAC bands (see SPCM Atlas). These are likely protostellar objects with the flux in the [8.0]-band strongly affected by an absorption from protostellar envelope, in the silicate band centered near 9.7~$\mu$m \citep[Figure 7a in][]{Povich2013}.

Optical photometric data are generally not as useful for identifying YSOs as IR/X-ray data. They often suffer the problems of source crowding and background nebulosity and are biased towards unobscured YSOs. Nevertheless, Figure~\ref{fig_z_vs_iz} shows the optical color-magnitude diagrams in the $i$ and $z$-bands for several SFiNCs SFRs with available SDSS  coverage \citep{Alam2015}. For all these SFRs, the SDSS coverage is only partial due to the presence of diffuse nebula background from heated gas. On these diagrams, the positions of SPCMs are generally consistent with the expected PMS loci, i.e., to the right from the 10 and 100~Myr (magenta and blue) PMS isochrones. For some disky SPCMs, their location to the left from the $10-100$~Myr PMS isochrones is inconsistent with their status of disky YSOs. This may point to discrepant optical magnitudes; for instance, some may possess disks at high inclination where the optical light is enhanced by scattering above the disk \citep{Guarcello2010}. It is interesting to note that for the NGC~2068-2071 SFR, both the NIR and optical color-magnitude diagrams point to the presence of an older, perhaps foreground, PMS population.

In all diagrams above, the loci of the previously identified YSOs (blue $+$s), when available, are in agreement with the loci of SPCMs. The more detailed comparison with the previous YSO catalogs is given in \S \ref{sfincs_vs_published_section}.

\subsection{Global Properties Of SPCMs} \label{spcm_global_section}

For the entire SFiNCs SPCM sample, Figure \ref{fig_global} (upper panels) shows a few important independently measured basic MIR, NIR, and X-ray source characteristics ($JHKs$ and [3.6] band magnitudes, X-ray net counts, and X-ray median energy), as well as derived using those and other quantities, X-ray luminosity, visual source extinction, absolute $J$-band magnitude, and stellar age. The relationships seen among these properties are in agreement with those of the MYStIX (bottom panels in Figure \ref{fig_global}) and YSO samples obtained in previous studies of SFRs and thus lend confidence in our data and membership methods and the resulted SFiNCs SPCM catalog. 

Specifically, panels (a) and (b) show bimodal distributions of the SFiNCs and MYStIX apparent IRAC SED slopes with peaks at around $\alpha_{IRAC} \sim -1.3$ and  $\alpha_{IRAC} \sim -2.7$ corresponding to disky and diskless star populations, respectively. Such bimodality was reported in the PMS populations of other SFRs, for instance, Cha~I cloud \citep{Luhman2008} and $\sigma$~Ori \citep{Hernandez2007}. This is consistent with the disk classification scheme of \citet{Lada1987}. 

But some outliers are present on these graphs. For instance, a few SPCMs classified as diskless have an unusually high slope, $\alpha_{IRAC} > -1$. These include sources \#\# 109, 118 in Mon~R2, \#\# 108, 144 in RCW~120, \# 918 in Cep~OB3b and \# 155 in OMC~2-3. All these sources are located in regions that are subject to a moderate MIR nebular background emission. All lack photometry measurements in the [8.0] band; the two sources in Mon~R2 lack photometry in both the [5.8] and [8.0] bands. The shapes of their IR SEDs in all but a single, longest IRAC band are consistent with stellar photospheres (see SPCM Atlas in Appendix \S \ref{sec_appendix_source_atlas}).  This marginal IR-excess can be attributed either to their physical evolutionary state as transition disk objects or to the problems in photometry measurements (inaccurate photometry in the longest band due to the prevalence of nebular background). We chose the latter case and classified these objects as diskless. At the other extreme, over two hundred disky SPCMs have a relatively lower slope, $\alpha_{IRAC} < -2$. A characteristic feature of their SEDs is a slight but consistent deviation of the SED shapes from stellar photospheres in two or more IR bands. The vast majority of these sources have been identified as disky YSOs in previous studies of the SFiNCs SFRs. For instance, a large number of such SPCMs is present in the NGC~2068-2071 SFR with the majority identified as disky by \citet{Megeath2012}: \#\# 50, 65, 75, 87, 112, 160, 255, 258, 270, 282, 299, and 302 (see SPCM Atlas in Appendix \S \ref{sec_appendix_source_atlas}).

On panel (a), at the faint end of [3.6] ([3.6]$ \ga 15.5$~mag), the points correspond to disky YSOs that are seen lying projected against the SFiNCs molecular structures; $76$\% of these faint, disky YSOs are associated with the very young Serpens South SFR (Figure \ref{fig_spcm_maps}). At the bright end of [3.6] ([3.6]$>7$~mag), many SFiNCs YSOs are identified as known OB-type stars; but a dozen more disky YSOs lie within this OB locus. The nearby OMC~2-3 SFR harbors the largest number of such objects (seven); these are SPCM sources \#\# 107, 115, 165, 170, 201, 438, and 441. Our visual inspection of the source's properties given in SPCM Atlas (Appendix \S \ref{sec_appendix_source_atlas}) suggests that all these objects lie projected against the OMC molecular filament. One half of them has IR SEDs that show a precipitous flux rise with increasing IRAC wavelength (characteristic of Class~I objects); the other half is characterized by shallower IRAC slopes (characteristic of Class~II objects). For most of them SIMBAD lists a class of ``Variable Star of Orion Type''; and for the Class~II objects, SIMBAD provides a spectral type of early K-type stars. The most distant SFINCs SFR, Sh~2-106, harbors three of such objects (\#\# 67, 231, 234); all listed as disky YSOs in previously published studies of the region. SPCM Atlas shows that these lie projected close to the center of the primary stellar cluster in Sh~2-106. While spectral types are not available for these objects, their bright IR magnitudes, spatial location close the giant molecular clumps, and steep IRAC slopes point towards the extreme youth and possibly high masses of these objects, especially for the X-ray emitting Class~I protostar \# 67. However, in the case of Sh~2-106 (an SFR close to the Galactic plane), where YSOs are observed against field population of AGB stars, a misclassification of an AGB as a bright IR YSO is also possible (\S \ref{source_contamination_section}). Specifically, the SPCM source \# 234 whose SED shape is characterized by a sharp rise of flux through the NIR bands followed by a flux flattening through IRAC bands might be also considered as an AGB stellar candidate with dust-rich winds.

The NIR $J-H$ color and X-ray median energy of PMS stars are excellent surrogates for line-of-sight obscuration by dust and gas, respectively \citep[e.g.,][]{Vuong2003,Getman2011}. Panels (c) and (d) show that the dust-to-gas absorption relationships for the SFiNCs and MYStIX SFRs look very similar. The SFiNCs SPCMs without disk classifications (``PMB''), many of which are found at the centers of the SFiNCs stellar clusters affected by MIR nebular emission, have gas-to-dust ratios similar to those of other SPCMs. Since the $J-H$ color is also a surrogate for spectral type, its values for many known lightly/moderately absorbed OB-type members are shifted downwards with respect to the PMS locus. In their MYStIX science study of new protostellar objects in the MYStIX regions, \citet{Romine2016} use X-ray median energy $>4.5$~keV as a strict criterion to discriminate Class~I protostars from Class~II-III systems. 
  
It is well known that PMS X-ray luminosities strongly correlate with stellar mass and bolometric luminosity. For instance, clear relationships are seen for YSOs in the Orion Nebula and Taurus SFRs \citep{Preibisch2005,Telleschi2007}. The astrophysical cause of this relationship is poorly understood but could be related to the scaling of X-ray luminosity with surface area and/or stellar convective volume. The spread and the slope of this relationship are also subject to the variability, accretion, X-ray saturation, and age effects \citep{Getman2014a}. NIR magnitudes, such as $J$ or $K$-band, and X-ray net counts of PMS stars are good surrogates for bolometric luminosity (and mass) and X-ray luminosity, respectively. Panels (e) and (f) show the correlation of these apparent quantities, $K$-band magnitude versus X-ray net counts, whereas panels (g) and (h) present the correlation between the intrinsic $J$-band and X-ray luminosities. The statistical significance of these correlations were evaluated by testing the null hypothesis that the Kendall's $\tau$ coefficient is equal to zero. This test was made using the {\it corr.test} program from the {\it R} statistical software system\footnote{The {\it corr.test} tool is part of the {\it psych} package. The description of the package is available on-line at \url{http://personality-project.org/r/psych/psych-manual.pdf}.}. The test shows statistically significant correlations for the data presented in all four panels, with the Kendall's $\tau$ $p$-values of $<0.0001$ for any of the panels. The observed $\tau$ values are $-0.42$, $-0.36$, $-0.46$, and $-0.38$ for the panels (e), (f), (g), and (h), respectively. Both the $K_{s}$-$\log(NC)$ and $M_{J}$-$\log(L_{tc})$ correlations are expressions of the same $\log(L_{bol})$-$\log(L_{X})$ relationship. From these plots, it is clearly seen that the OB-type stars do not follow this PMS relationship, in part due to the difference in X-ray production mechanisms. On the other hand, the location of the ``PMB'' YSOs on these plots are consistent with that of the dikless and disky PMS stars. On panels (g) and (h), the SFiNCs X-ray luminosities are systematically below the MYStIX luminosities due to closer distances with similar {\it Chandra} exposures.

Due to the depletion of molecular clouds and stellar kinematic drift, older YSOs are expected to exhibit lower interstellar absorptions than younger YSOs. \citet{Getman2014a} find that the stellar ages of the MYStIX YSOs are anticorrelated  with the source extinction $A_V$ (panel j here and Figure~4 in Getman et al.). Similar relationship was found for the YSOs in the Rosette SFR \citep{Ybarra2013}. And again, similar $AgeJX - A_V$ relationship is seen for the YSOs in the SFiNCs SFRs (panel i). The SFiNCs disky YSOs are found to be on average younger than the SFiNCs diskless YSOs.

\subsection{Comparison Between Bright And Faint X-ray SPCMs} \label{xray_bright_vs_faint_section}

As it is mentioned in \S \ref{acis_catalog_section}, our data reduction procedures produce very sensitive X-ray source catalogs that are undoubtedly subject to contamination by faint spurious X-ray sources. In this section we demonstrate that the great majority of faint SPCMs are not spurious but are real YSOs.

Since the YSO selection for the vast majority of faint X-ray SPCMs relies on the presence of an IR counterpart (\S \ref{yso_selection_section}), it is important to verify that the faint X-ray SPCMs are not background fluctuations with spurious IR matches. The major steps of our test analysis here are as follows.

First, the X-ray SPCM source sample is divided into the bright and faint X-ray sub-samples, as it is shown on the X-ray flux versus median energy diagram (analogous to an IR color-magnitude diagram; Figure \ref{fig_fx_vs_me}a). For the SFiNCs SFRs with rich X-ray SPCM populations, the X-ray flux threshold for separating bright and faint X-ray SPCMs is set at the photometric flux level of $\log(PF_{lim}) = -6.4$ photons~cm$^{-2}$~s$^{-1}$; but for the sparser populations this threshold is raised to accumulate at least $\sim 20$ SPCMs (if possible) within the faint sub-sample. The entire sample of faint X-ray SPCMs across 22 SFiNCs SFRs comprises 889 sources. Disregarding the faint ``PMB'' SPCMs located in the cluster centers of Mon~R2, RCW~120, Sh~2-106, LkH$\alpha$, and Cep~A (about 50 sources in Figure \ref{fig_fx_vs_me}a), which were shown to be likely cluster members using the independent analysis in \S \ref{yso_selection_section}, among the remaining faint X-ray SPCMs 94\% and 99\% have 2MASS and IRAC counterparts, respectively. It is important to note here that the non-SPCM sample (shown in black) is expected to be composed of different classes of X-ray sources, including AGN, stellar background/foreground contaminants, as well as spurious X-ray sources.

Second, simulated sets of spurious ACIS-2MASS matches are constructed via 100 Monter-Carlo draws by randomly shifting the positions of all real X-ray ACIS-I sources (including both SPCM and non-SPCM X-ray sources) within $r = [20-50]\arcsec$ distance of the true source's positions and by applying the trivial matching of source pairs within the constant $2\arcsec$ radius (first step in \S \ref{source_ir_matching_section}). Across the 100 simulated sets, the faint X-ray sources (with their X-ray fluxes below $PF_{lim}$) that appear to have ``spurious'' 2MASS matches in such simulations are combined into the ``simulated faint'' source sub-sample. The cumulative distributions of the X-ray median energy (an excellent surrogate for the line-of-sight obscuration) are further compared among different source sub-samples.

In Figure \ref{fig_fx_vs_me}b, the cumulative distributions of the X-ray median energy ($ME$) are compared among the real X-ray SPCMs (bright and faint given in green color), non-SPCMs (black), and simulated faint X-ray sources with spurious 2MASS matches (orange). There are two independent supporting lines of evidence suggesting that the vast majority of the faint SPCMs are not spurious but true YSO sources. 

First, judging from the source numbers (provided in the figure legends), the numbers of the expected (simulated) faint sources with spurious matches typically comprise only several percent of the real faint SPCM sources. For instance, in the case of Be59, we expect $178/100$ spurious sources among 29 faint SPCMs; that is only 6\%. Across the entire set of 22 SFiNCs SFRs, the median fraction of spurious sources among the faint SPCMs is only 7\% with interquartile range of $5-10$\%. 

Second, it is clearly seen that the $ME$ distributions of the simulated faint sources with spurious X-ray-IR matches are similar to those of the real X-ray non-SPCM sources; whereas the $ME$ distributions of the faint SPCMs are found to be significantly lower and often (but not always) consistent with those of the bright SPCMs. For the cases when the $ME$ distributions of the faint SPCMs are inconsistent with (systematically higher than) those of bright SPCMs, such as, OMC~2-3, Mon~R2, Serpens South, IRAS~20050+2720, IC~5146, and Cep~A, an additional analysis comparing the distributions of right ascension and declination for the same source samples was performed (graphs are not shown). The results of this analysis indicate that the spatial distributions of both the faint and bright SPCMs are inconsistent with the relatively uniform distributions of non-SPCMs, but are either consistent with each other or the fainter SPCMs are found to be more clustered in/around the SFiNCs molecular clouds, suggesting that the fainter SPCMs are more absorbed YSOs. 

Considering that all faint SPCMs have passed through our rigorous YSO selection (\S \ref{yso_selection_section}), we believe that the vast majority ($\ga 90$\%) of the faint X-ray SPCMs are real YSOs.

\section{Comparison With Previously Published YSO Catalogs} \label{sfincs_vs_published_section}

The SFiNCs SPCM catalog can be compared to the YSO catalogs independently derived in earlier published studies. The previously published YSO catalogs are abbreviated here as Pub.  Here we are not principally interested in evaluating properties of individual YSOs that are uncommon between the SPCM and Pub catalogs (properties of individual unique SPCMs can be found in the SPCM source atlas, \S \ref{sec_appendix_source_atlas}). Our principal interest here is the differences between the total numbers, spatial distributions, and IR magnitudes of the entire SPCM and Pub YSO populations.

For this analysis we choose previous YSO studies from the literature that clearly define membership lists and include YSO selection using MIR {\it Spitzer} data, often with the addition of X-ray {\it Chandra} and optical data. Summary of such studies is given in Table \ref{tbl_spcm_vs_previous}. 

Table \ref{tbl_spcm_vs_previous} shows that previous {\it Spitzer} and {\it Chandra} YSO catalogs are available for 20 and 12 SFiNCs SFRs, respectively (see Column 5 in the table). For ten SFiNCs SFRs (Be~59, SFO~2, NGC~2068-2071, GGD12-15, RCW~120, Serpens South, IC~5146, NGC~7160, LDN~1251B, Cep~C) X-ray YSO catalogs are published here for the first time. For RCW~120 and Be~59, both MIR-{\it Spitzer} and ACIS-{\it Chandra} YSO catalogs are published here for the first time.

Column 8 in Table \ref{tbl_spcm_vs_previous} gives the contribution of SPCM to the increase in the census of YSOs in the SFiNCs SFRs. This number ranges widely: several percent in IC~348, Cep~OB3b, and ONC Flank S; from $\sim 30$\% to $80$\% for the majority of the regions; to more than 100\% in LDN~1251B, Cep~A, Be~59, and RCW~120.  The total census increase in 20 SFiNCs SFRs (disregarding Be~59 and RCW~12), relative to the previously published {\it Spitzer} and {\it Chandra} YSO catalogs, is 26\%. The total census increase for all 22 SFiNCs SFRs (including Be~59 and RCW~120) is 40\%.

The Figure \ref{fig_spcm_pub_maps} shows the spatial distributions of all SPCM and Pub YSOs. For most of the SFiNCs SFRs, the SPCM and Pub surveys have comparable spatial coverage sizes (Column 7 in Table \ref{tbl_spcm_vs_previous}). The SPCM field of view is significantly larger for Mon~R2 and Cep~A, and is noticeably smaller for NGC~1333, NGC~7160, and Cep~OB3b. The spatial distributions of SPCMs are generally consistent with those of Pubs. Significant differences in YSO distributions are readily noticed for the SFiNCs SFRs that lack Pub X-ray YSOs, such as SFO2, NGC~2068-2071, GGD~12-15, IC~5146, NGC~7160, LDN~1251B, and Cep~C. In these SFRs, many diskless SPCMs, which often lie projected outside the SFiNCs clouds, represent newly discovered YSO populations.

While Figures \ref{fig_j_vs_jh} through \ref{fig_ch12_vs_ch34} present IR magnitudes and colors for all SPCMs, with and without Pub counterparts, Figure \ref{fig_ch1_vs_ch12_nomatch} shows MIR color-magnitude diagrams for the YSOs that are uncommon between the SPCM and Pub catalogs. These color-color and color-magnitude diagrams suggest that the vast majority of the unique SPCM and Pub YSOs are relatively faint MIR sources, possibly either lower mass and/or higher absorbed members of the regions. In several  exceptional cases (NGC~1333, IC~348, sub-regions in Orion, Serpens Main, NGC~7160, and Cep~OB3b), some of the unique Pub YSOs are relatively bright MIR sources; these are located mainly outside the boundaries of the SPCM-ACIS fields.

\section{Limitations And Advantages Of The SFiNCs SPCM Sample} \label{sfincs_limitations}

The SFiNCs SPCM sample is prone to the following known limitations. \begin{enumerate}

\item As in MYStIX, here we omitted the usage of the MIPS data \S \ref{yso_selection_section}. This leads to the inability to identify some fraction of protostellar objects, especially those that lack [3.6] and [4.5] measurements. This also results in the loss of some fraction of transition disk objects, especially those that were not detected in X-rays.

\item While extragalactic background objects and the bulk of foreground stars were successfully removed from the SPCM sample, the X-ray part of the sample is still subject to contamination from Galactic field stars, mainly background stars.  It is hard to identify background stars since their IR and X-ray properties considerably overlap with those of YSOs \citep{Broos2013}. Nevertheless, based on the results of the contamination simulations given in \S \ref{source_contamination_section}, one can reasonably guess that the typical number of expected Galactic field X-ray stars with 2MASS counterparts should not exceed several-10 stars per $17\arcmin \times 17\arcmin$ field. The median number of SPCMs per $17\arcmin \times 17\arcmin$ SFiNCs field is 250 YSOs with interquartile range $[165-400]$ YSOs (Table \ref{tbl_spcm_irprops} or Figure \ref{fig_spcm_maps}). A few to several foreground stellar contaminants per $17\arcmin \times 17\arcmin$ SFiNCs field have already been identified and removed (\S \ref{yso_selection_section}). This suggests that less than a few percent of the SFiNCs SPCMs could be Galactic field contaminants.

\item The IRAC selected sub-sample of the SFiNCs SPCMs is generally deeper towards lower-mass YSOs than the ACIS selected sub-sample (\S \ref{irac_catalog_section}). In the future SFiNCs studies, care must be taken to account for the unseen YSO populations using the XLF/IMF analyses similar to those of \citet{Kuhn2015a}. 

\end{enumerate}

Compared to the previously published YSO samples for the SFiNCs SFRs, our SFiNCs SPCM sample has the following two main advantages.

\begin{enumerate}

\item SFiNCs offers a consistent treatment of IR and X-ray datasets across 22 SFRs (\S\S \ref{data_section}, \ref{membership_section}). It provides a uniform and comprehensive database of SFiNCs cluster members and their properties, which is suitable for further comparison across all SFiNCs SFRs and with MYStIX. 

\item As in MYStIX, the SFiNCs samples are advantageous by inclusion of YSOs both with and without disks. For ten SFiNCs SFRs the lists of X-ray YSOs are published here for the first time (\S \ref{sfincs_vs_published_section}). These include a thousand new diskless PMS stars. For two of these regions (RCW~120 and Be~59), both MIR-{\it Spitzer} and ACIS-{\it Chandra} YSO catalogs are published here for the first time.

\end{enumerate}

\section{Summary} \label{summary_sec}

This paper presents the homogeneous data and YSO membership analyses, and tabulated results for a large number of {\it Chandra}-ACIS and {\it Spitzer}-IRAC sources across the 22 nearby SFiNCs SFRs, for further comparison with our earlier MYStIX survey of richer and more distant regions.

The MYStIX-based data reduction and catalog production methods were applied to the sixty five {\it Chandra}-ACIS and four hundred and twenty three {\it Spitzer}-IRAC observations of the SFiNCs SFRs, resulting in the tables of the source properties for over 15300 X-ray and 1630000 IR point sources (\S \ref{data_section} and Tables \ref{tbl_acis_src_properties}, \ref{tbl_acis_src_fluxes}, and \ref{tbl_irac_sources}).

Unlike the MYStIX SFRs, most of the SFiNCs SFRs are high Galactic latitude regions, where the Galactic field star contamination is greatly reduced. This allowed us to use simpler (than MYStIX) IR and X-ray YSO classification schemes based on the approaches of \citet{Gutermuth2009} and \citet{Getman2012}. These classifications yield 8492 SFiNCs probable cluster members (SPCMs). The properties of the SPCMs are reported here in the form of the tables (\S \ref{membership_section}, Tables \ref{tbl_spcm_irprops} and \ref{tbl_spcm_otherprops}), the maps of sources distributions (Figure \ref{fig_spcm_maps}), and the visual atlas with various source's characteristics (Appendix \ref{sec_appendix_source_atlas}). Due to both the closer distances and reduced field star crowding, the 2MASS survey provides sufficient depth and resolution for SFiNCs PMS stars, whereas the deeper and higher resolution UKIDSS survey was often needed for MYStIX. Comparison with the previously published {\it Spitzer} and {\it Chandra} YSO catalogs shows that the SPCM list increases the census of the IR/X-ray member populations by 6--200\% for individual SFRs and by $40$\% for the merged sample of all 22 SFiNCs SFRs (\S \ref{sfincs_vs_published_section}).

The uniform and comprehensive database of SFiNCs probable cluster members serves as a foundation for various future SFiNCs/MYStIX-related studies, including such topics as identification and apparent properties of SFiNCs subclusters, age and disk fraction gradients, intrinsic physical properties of the SFiNCs subclusters and their comparison with those of MYStIX, star formation histories in the SFiNCs and MYStIX SFRs, and dynamical modeling of the SFiNCs/MYStIX subclusters.

\acknowledgements

We thank the anonymous referee for his time and many useful comments that improved this work. We thank Kevin Luhman for assistance with data analysis. The SFiNCs project is supported at Penn State by NASA grant NNX15AF42G, {\it Chandra} GO grant SAO AR5-16001X, {\it Chandra} GO grant GO2-13012X, {\it Chandra} GO grant GO3-14004X, {\it Chandra} GO grant GO4-15013X, and the {\it Chandra} ACIS Team contract SV474018 (G. Garmire \& L. Townsley, Principal Investigators), issued by the {\it Chandra} X-ray Center, which is operated by the Smithsonian Astrophysical Observatory for and on behalf of NASA under contract NAS8-03060. The Guaranteed Time Observations (GTO) included here were selected by the ACIS Instrument Principal Investigator, Gordon P. Garmire, of the Huntingdon Institute for X-ray Astronomy, LLC, which is under contract to the Smithsonian Astrophysical Observatory; Contract SV2-82024. This research made use of data products from the {\it Chandra} Data Archive and the {\it Spitzer Space Telescope}, which is operated by the Jet Propulsion Laboratory (California Institute of Technology) under a contract with NASA. This research used data products from the Two Micron All Sky Survey, which is a joint project of the University of Massachusetts and the Infrared Processing and Analysis Center/California Institute of Technology, funded by the National Aeronautics and Space Administration and the National Science Foundation. This research has also made use of NASA's Astrophysics Data System Bibliographic Services and SAOImage DS9 software developed by Smithsonian Astrophysical Observatory, and the SIMBAD database (operated at CDS, Strasbourg, France).

\begin{figure}
\centering
\includegraphics[angle=0.,width=7.0in]{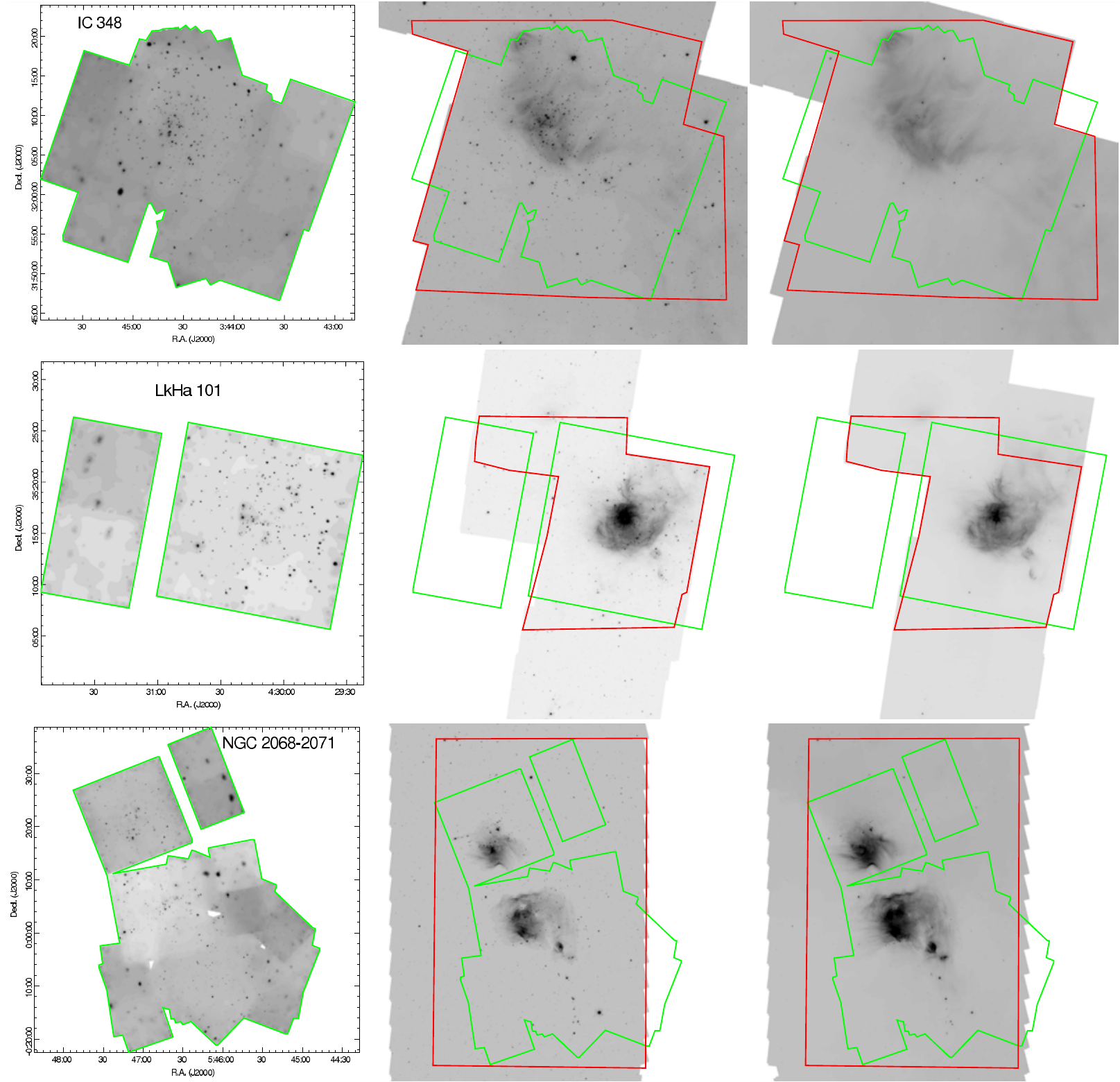}
\end{figure}

\begin{figure*}
\centering
\includegraphics[angle=0.,width=7.0in]{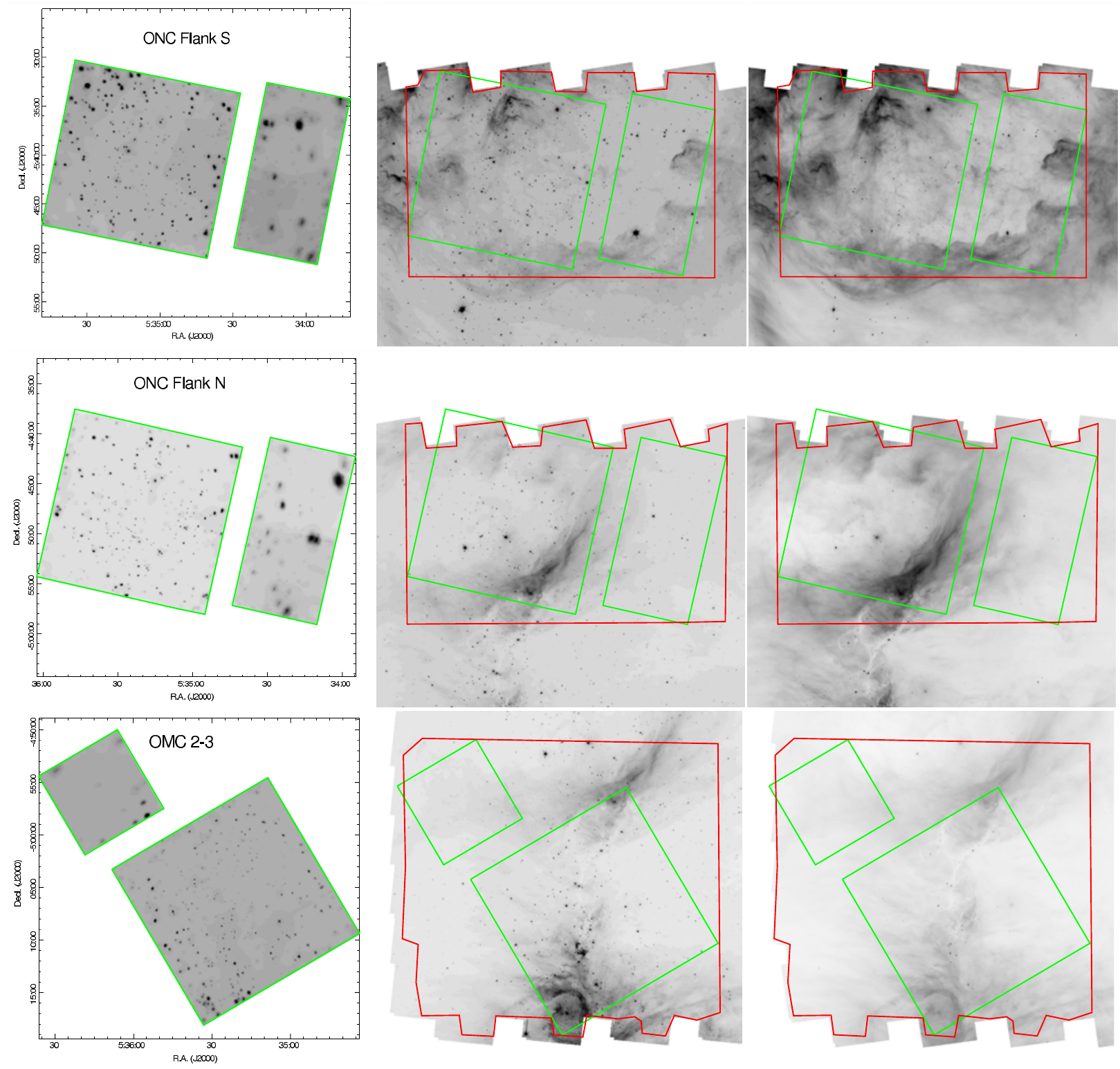}
\end{figure*}

\begin{figure*}
\centering
\includegraphics[angle=0.,width=7.0in]{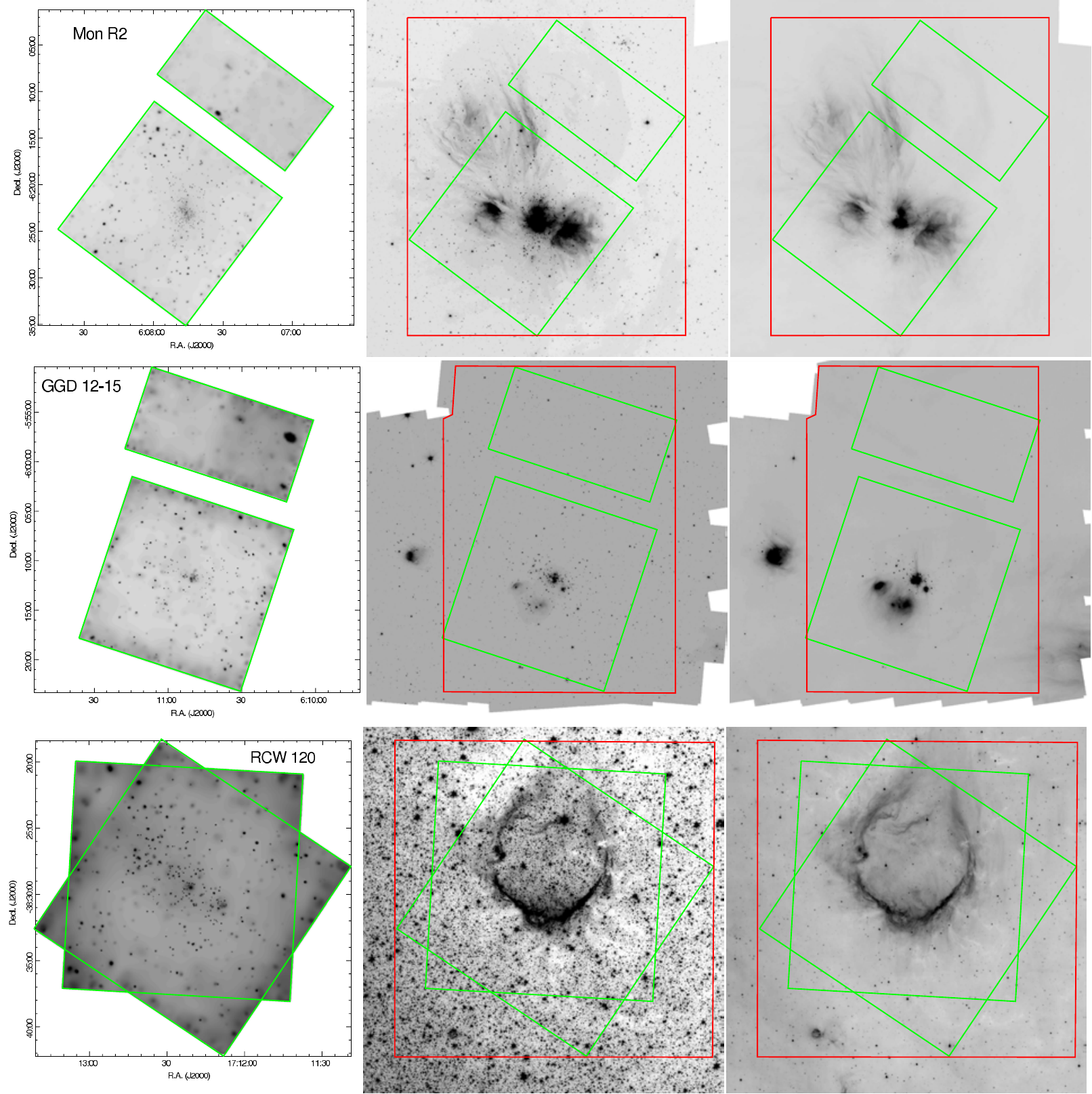}
\end{figure*}

\begin{figure*}
\centering
\includegraphics[angle=0.,width=7.0in]{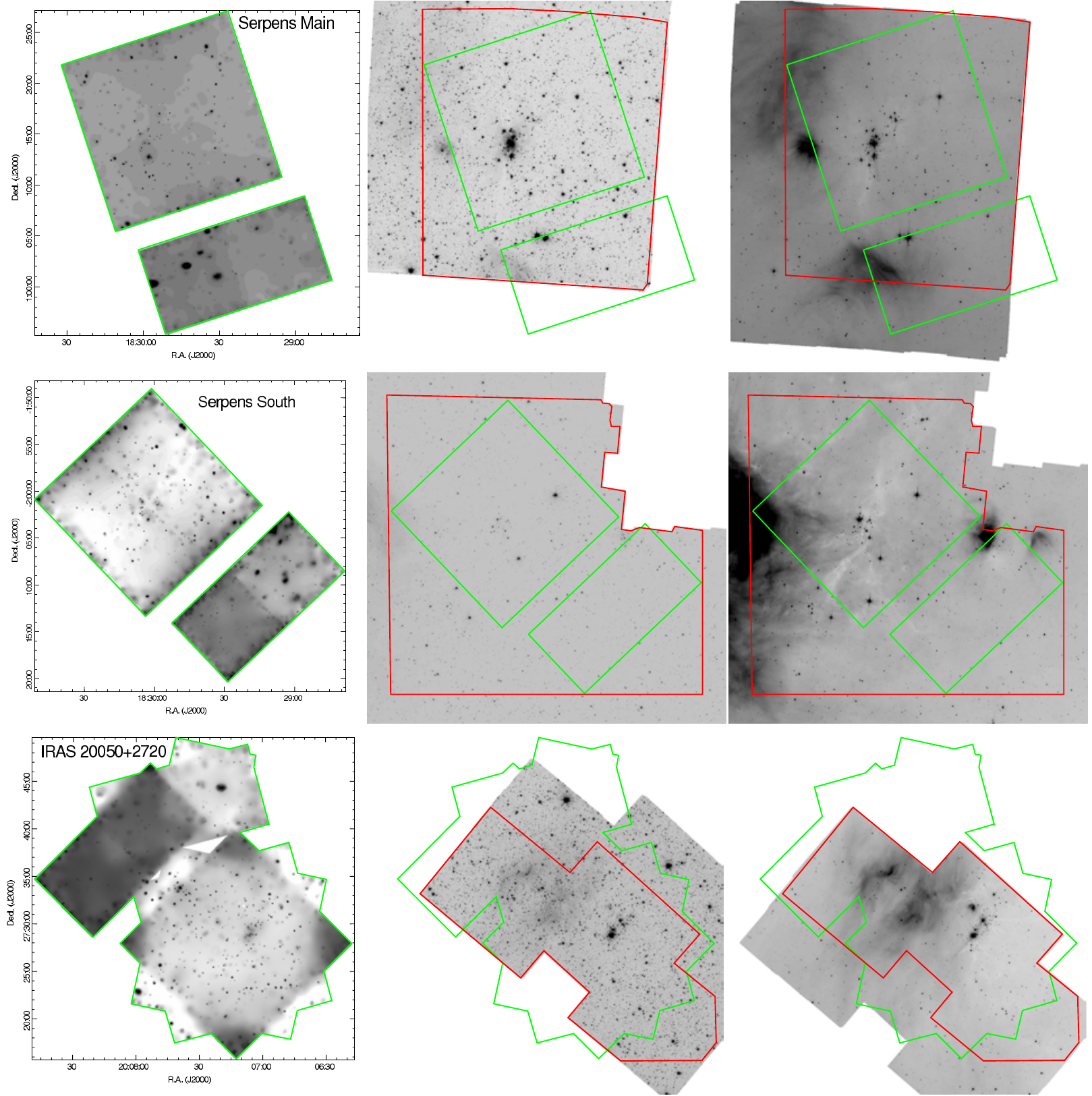}
\end{figure*}

\begin{figure*}
\centering
\includegraphics[angle=0.,width=7.0in]{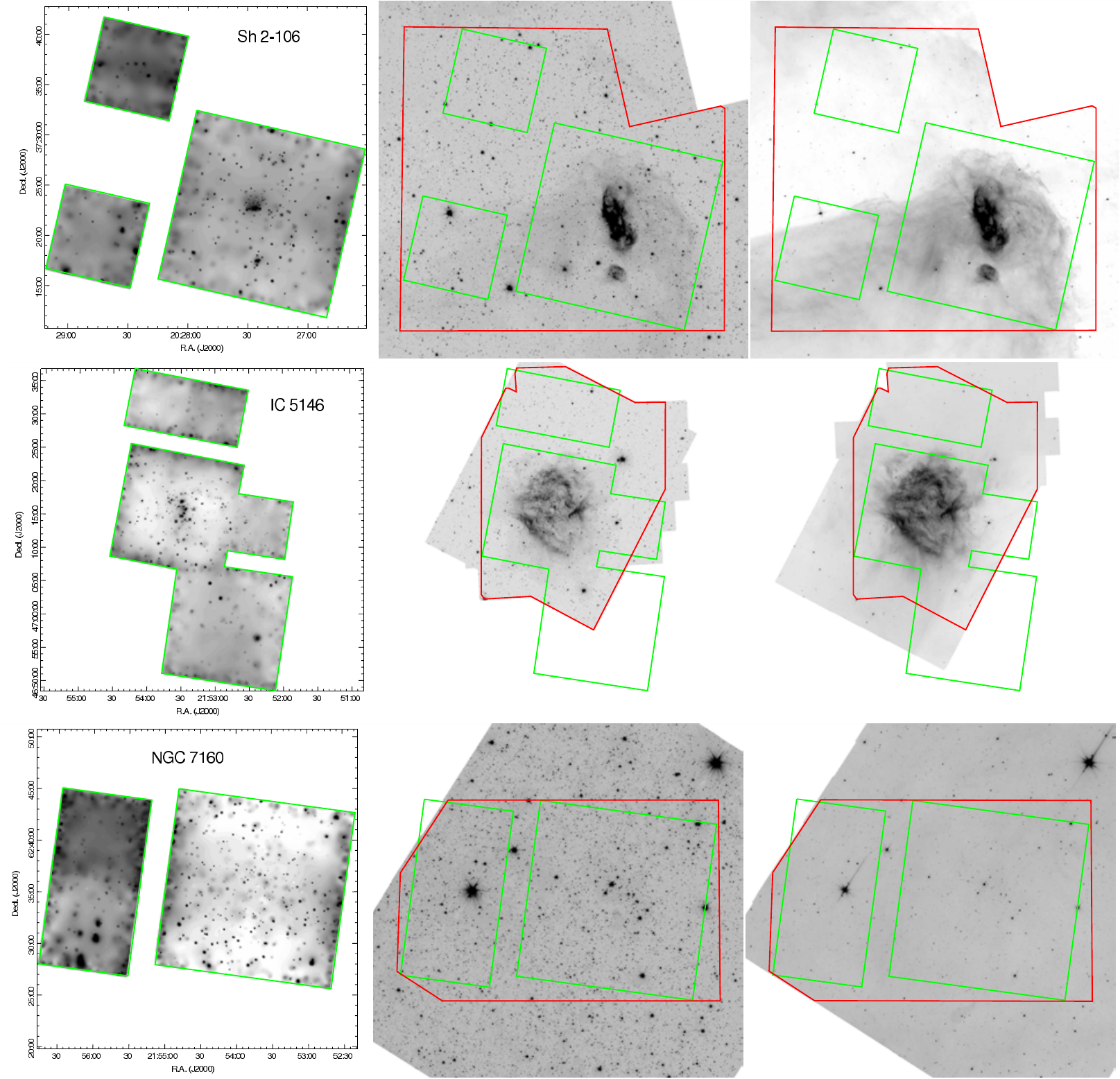}
\end{figure*}

\begin{figure*}
\centering
\includegraphics[angle=0.,width=7.0in]{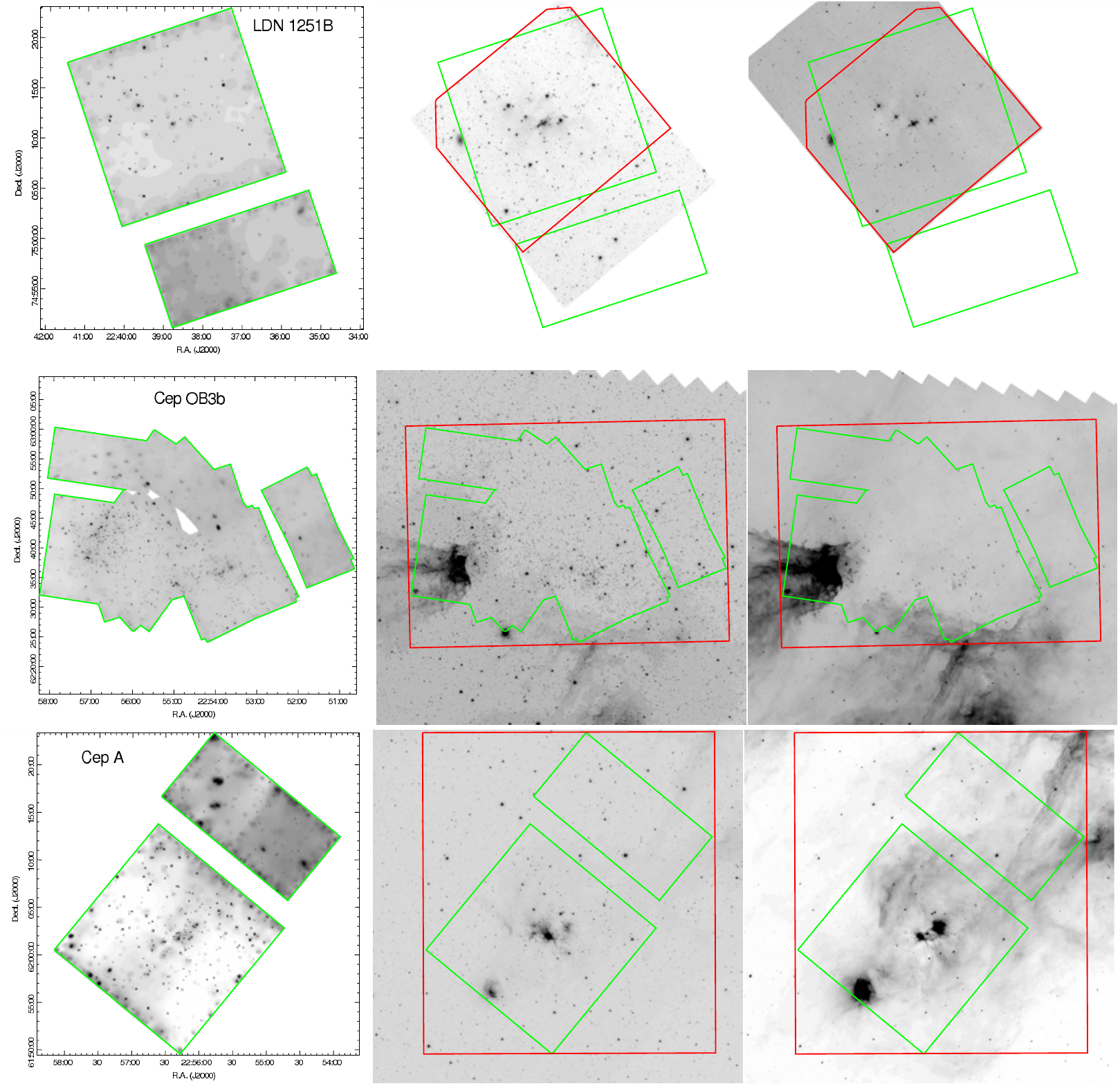}
\end{figure*}

\begin{figure*}
\centering
\includegraphics[angle=0.,width=7.0in]{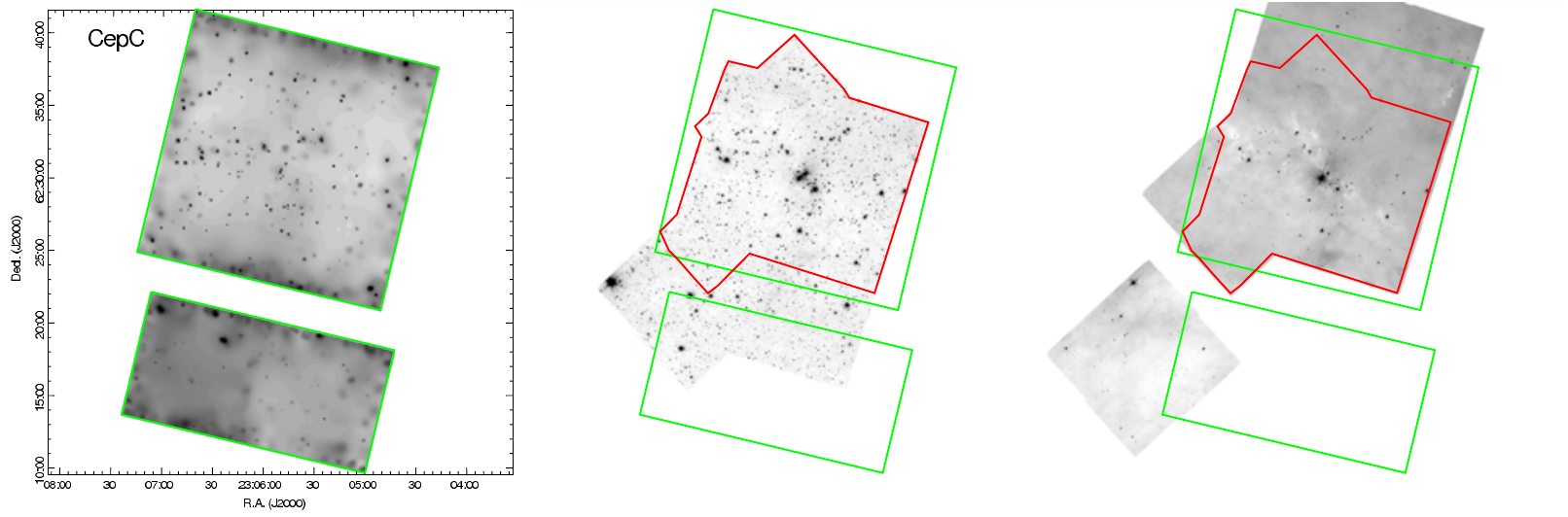}
\end{figure*}

\begin{figure*}
\centering
\includegraphics[angle=0.,width=7.0in]{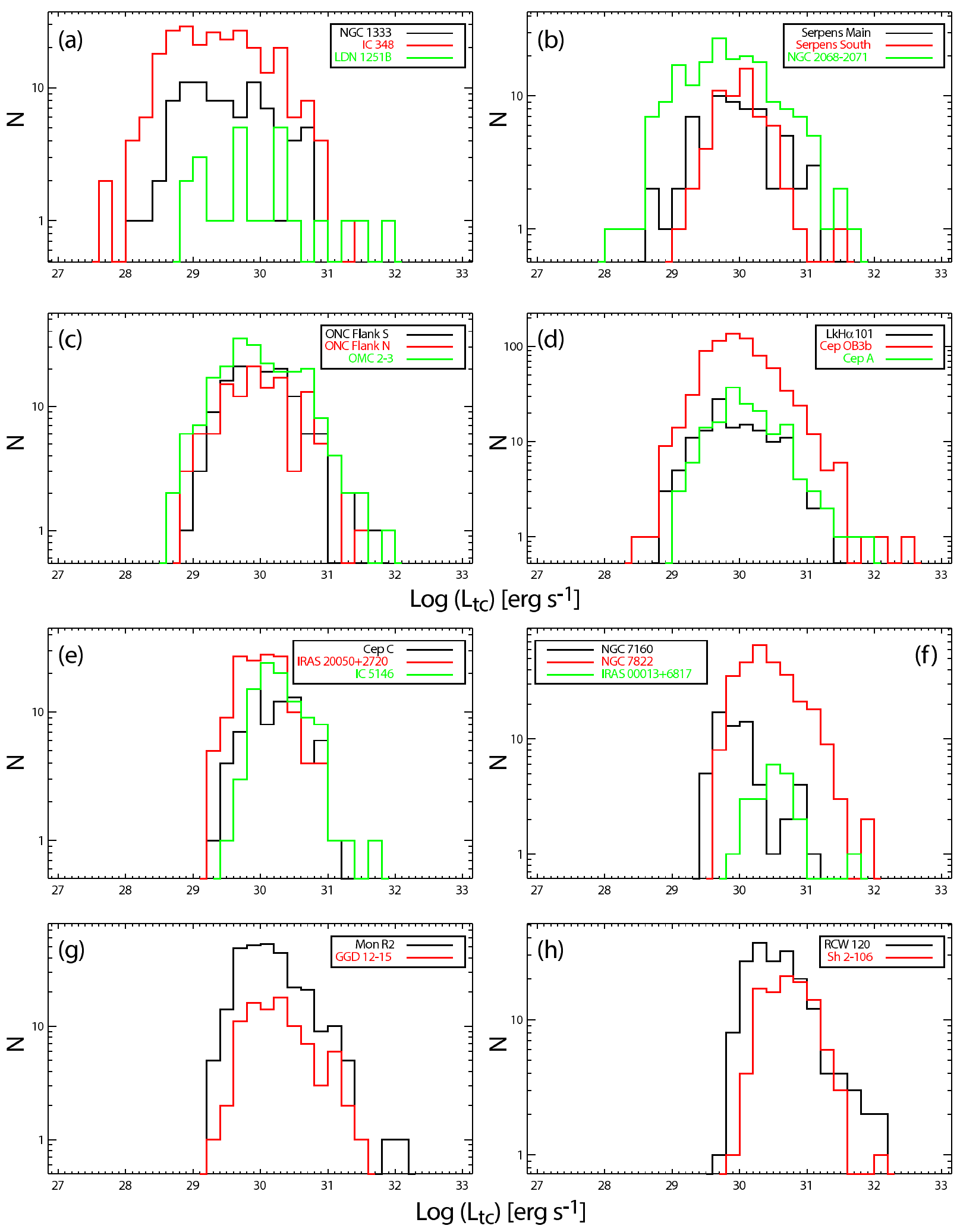}
\caption{X-ray luminosity functions (XLFs) for the SFiNCs cluster members with available $L_{tc}$ estimates. The XLFs are arranged in figure panels based on the SFiNCs SFRs' distances: from the nearest SFRs (panel a) to the most distant SFRs (panel h).  \label{fig_xlf_bright}}
\end{figure*}


\begin{figure*}
\centering
\includegraphics[angle=0.,width=5.5in]{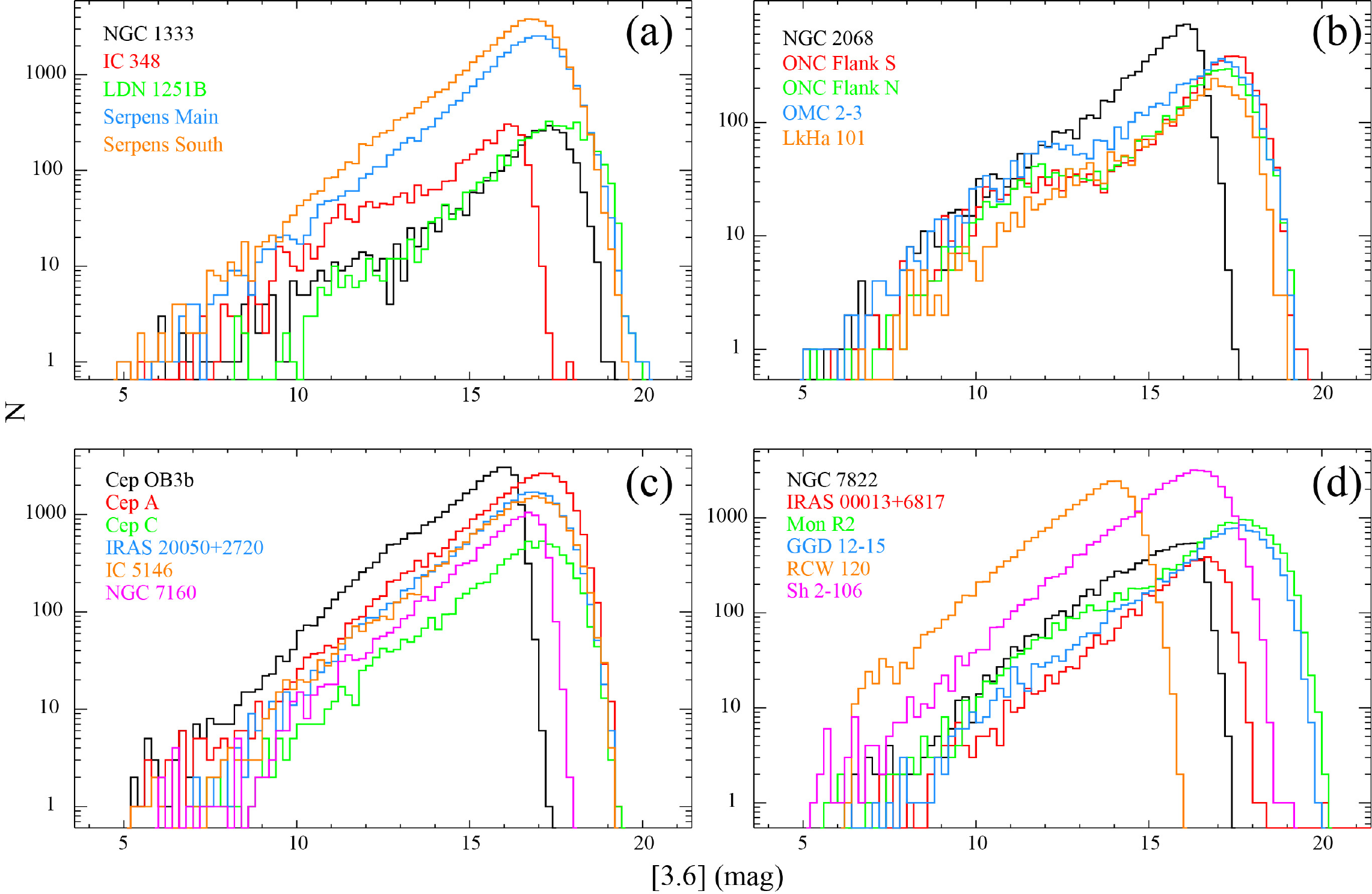}
\caption{Histograms of the [3.6]-band magnitude for the SFiNCs IRAC ``cut-out'' source sample. The histograms are arranged in figure panels based on the SFiNCs SFRs' distances: from the nearest SFRs (panel a) to the most distant SFRs (panel d).  \label{fig_mlf}}
\end{figure*}

\begin{figure*}
\centering
\includegraphics[angle=0.,width=4.5in]{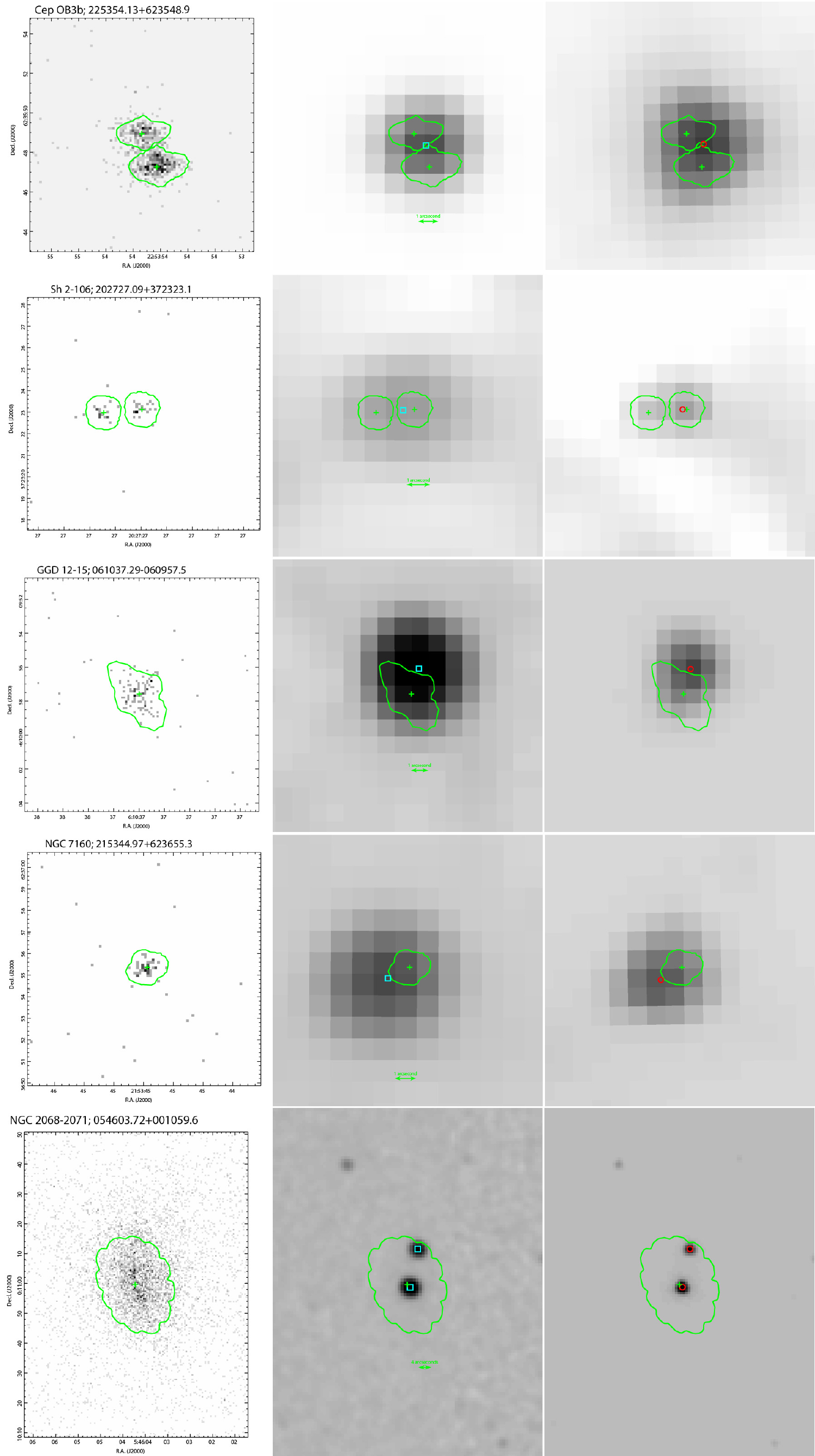}
\caption{Examples of SFiNCs X-ray-IR pairs likely affected by the presence of double sources. {\it Chandra}-ACIS images (left), 2MASS $K$-band images (middle), and {\it Spitzer}-IRAC 3.6~$\mu$m images (right). Cataloged source positions of X-ray, 2MASS, and {\it Spitzer}-IRAC sources are given in green, cyan, and red colors, respectively. X-ray source extraction regions are outlined by the green polygons. \label{fig_acis_ir_matches}}
\end{figure*}

\begin{figure*}
\centering
\includegraphics[angle=0.,width=4.5in]{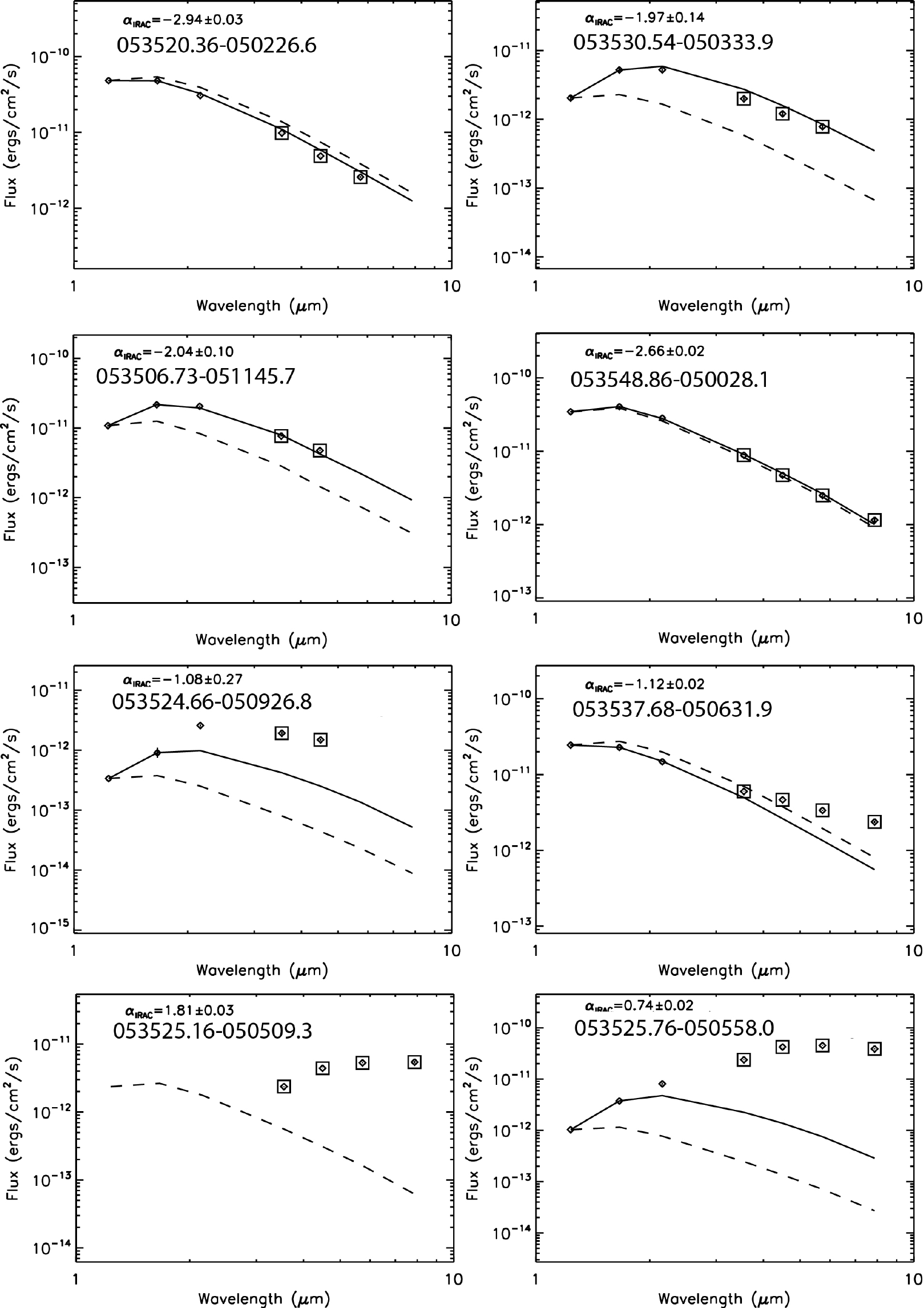}
\caption{Examples of the IR SEDs for the diskless (upper 4 panels) and disky (bottom 4 panels) X-ray young objects in OMC~2-3. $JHK_s$ (diamonds without squares) and IRAC-band (diamonds outlined by squares) flux points with usually small errors. The dashed and solid lines give the original and (de)reddened median SEDs for the IC~348 YSOs \citep{Lada2006} fitted to the SFiNCs data. The panel legends give information on the SFiNCs source' IAU name and apparent IRAC SED slope. \label{fig_seds}}
\end{figure*}

\begin{figure*}
\centering
\includegraphics[angle=0.,width=6.5in]{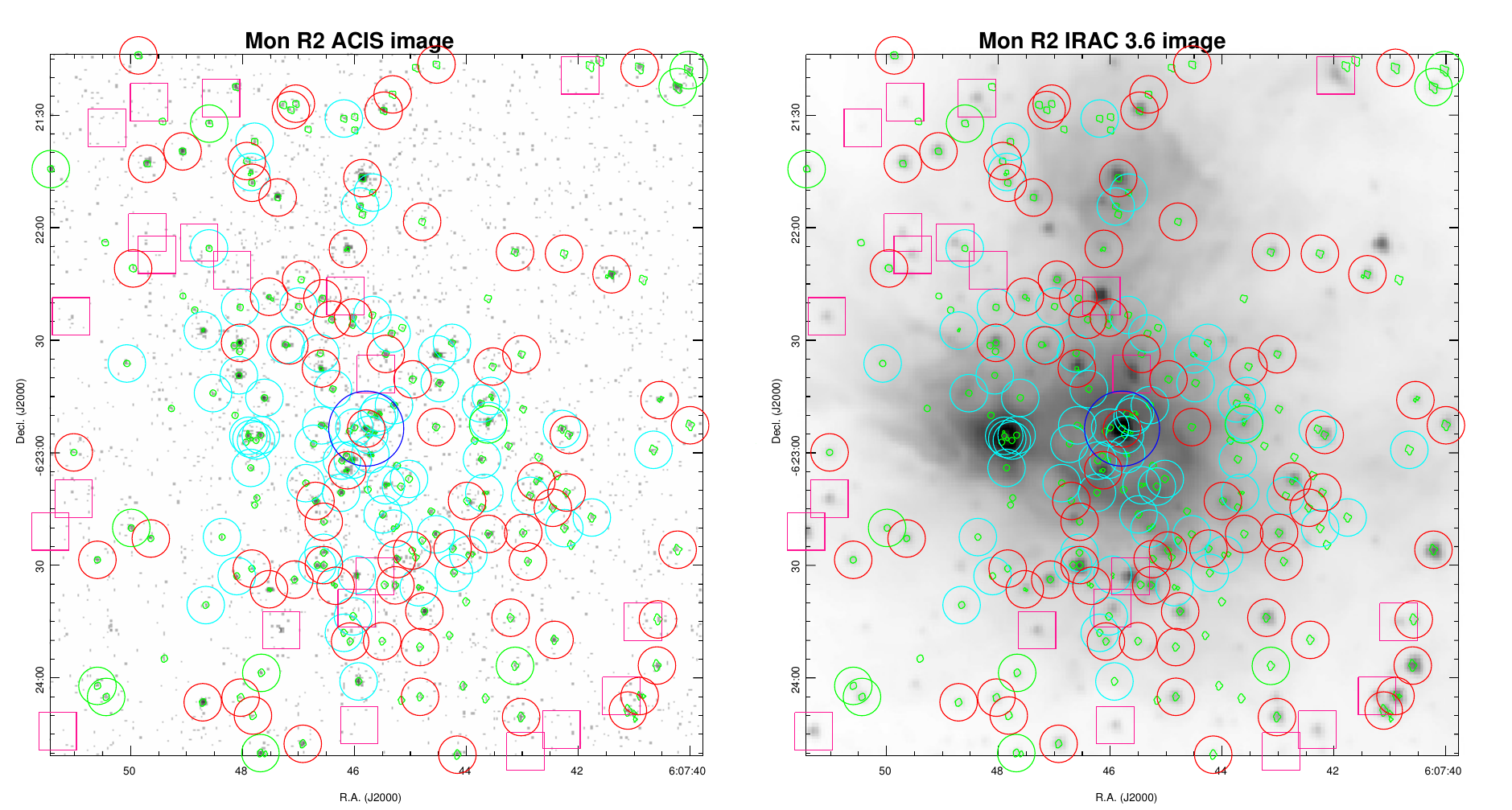}
\caption{A close-up view of the centers of the YSO clusters/groups in LkH$\alpha$~101, Mon~R2, RCW~120, Sh 2-106, and Cep~A that are subject to high MIR diffuse nebular background. An example is given for Mon~R2. The figure set presenting all five SFRs is available in the on-line journal. To assist with the review process a single pdf file comprising the entire figure set is also provided (f6\_figureset\_merged\_pmb\_maps.pdf). {\it Chandra}-ACIS at $(0.5-8)$~keV band  and {\it Spitzer}-IRAC 3.6~$\mu$m images are shown on the left and right panels, respectively. The SFiNCs YSO members, as a final outcome of the YSO selection procedures in \S \ref{yso_selection_section}, are color-coded as: X-ray diskless YSOs (green circles), X-ray disky YSOs (red circles), non X-ray disky YSOs (pink squares), ``PMB'' YSO members (cyan circles), and OB-type stars (blue circles). ACIS source extraction regions are outlined by green polygons. \label{fig_pmb_map}}
\end{figure*}

\begin{figure}
\centering
\includegraphics[angle=0.,width=7.0in]{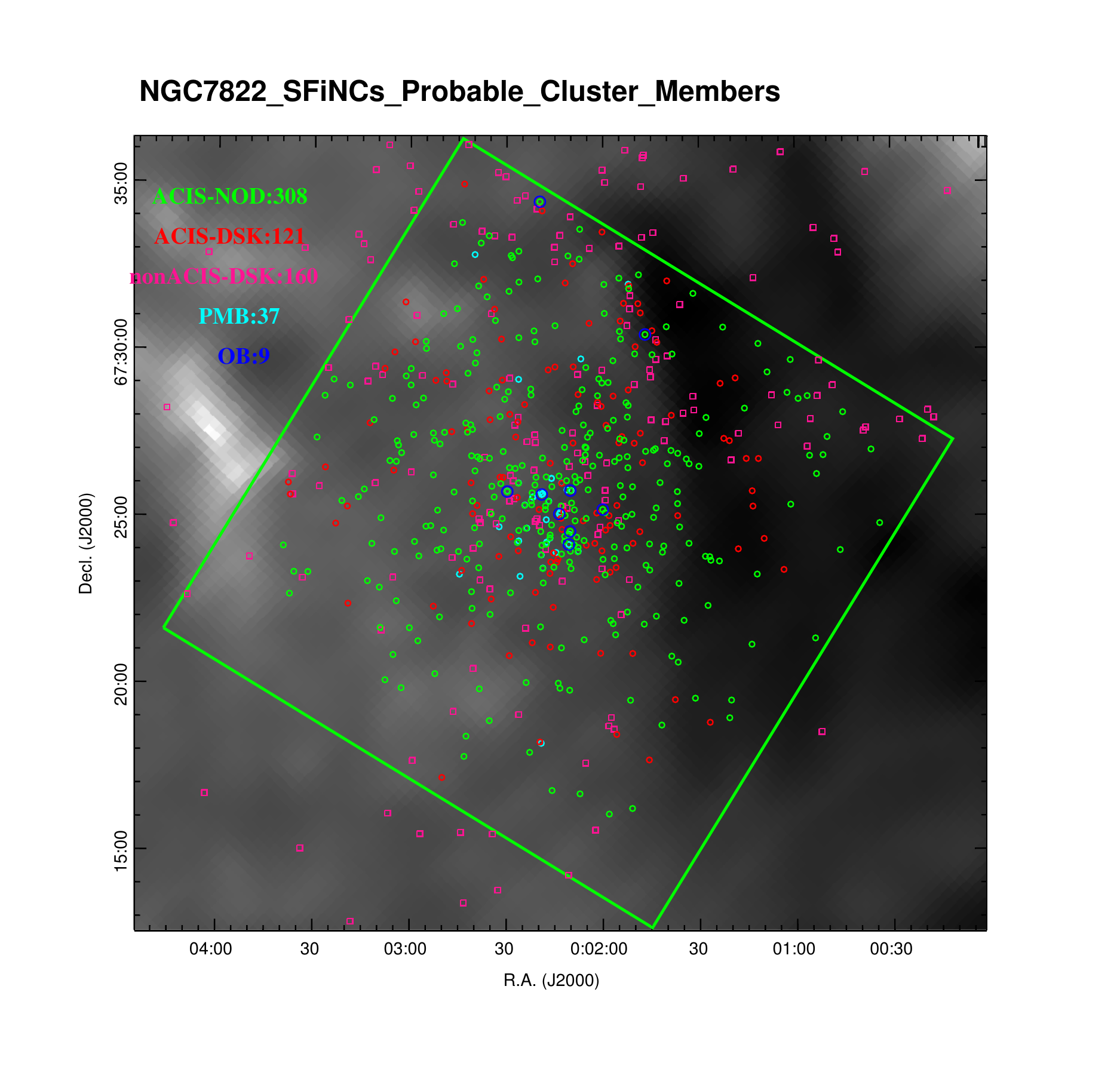}
\caption{SFiNCs probable cluster members superimposed on the images taken by $AKARI$-FIS at 160~$\mu$m (for NGC~7822, IRAS~00013+6817, IRAS~20050+2720, NGC~7160, CepOB3b) and $Herschel$-SPIRE at 500~$\mu$m (for the rest of the SFiNCs SFRs). These images trace the locations of the SFiNCs molecular clouds. The images are shown in inverted colors with logarithmic scales (denser clouds look darker, except for NGC~7160 where there is no cloud left). X-ray diskless YSOs, X-ray disky YSOs, non X-ray disky YSOs, additional possible YSO members, and OB-type stars are marked as green circles, red circles, pink squares, cyan circles, and blue circles, respectively. {\it Chandra}-ACIS-I field of view is outlined by the green polygons. The figure legends provide information on the numbers of the SFiNCs YSOs. Since many ``PMB'' YSOs (cyan) are parts of visual double/multiple X-ray systems, their symbols are often covered up by the nearby ACIS-DSK/NOD symbols. For NGC~2068-2071, Mon~R2, GGD~12-15, Cep OB3b, and CepC, a handful of non X-ray disky SPCMs located outside the boundaries of the current FIR images are not shown in these figures. The SPCM spatial distributions with expanded FOVs can be seen in Figure \ref{fig_spcm_pub_maps}. \label{fig_spcm_maps}}
\end{figure}
\clearpage

\begin{figure}
\centering
\includegraphics[angle=0.,width=7.0in]{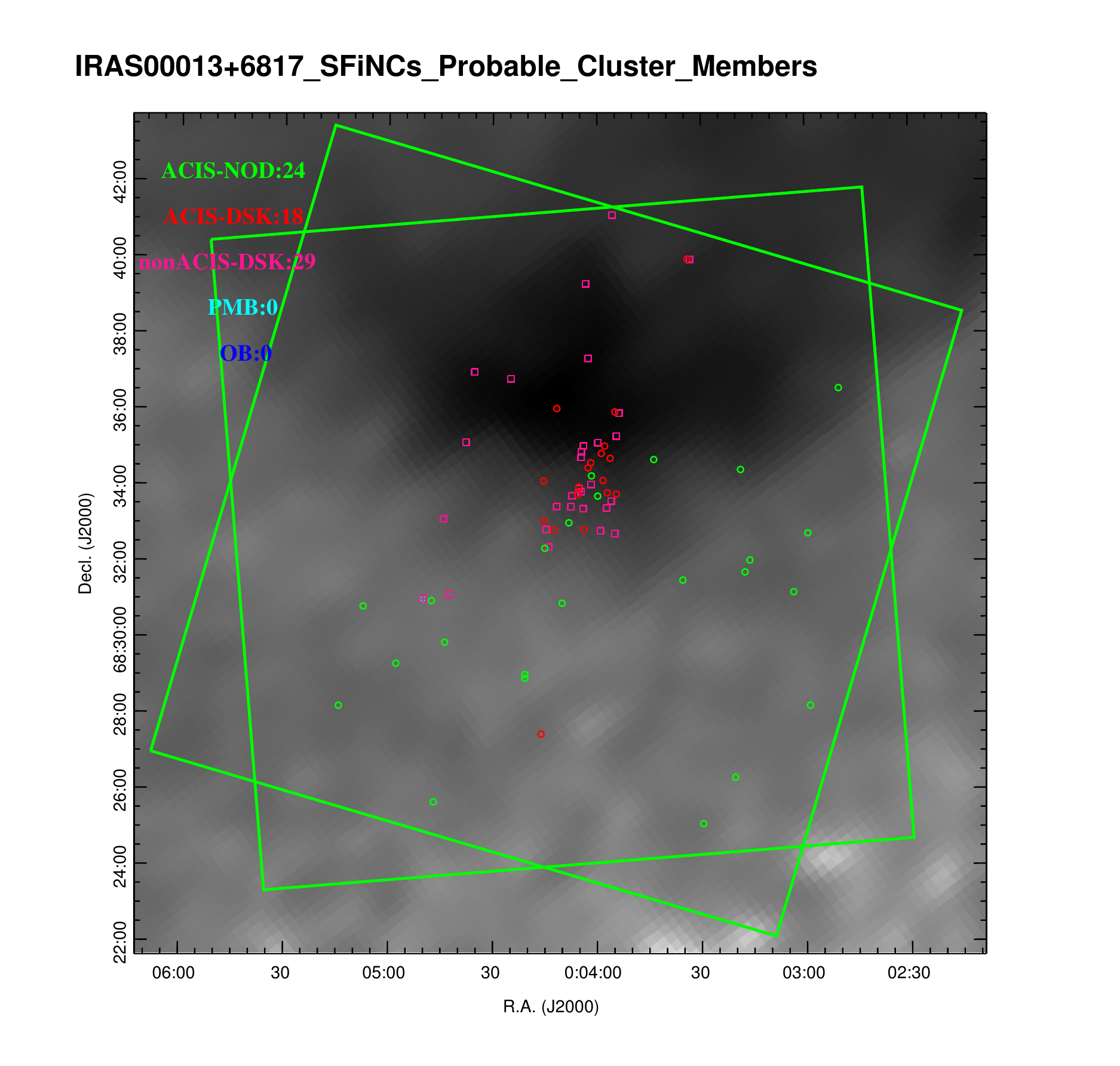}
\end{figure}
\clearpage

\begin{figure}
\centering
\includegraphics[angle=0.,width=7.0in]{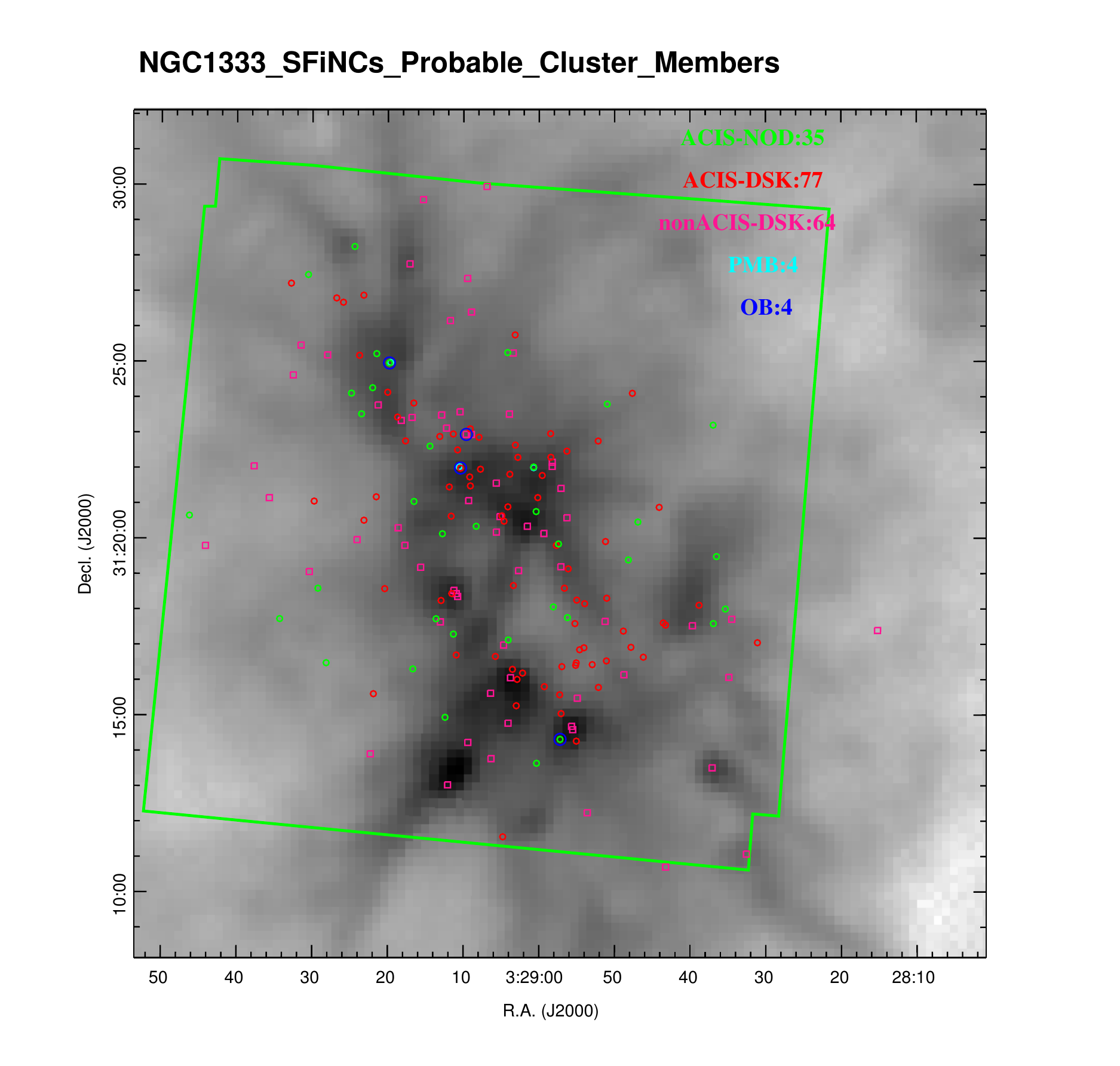}
\end{figure}
\clearpage

\begin{figure}
\centering
\includegraphics[angle=0.,width=7.0in]{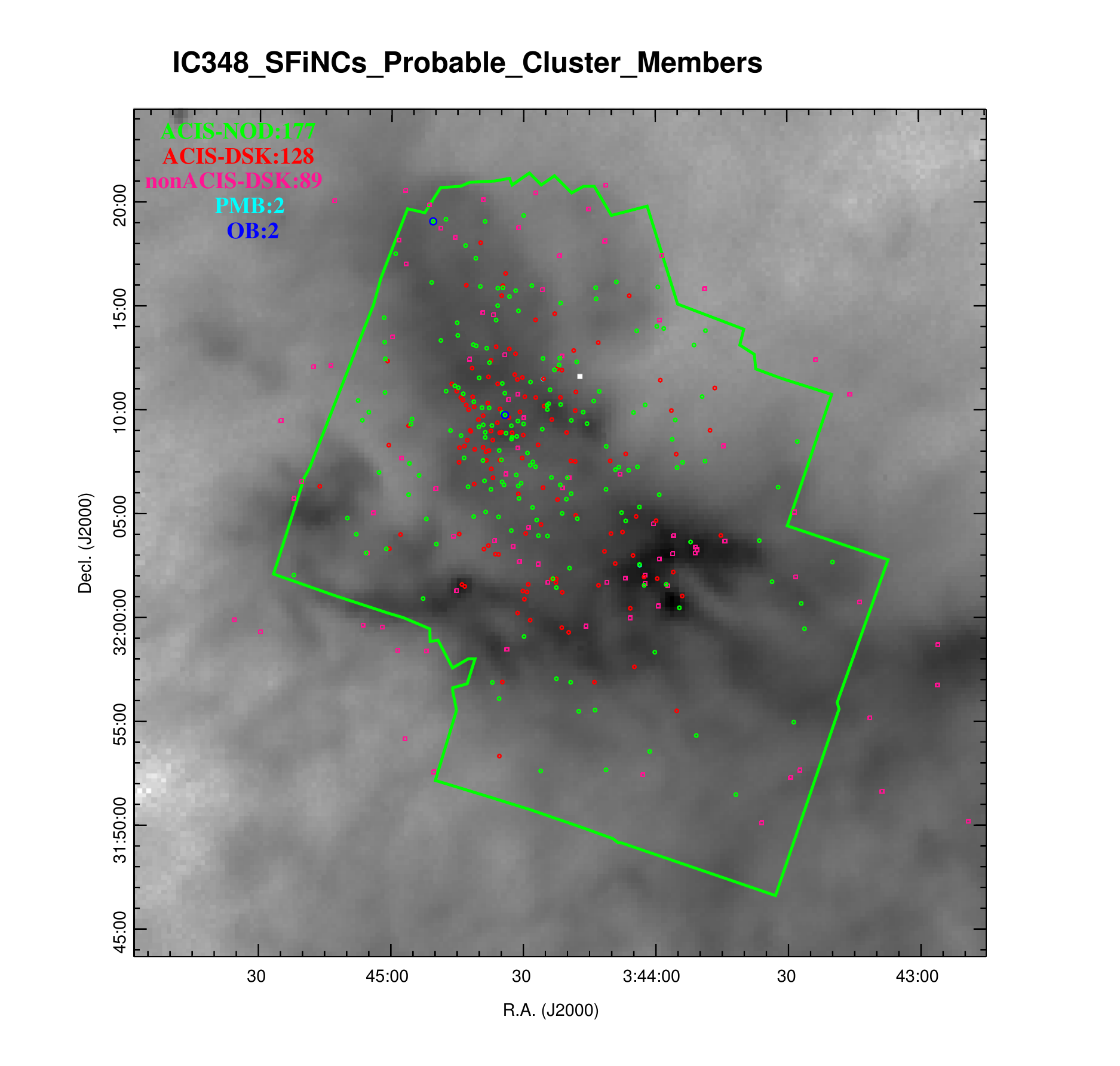}
\end{figure}
\clearpage

\begin{figure}
\centering
\includegraphics[angle=0.,width=7.0in]{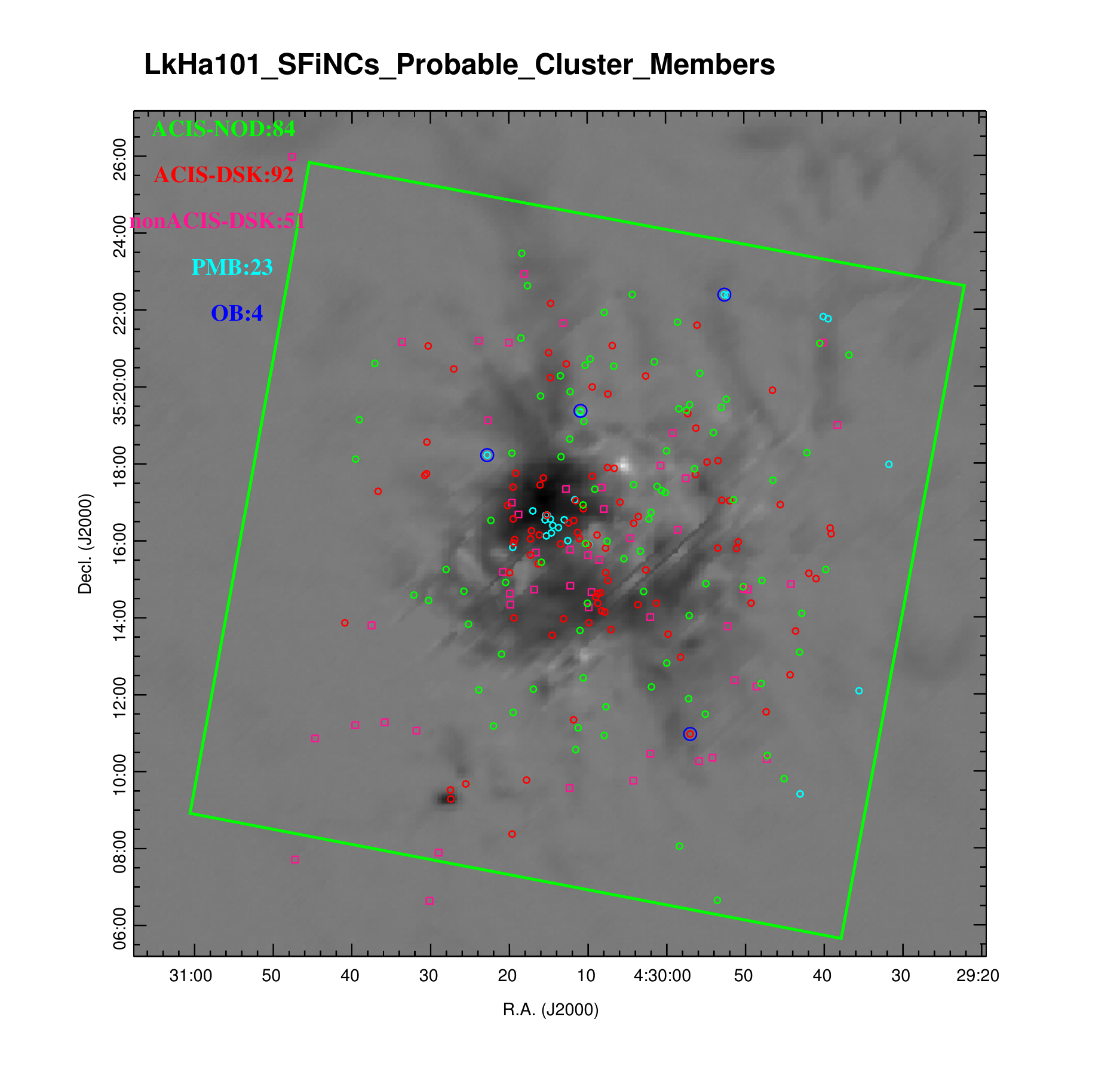}
\end{figure}
\clearpage

\begin{figure}
\centering
\includegraphics[angle=0.,width=7.0in]{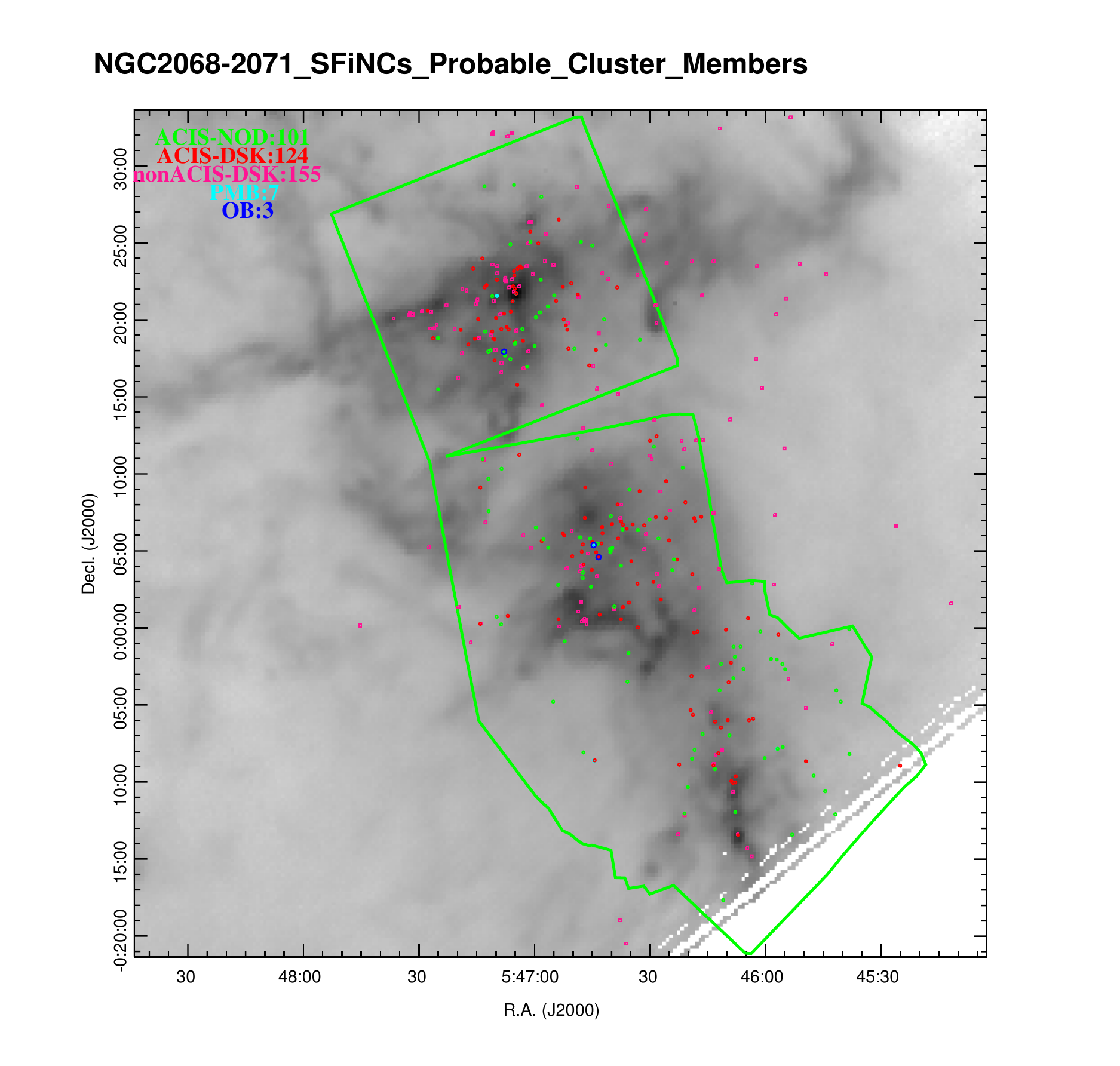}
\end{figure}
\clearpage

\begin{figure}
\centering
\includegraphics[angle=0.,width=7.0in]{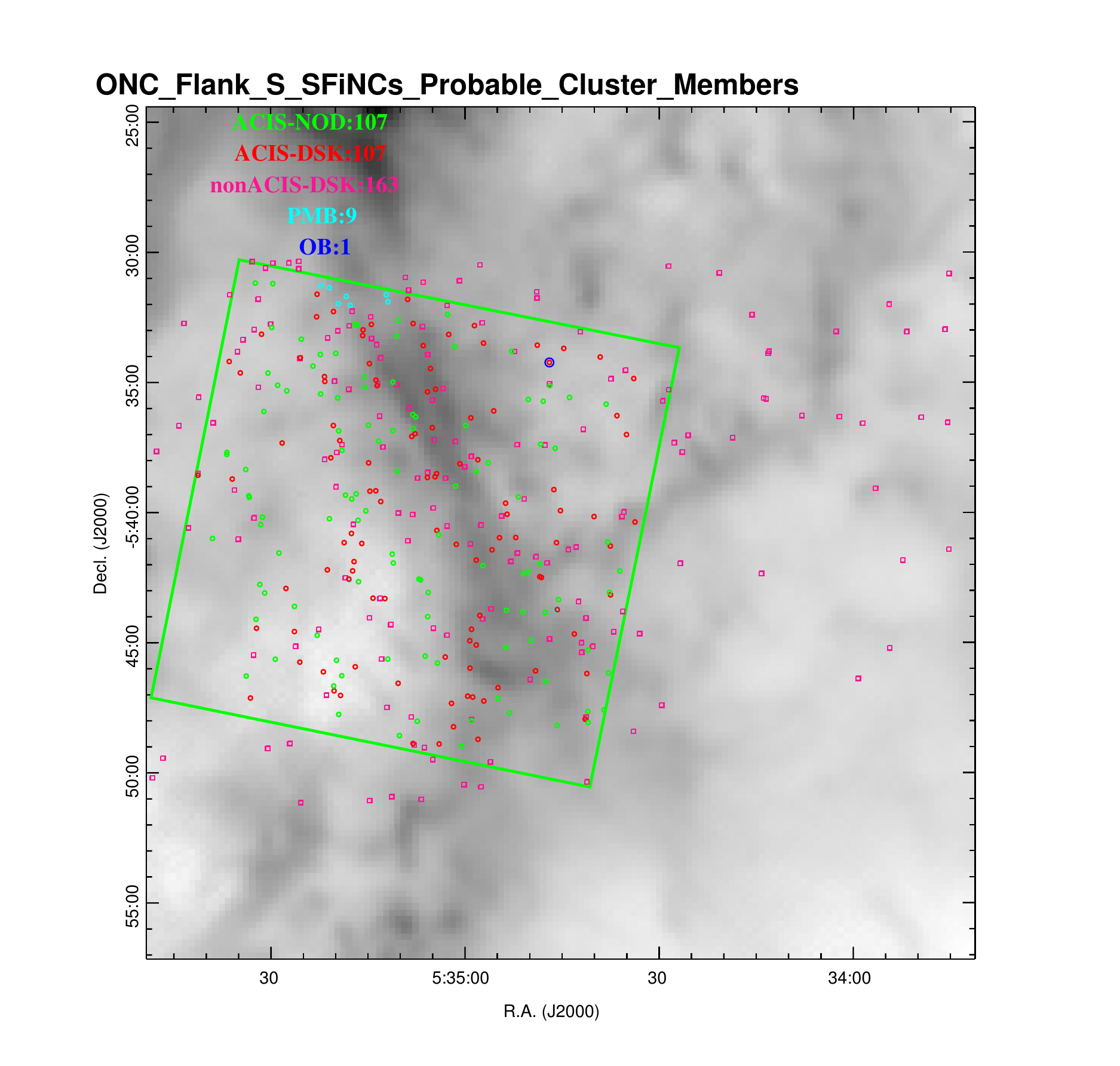}
\end{figure}
\clearpage

\begin{figure}
\centering
\includegraphics[angle=0.,width=7.0in]{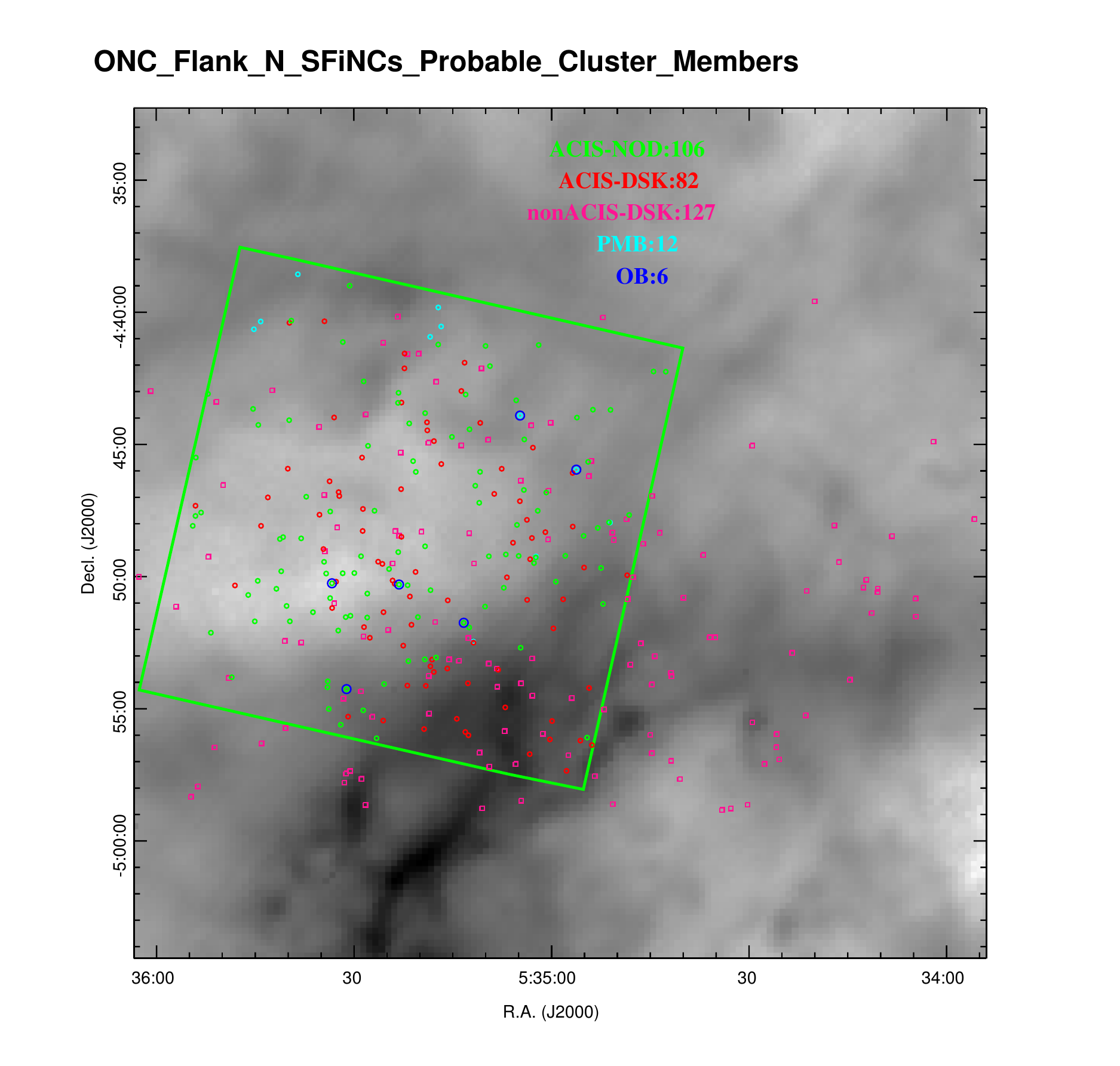}
\end{figure}
\clearpage

\begin{figure}
\centering
\includegraphics[angle=0.,width=7.0in]{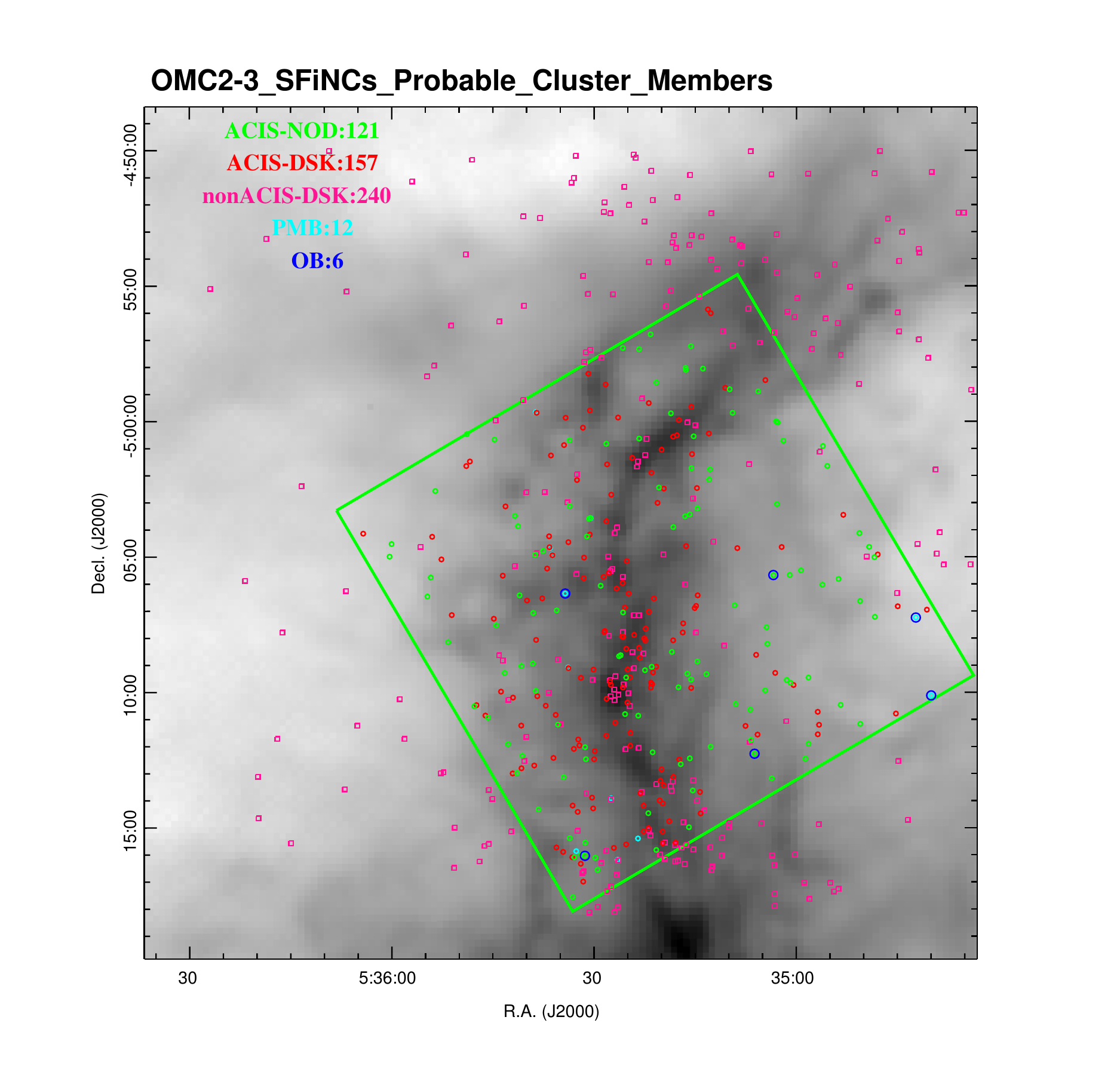}
\end{figure}
\clearpage

\begin{figure}
\centering
\includegraphics[angle=0.,width=7.0in]{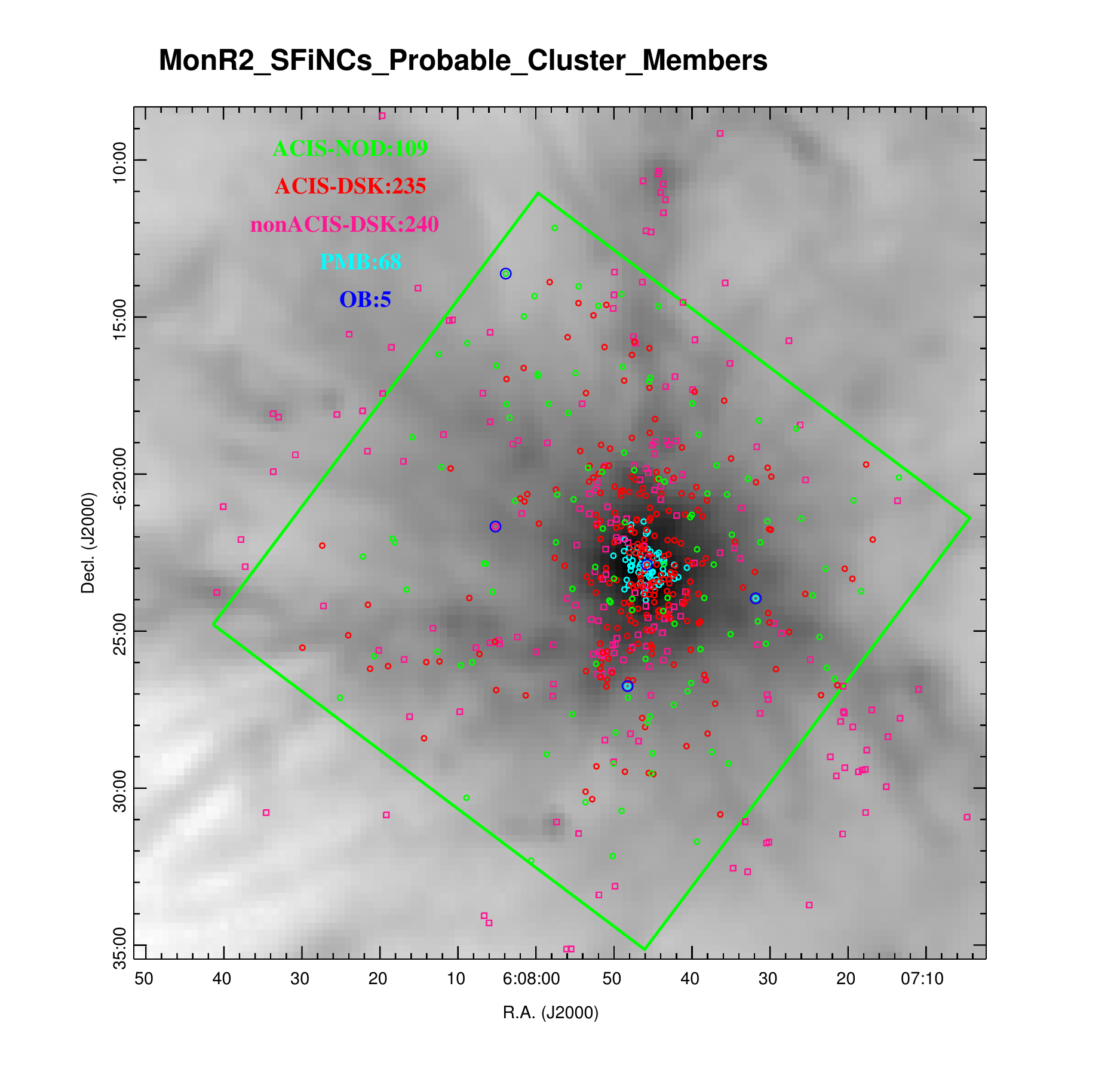}
\end{figure}
\clearpage

\begin{figure}
\centering
\includegraphics[angle=0.,width=7.0in]{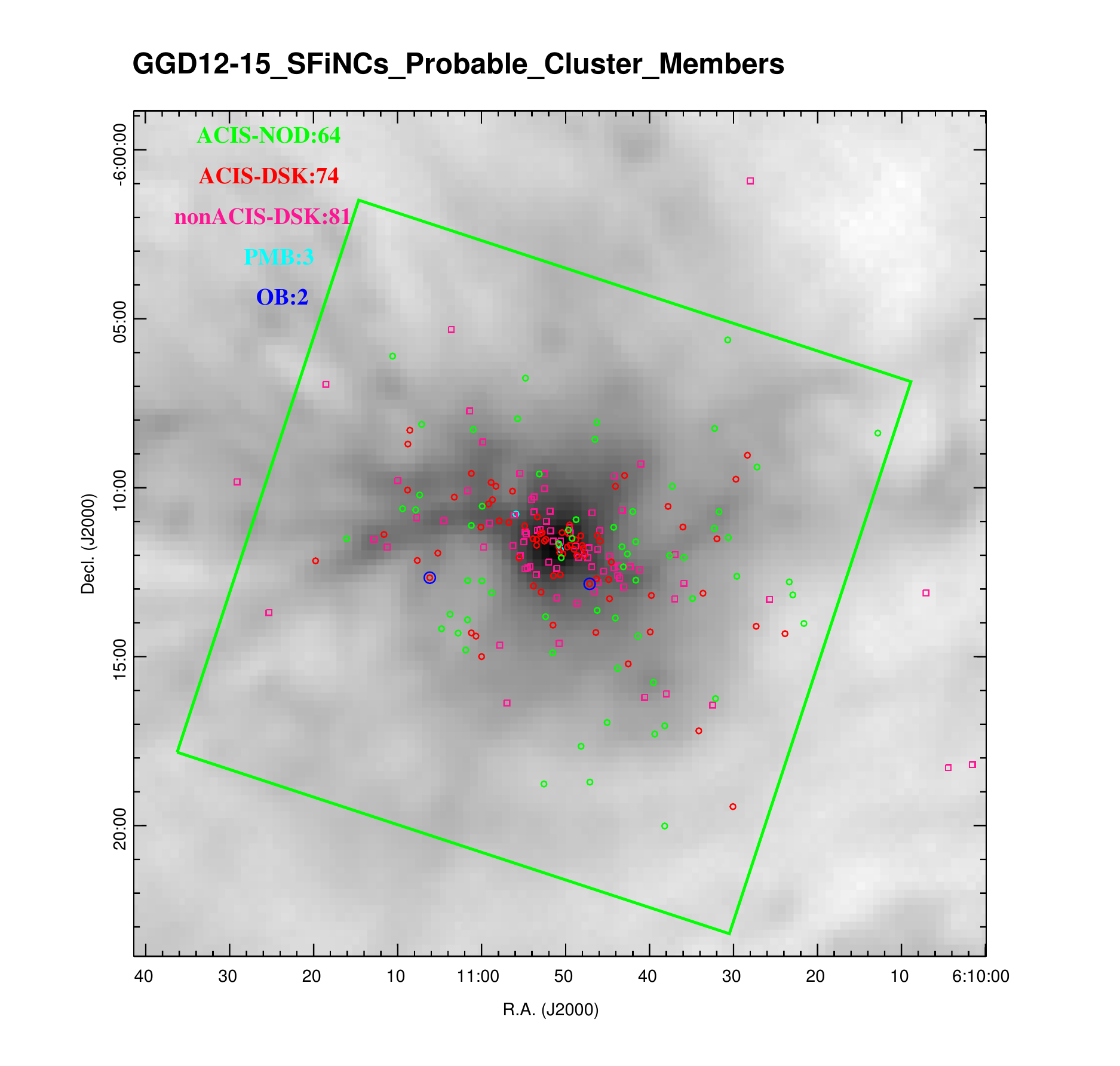}
\end{figure}
\clearpage

\begin{figure}
\centering
\includegraphics[angle=0.,width=7.0in]{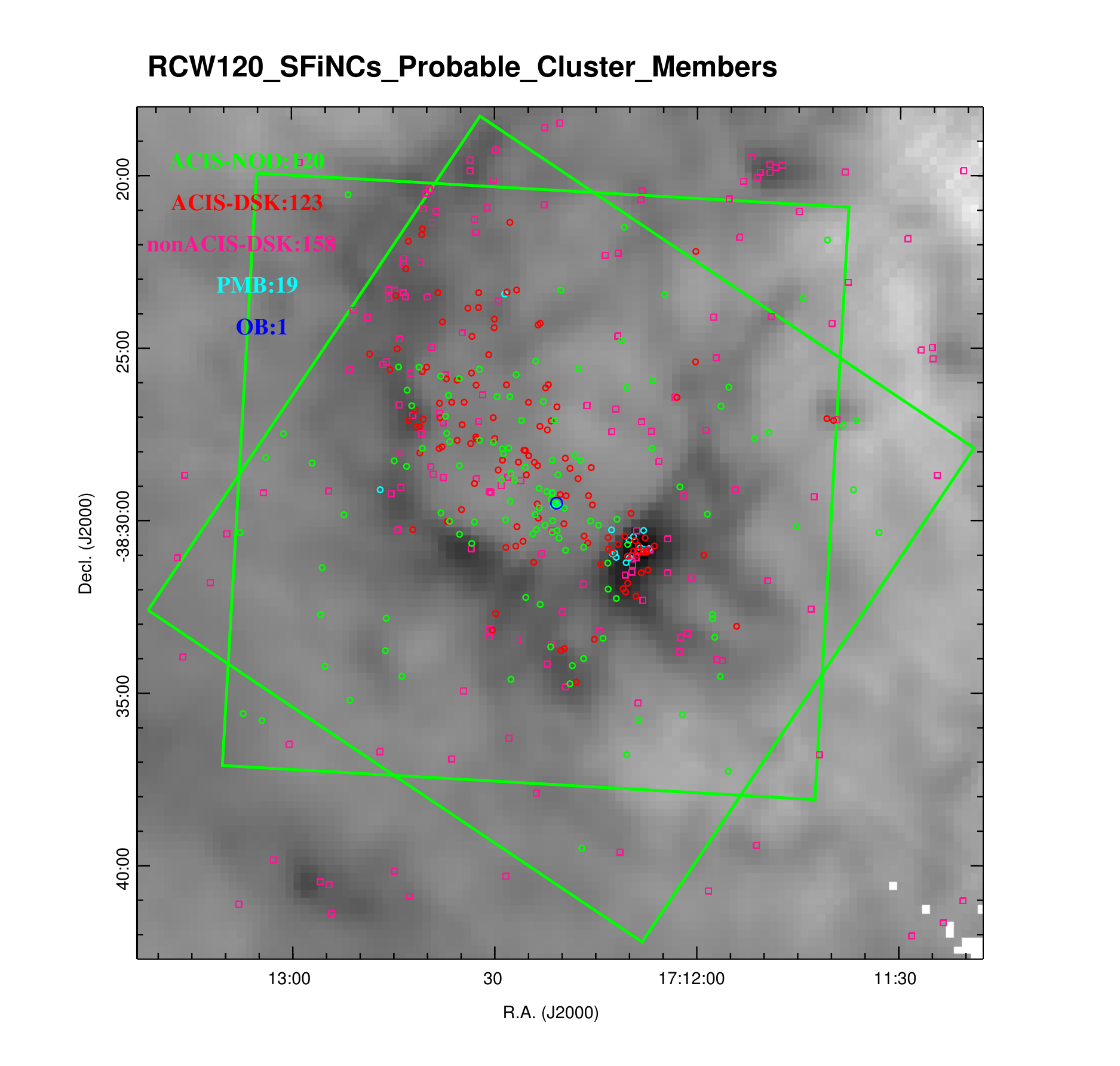}
\end{figure}
\clearpage

\begin{figure}
\centering
\includegraphics[angle=0.,width=7.0in]{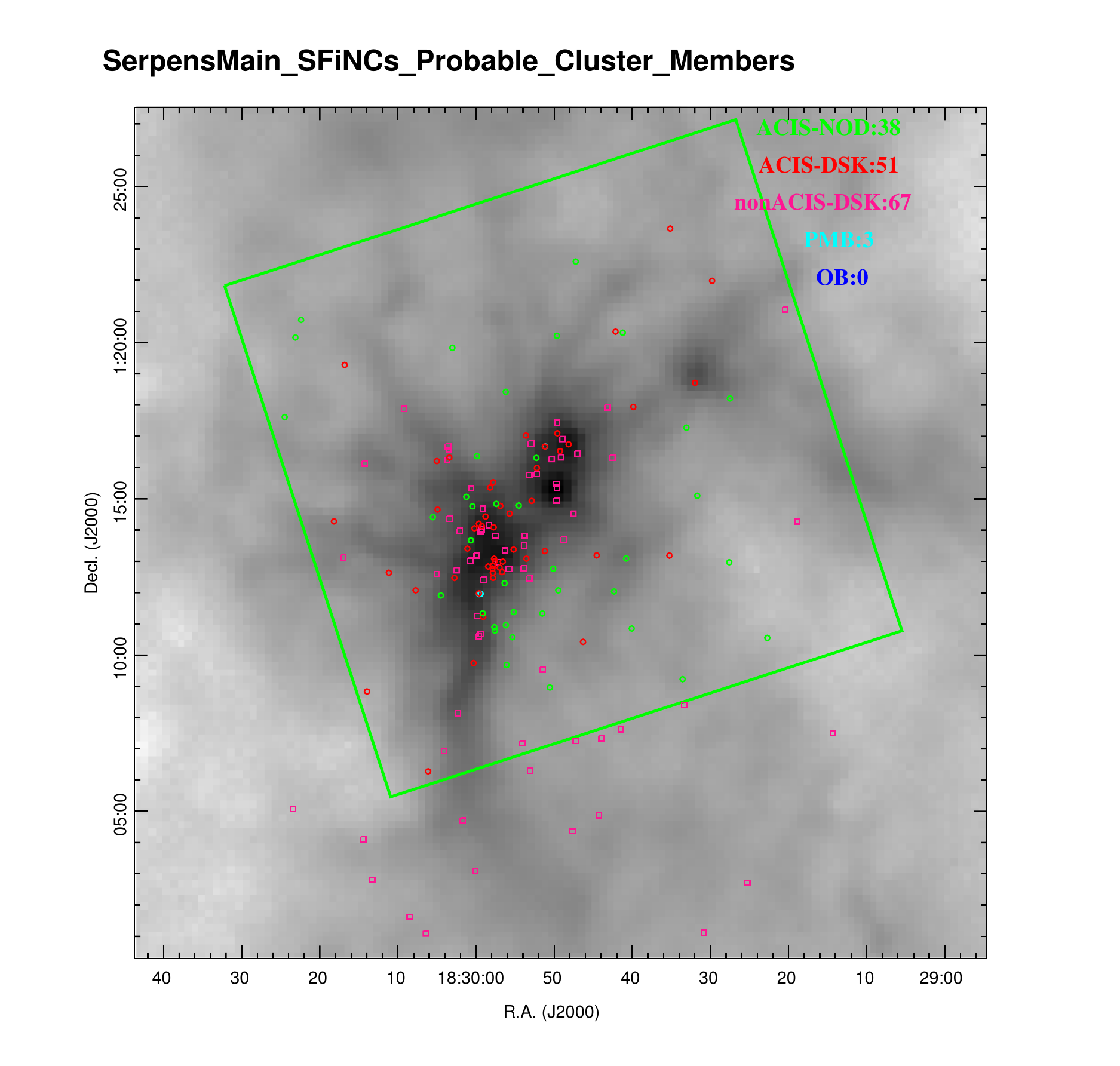}
\end{figure}
\clearpage

\begin{figure}
\centering
\includegraphics[angle=0.,width=7.0in]{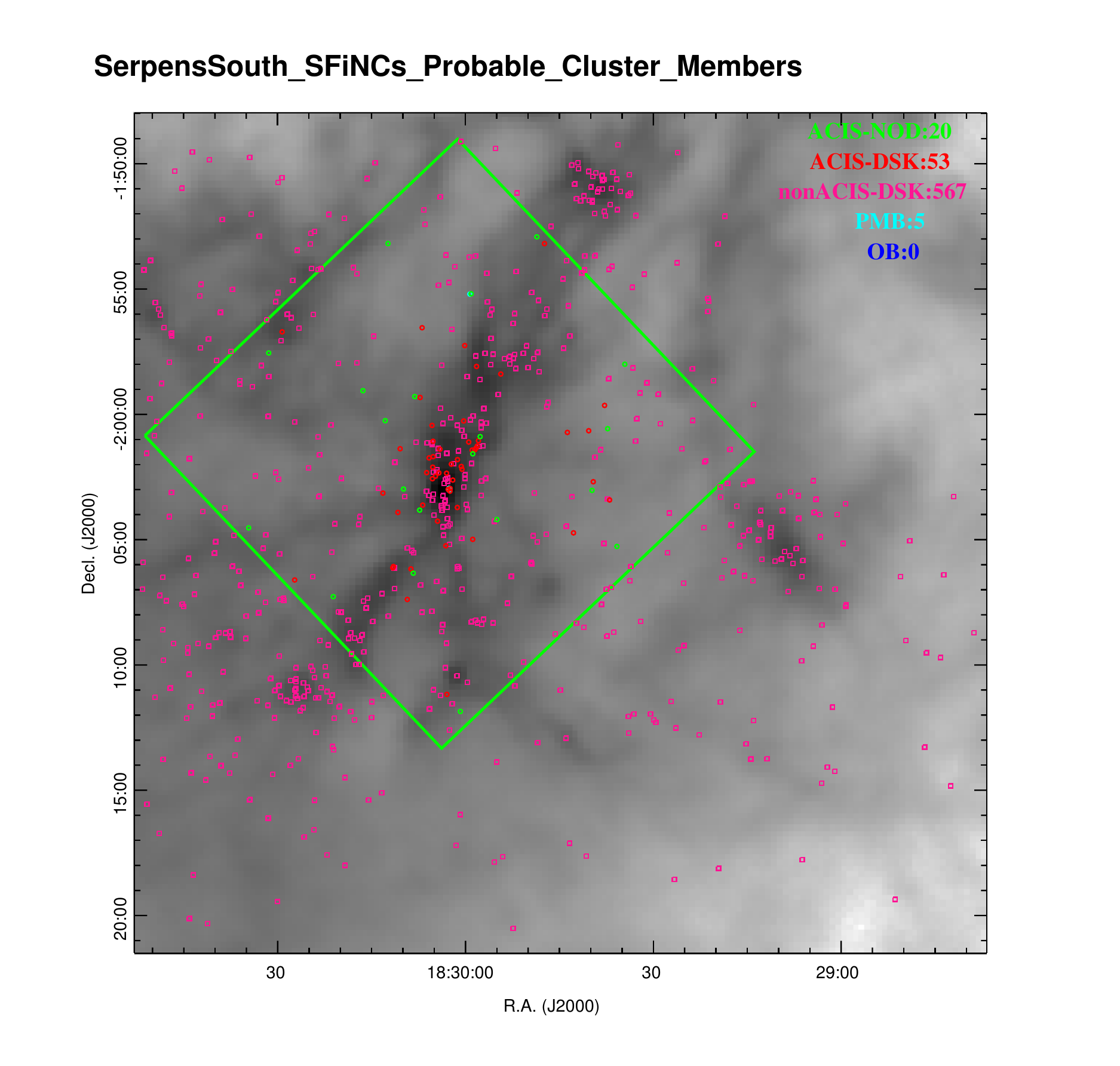}
\end{figure}
\clearpage

\begin{figure}
\centering
\includegraphics[angle=0.,width=7.0in]{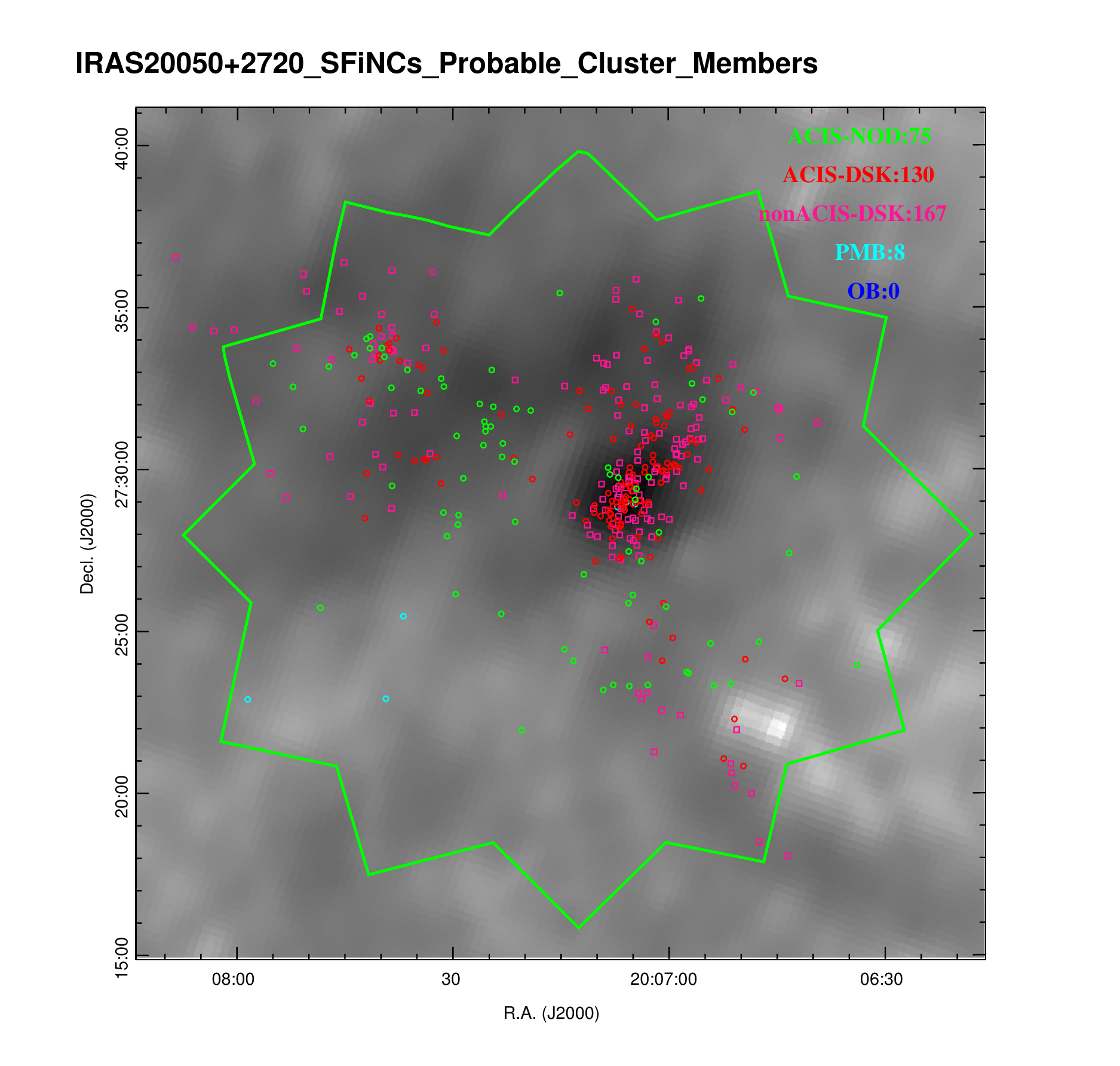}
\end{figure}
\clearpage

\begin{figure}
\centering
\includegraphics[angle=0.,width=7.0in]{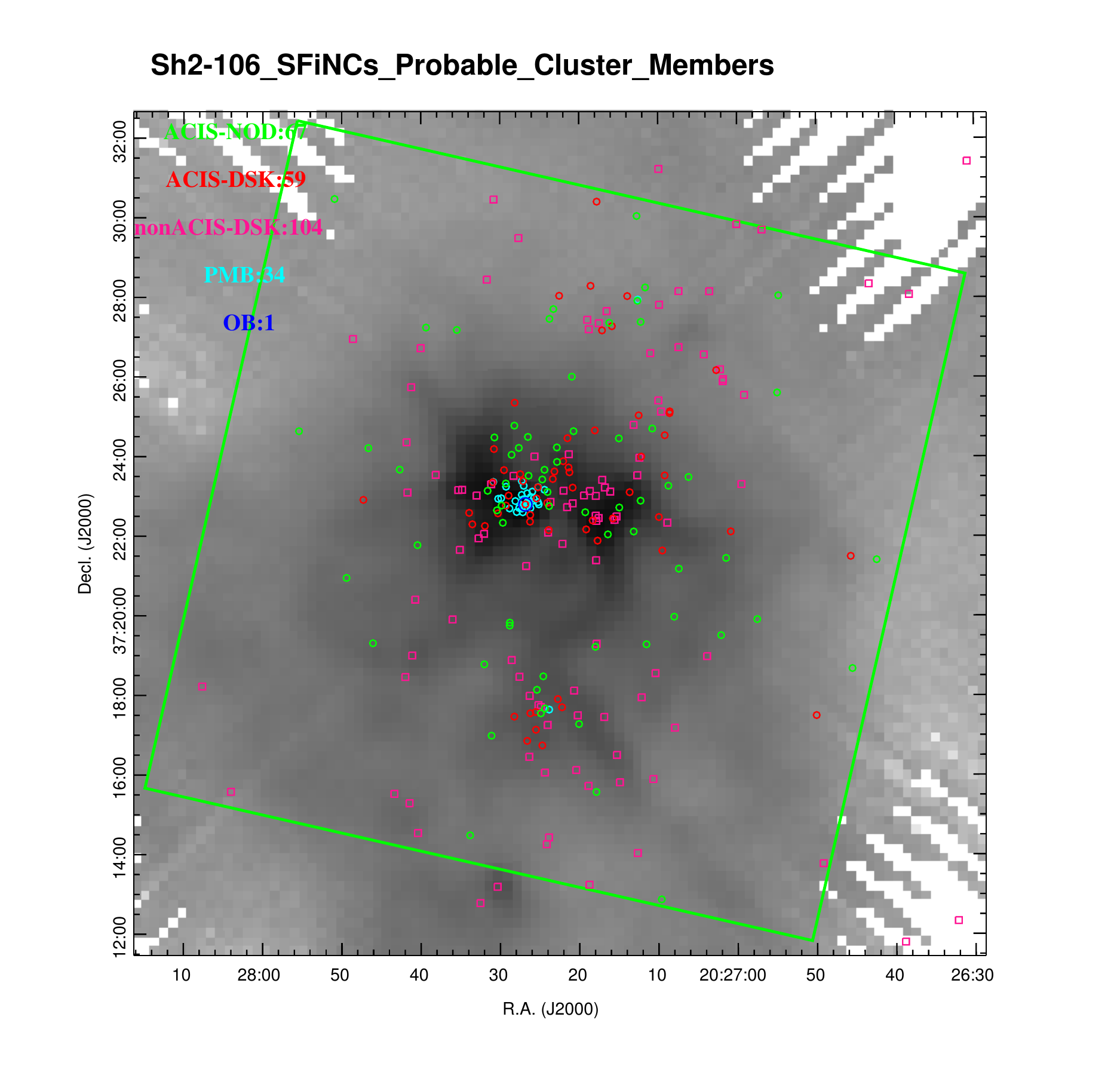}
\end{figure}
\clearpage

\begin{figure}
\centering
\includegraphics[angle=0.,width=7.0in]{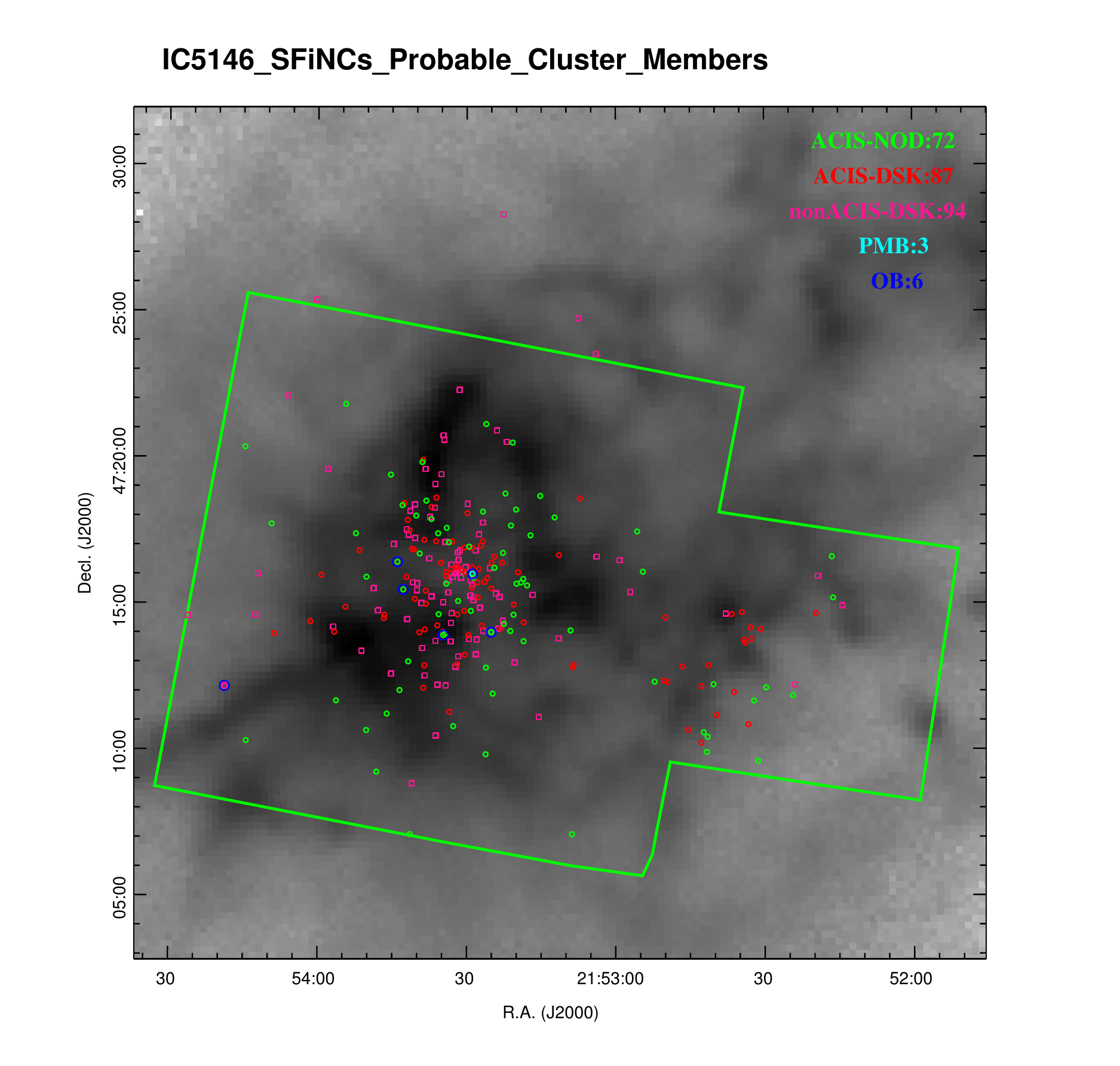}
\end{figure}
\clearpage

\begin{figure}
\centering
\includegraphics[angle=0.,width=7.0in]{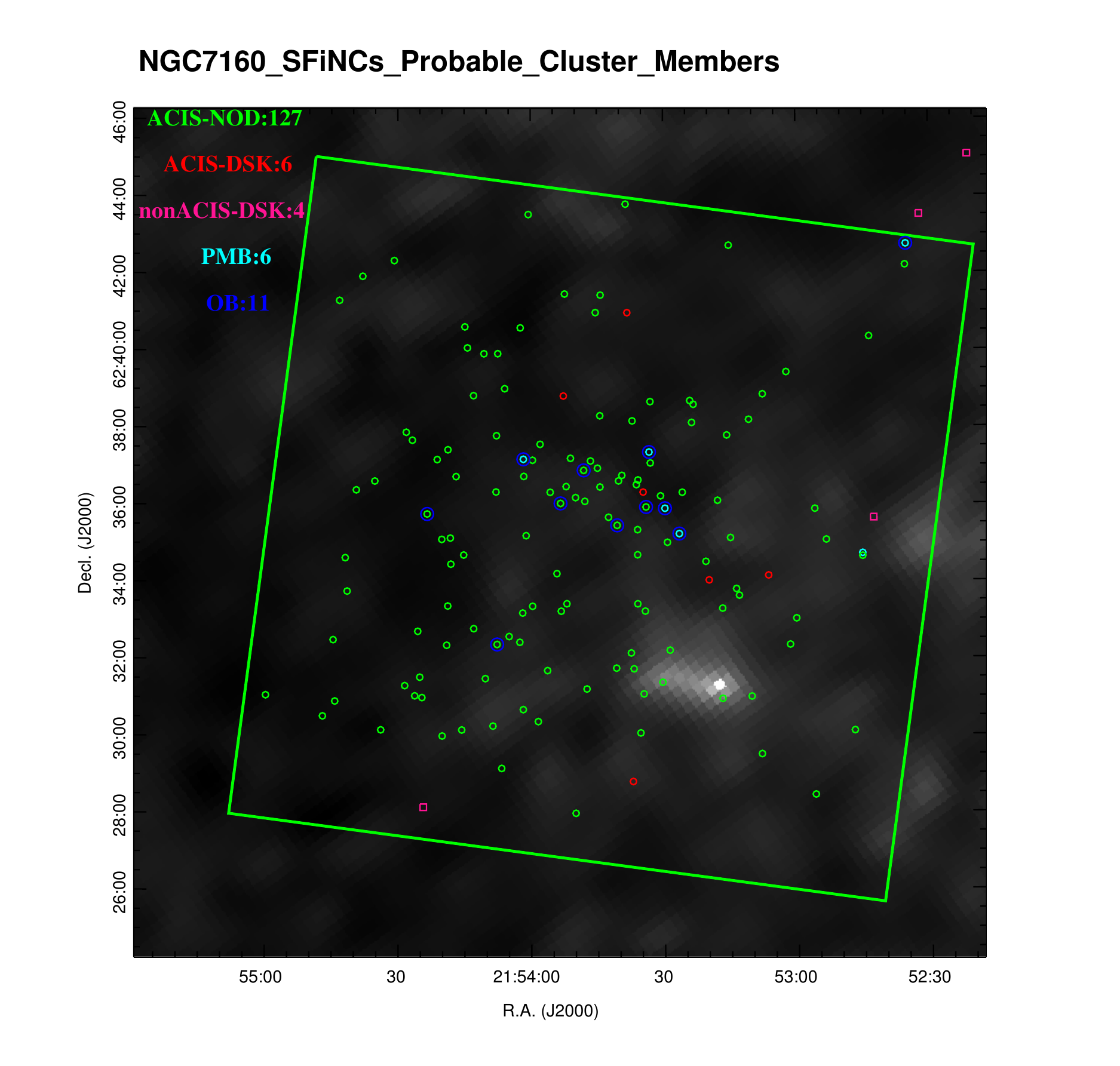}
\end{figure}
\clearpage

\begin{figure}
\centering
\includegraphics[angle=0.,width=7.0in]{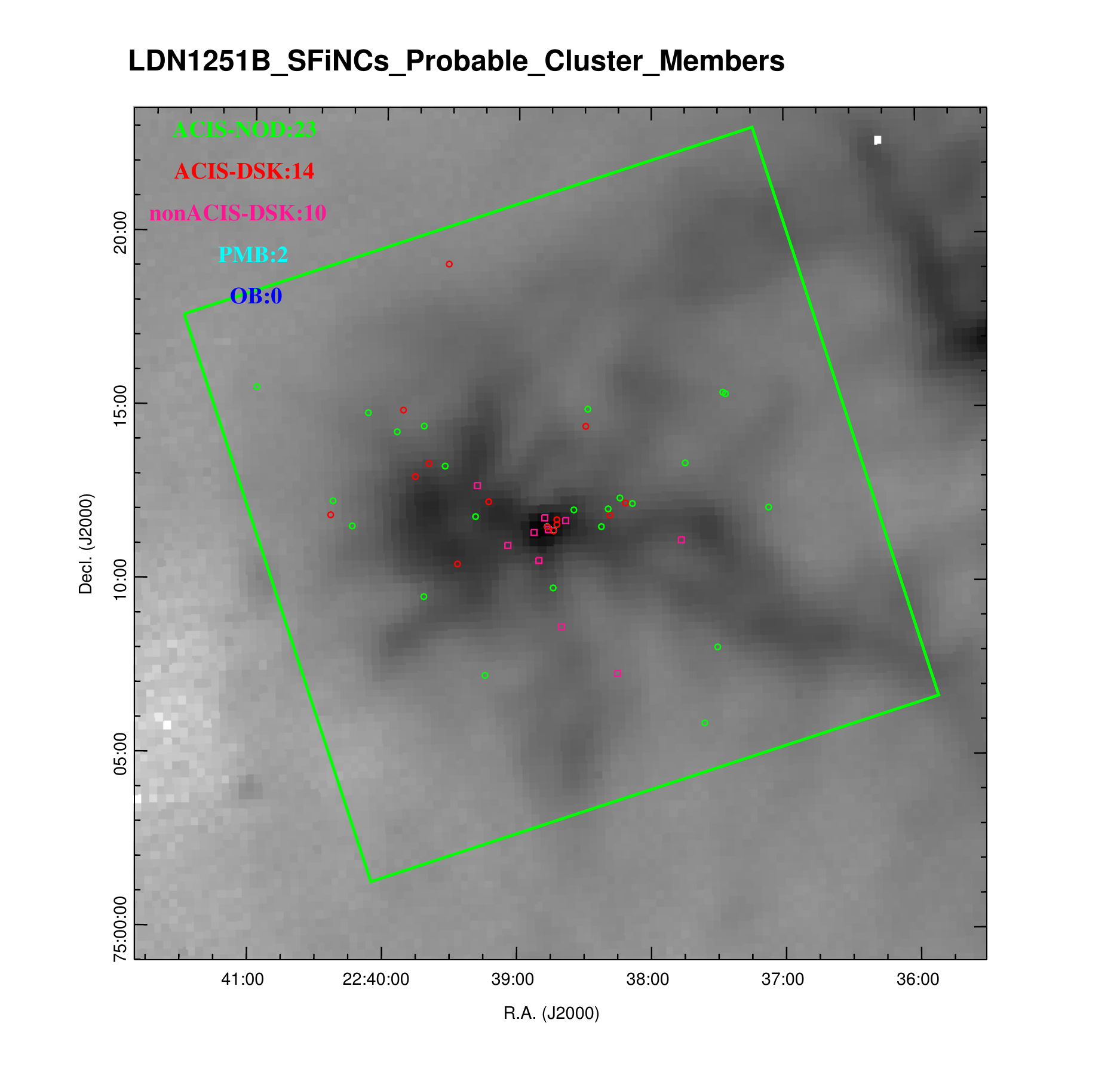}
\end{figure}
\clearpage

\begin{figure}
\centering
\includegraphics[angle=0.,width=7.0in]{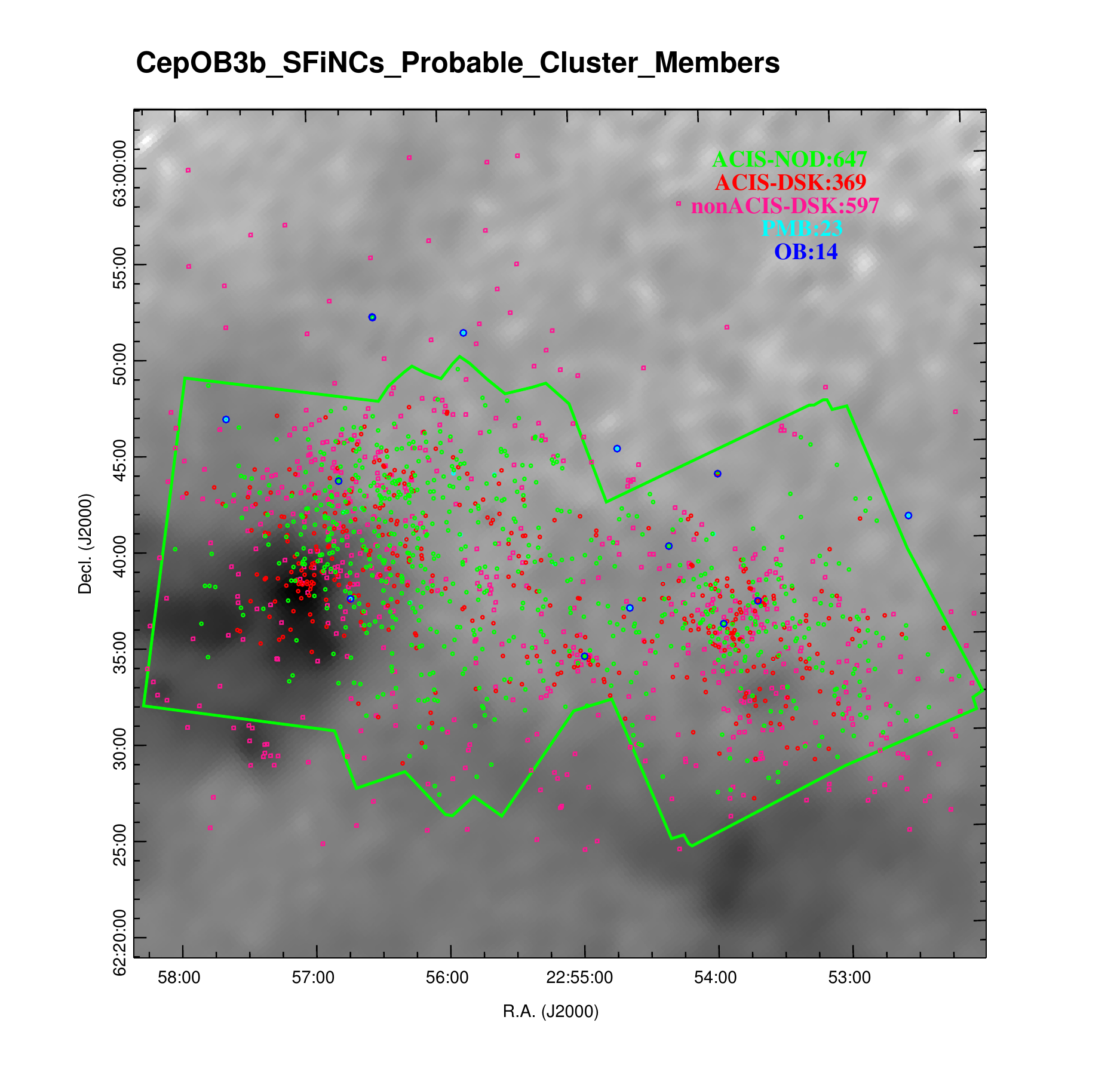}
\end{figure}
\clearpage

\begin{figure}
\centering
\includegraphics[angle=0.,width=7.0in]{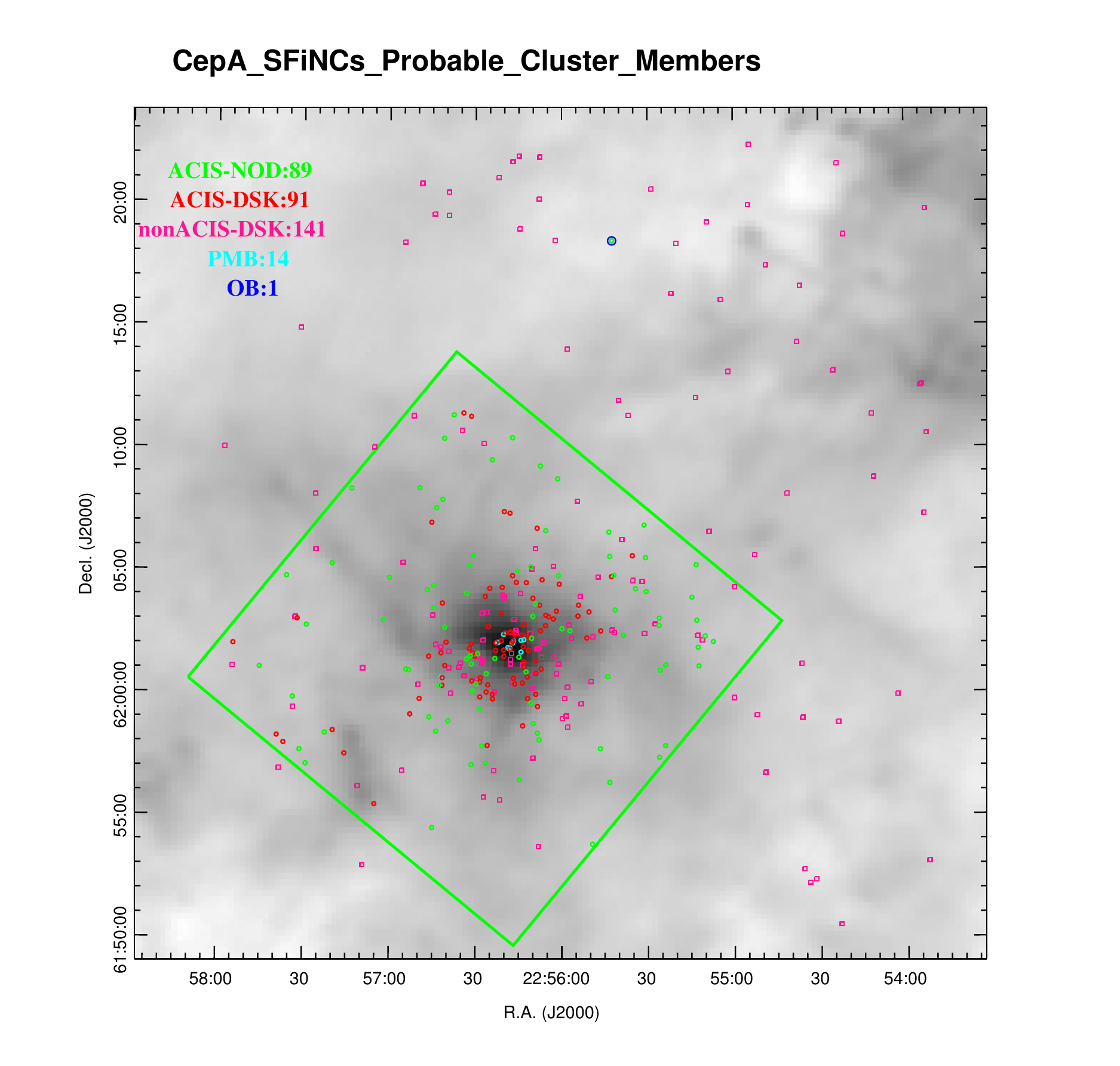}
\end{figure}
\clearpage

\begin{figure}
\centering
\includegraphics[angle=0.,width=7.0in]{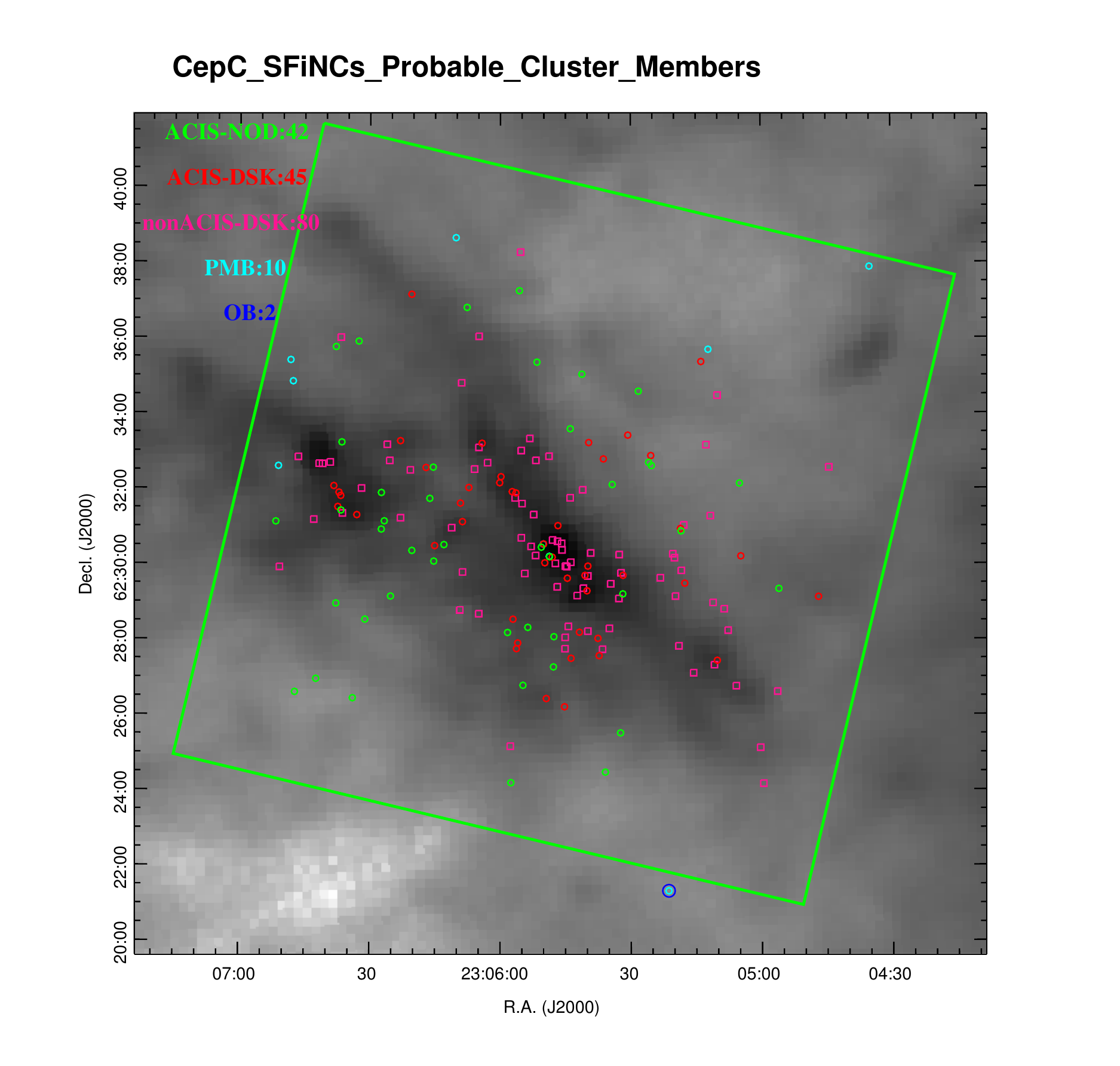}
\end{figure}
\clearpage

\begin{figure*}
\centering
\includegraphics[angle=0.,width=6.in]{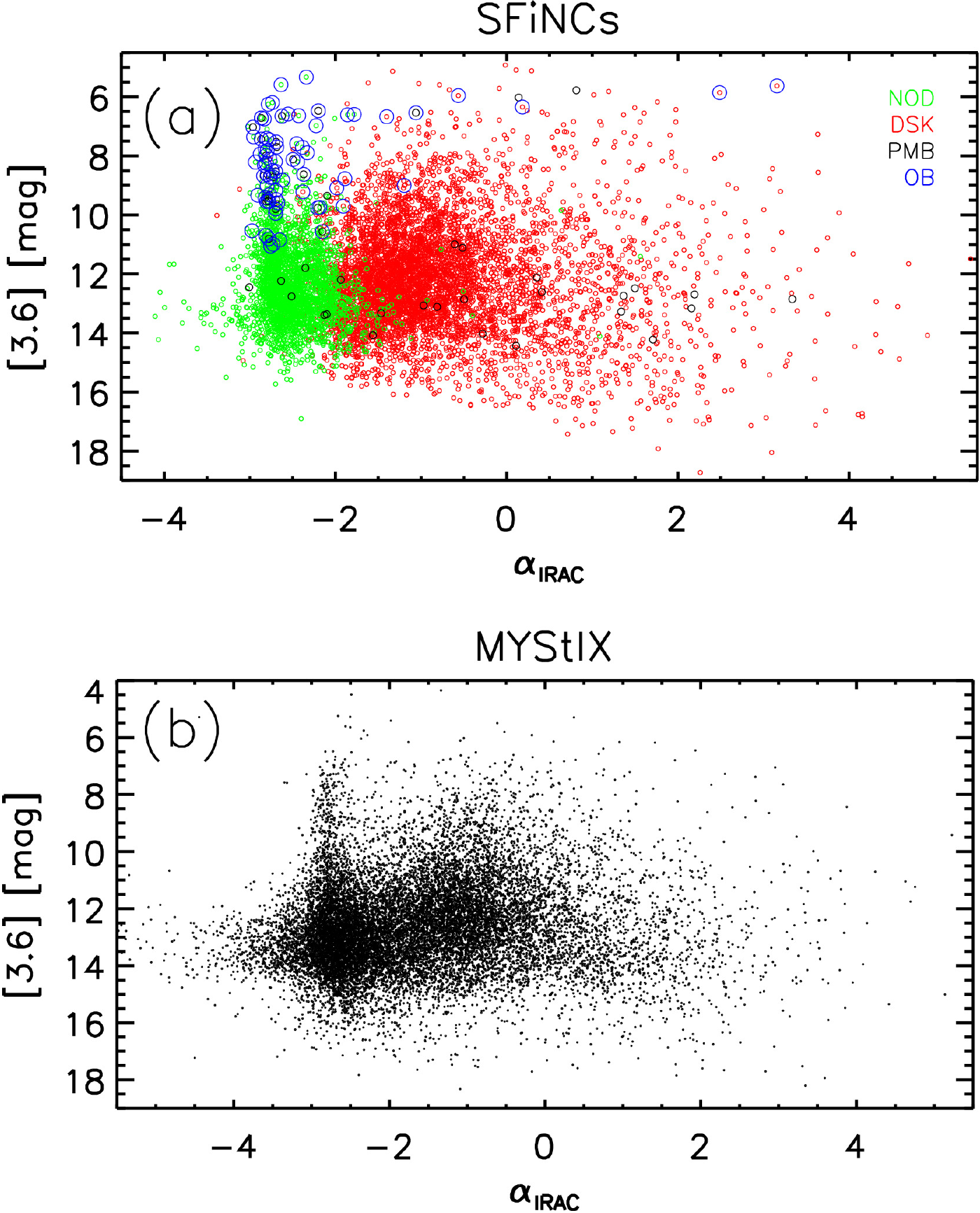}
\caption{Examples of some of the global apparent and intrinsic MIR, NIR, and X-ray properties of the SFiNCs (upper panels) and MYStIX (lower panels) probable cluster members. These MYStIX quantities are tabulated in \citet{Broos2013,Getman2014a}.  The SFiNCs SPCMs are color-coded according to their classes: diskless YSOs in green, disky YSOs in red, YSOs without disk classification in black, and OB-type stars in blue. (a,b) MIR [3.6] magnitude versus apparent IRAC SED slope. (c,d) NIR $J-H$ color versus X-ray median energy. (e,f) NIR $K$-band  magnitude versus X-ray net counts. (g,h) NIR absolute $J$-band magnitude corrected for source extinction versus X-ray intrinsic source luminosity. (i,j) Stellar age versus visual source extinction. \label{fig_global}}
\end{figure*}

\begin{figure*}
\centering
\includegraphics[angle=0.,width=7.0in]{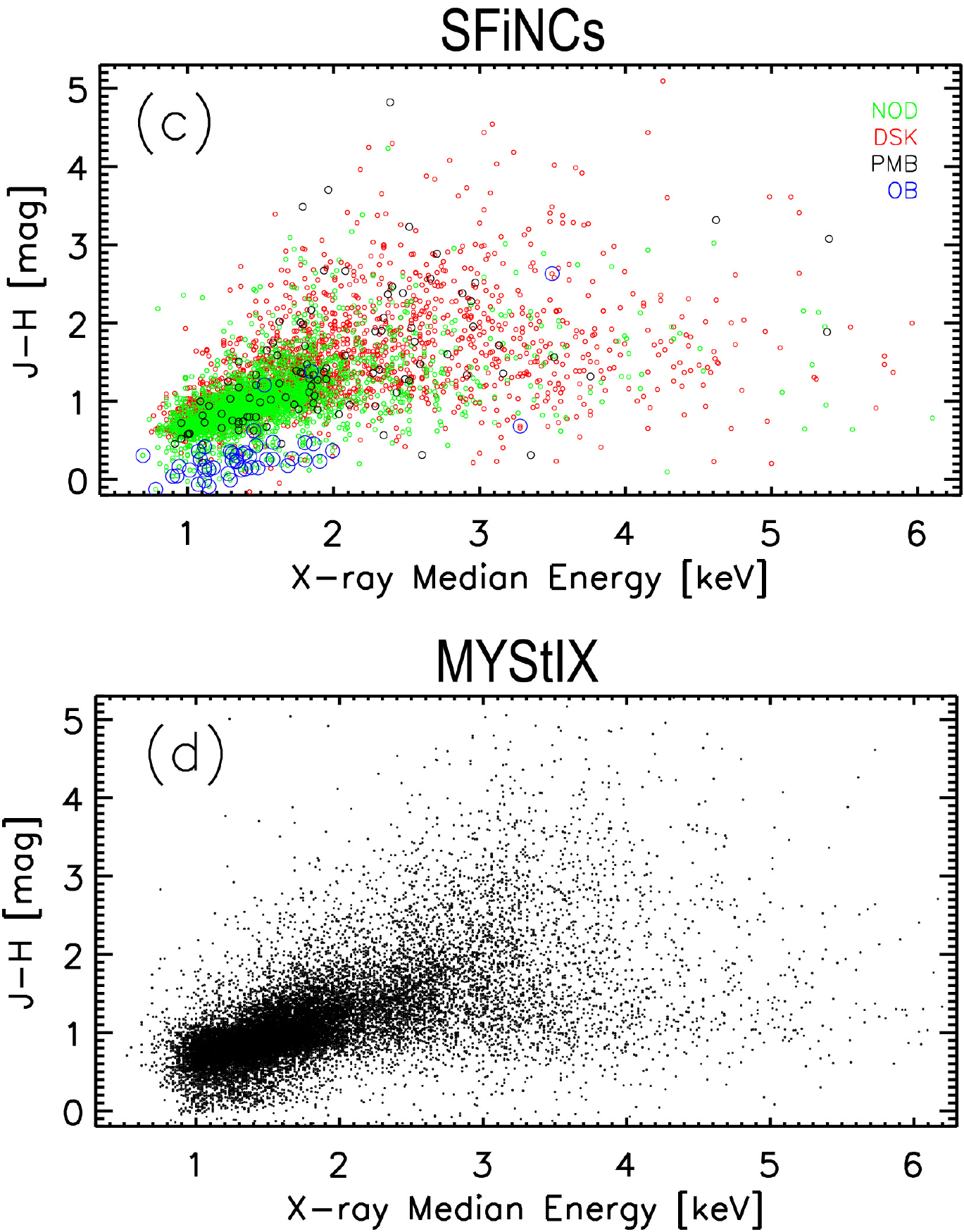}
\end{figure*}

\begin{figure*}
\centering
\includegraphics[angle=0.,width=7.0in]{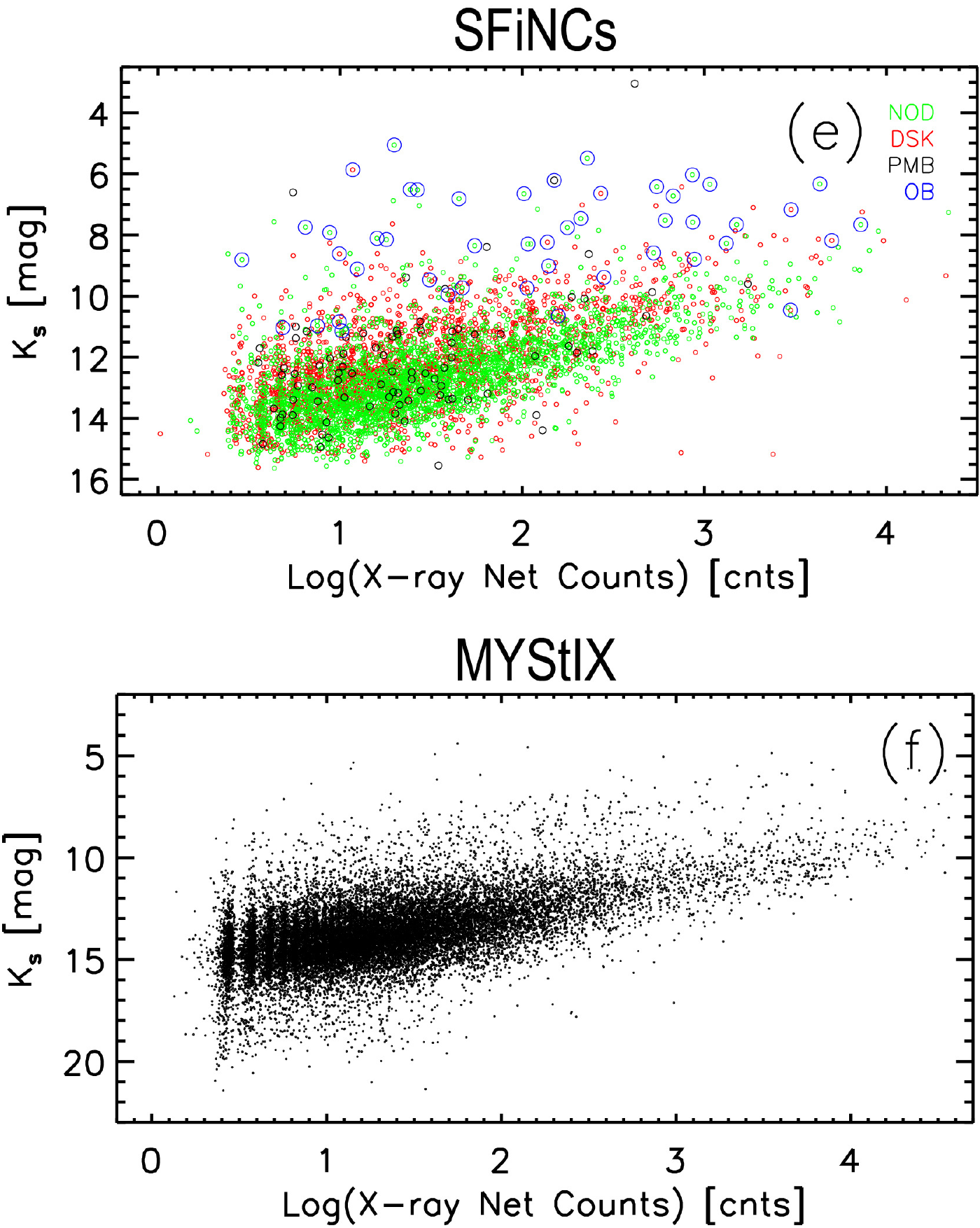}
\end{figure*}

\begin{figure*}
\centering
\includegraphics[angle=0.,width=7.0in]{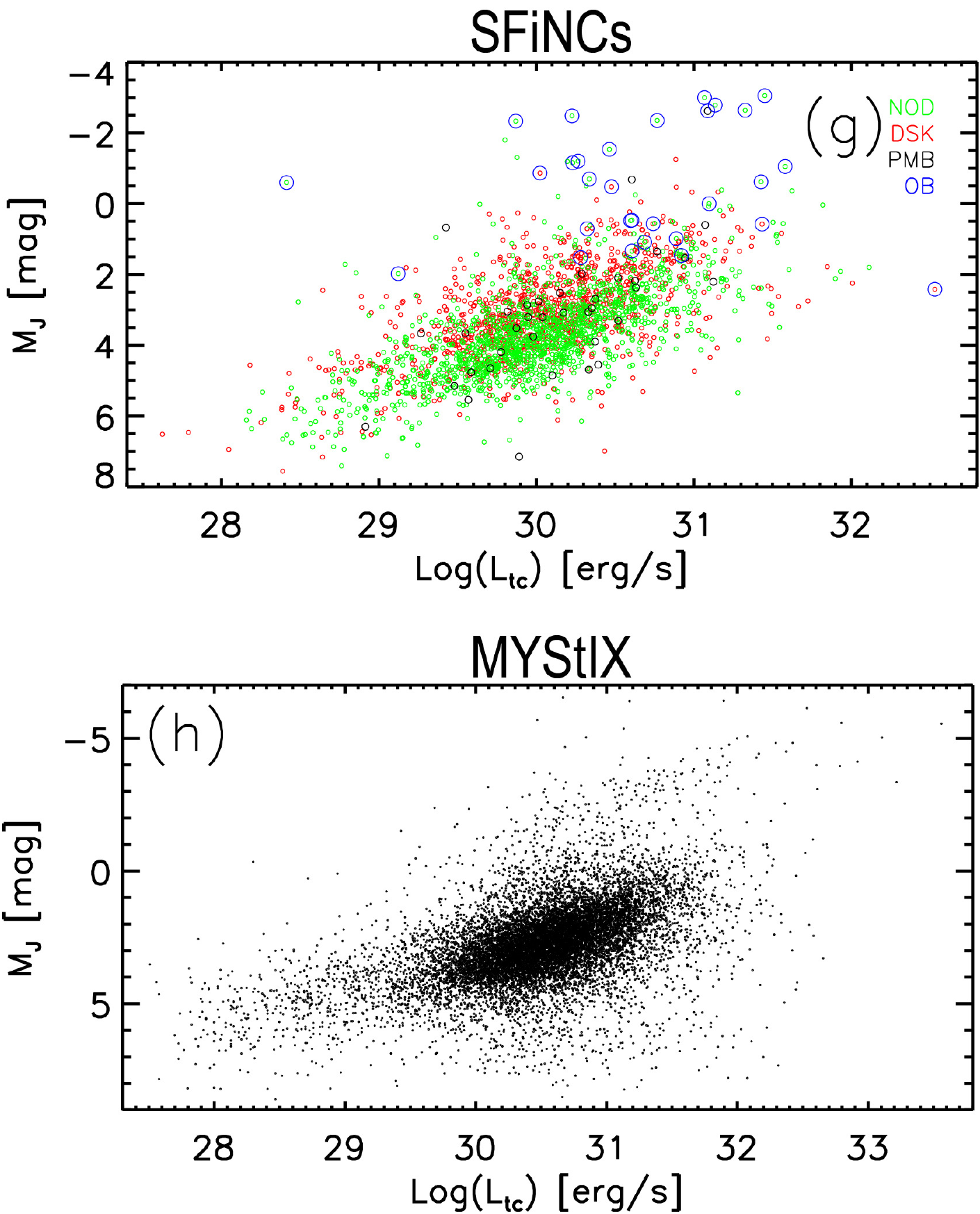}
\end{figure*}

\begin{figure*}
\centering
\includegraphics[angle=0.,width=7.0in]{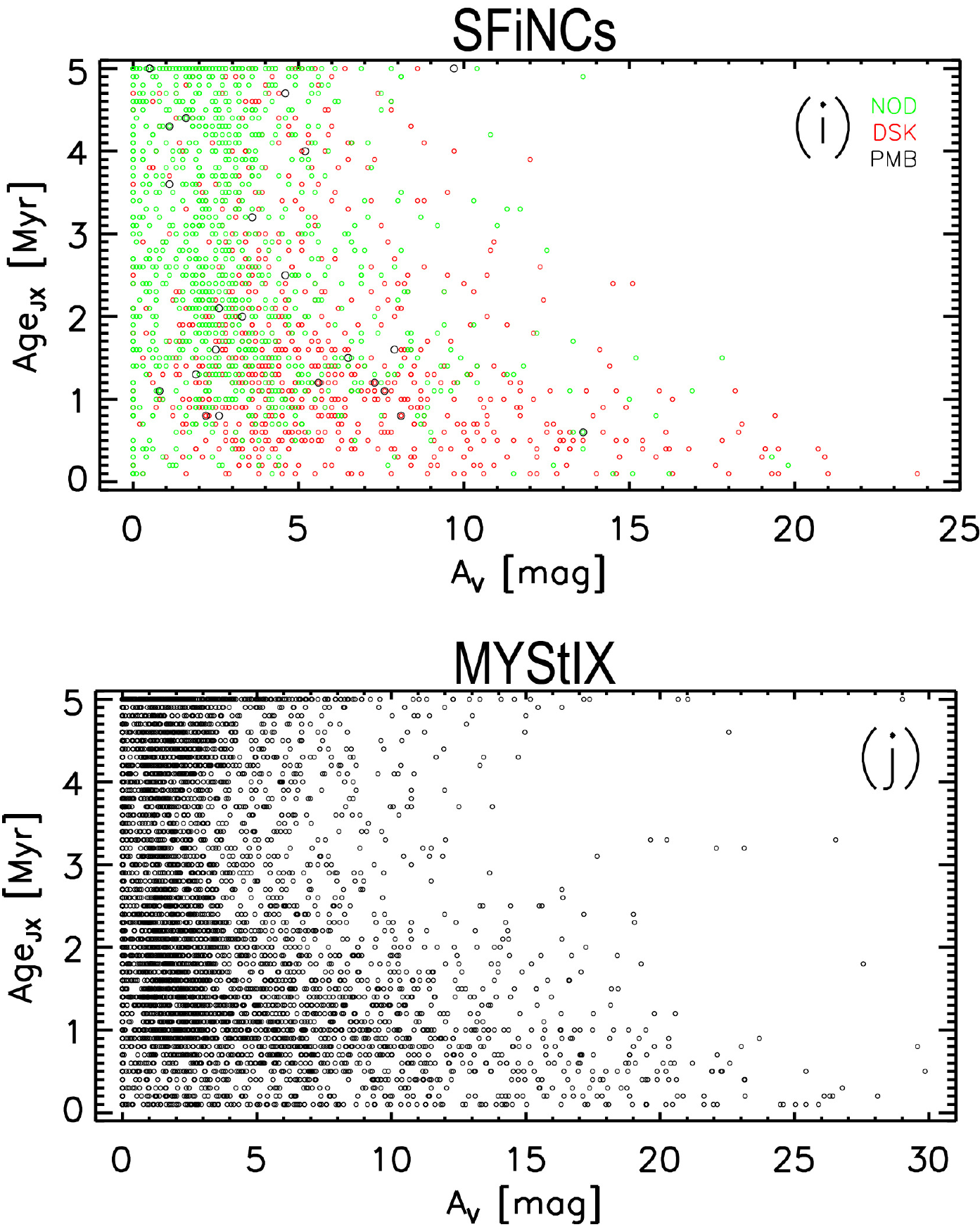}
\end{figure*}

\begin{figure*}
\centering
\includegraphics[angle=0.,width=5.in]{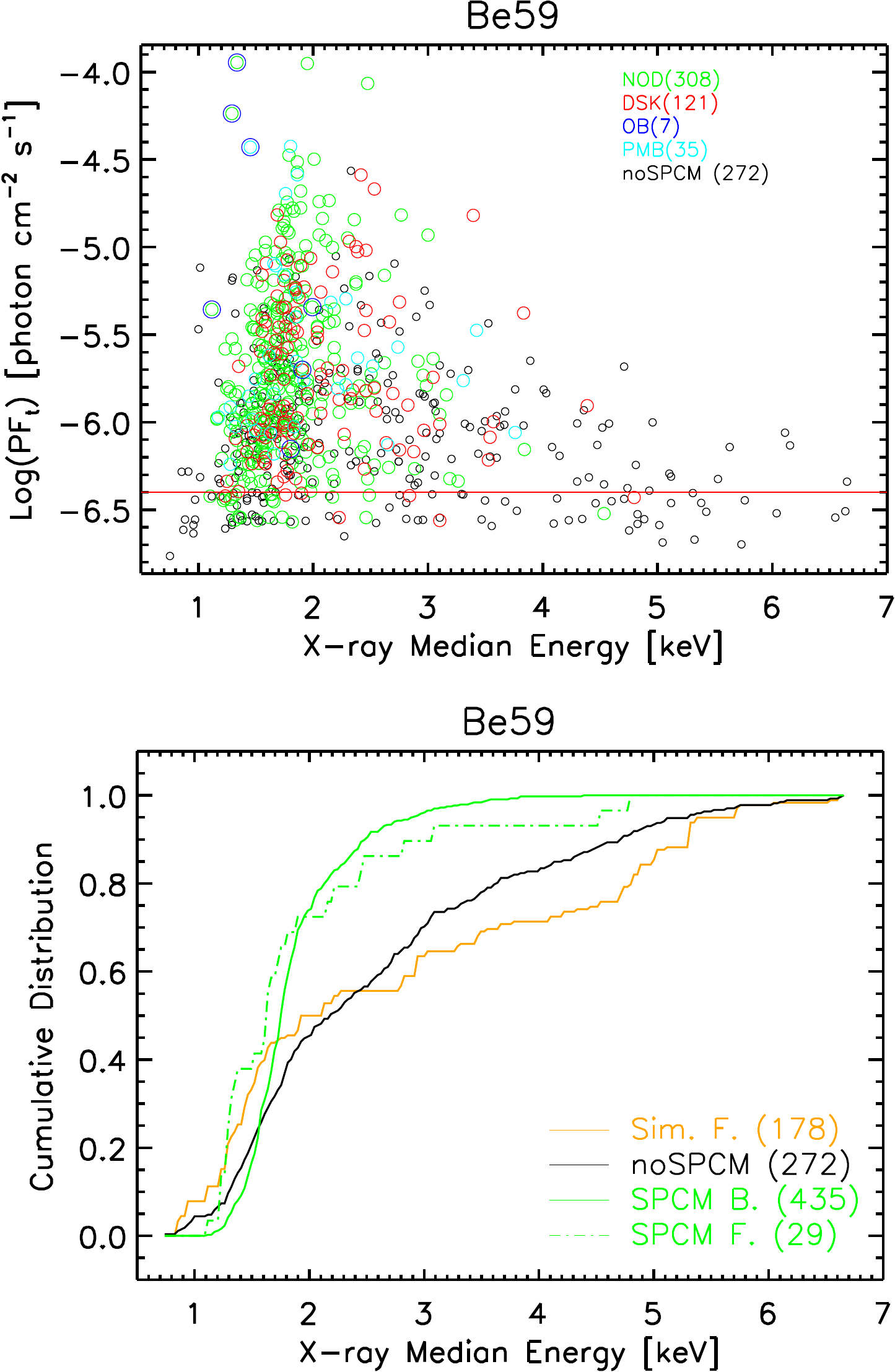}
\caption{Comparison between the bright and faint X-ray SPCMs. An example is given for the Be59 SFR. The figure set presenting all SFiNCs SFRs is available in the on-line journal. To assist with the review process a single pdf file comprising the entire figure set is also provided (f9\_figureset\_merged\_fx\_me.pdf)  (a) Apparent X-ray photometric flux versus X-ray median energy. All ACIS-I X-ray SPCM and non-SPCM sources are plotted.  Diskless, disky, OB-type, and PMB SPCMs are marked as green, red, blue, and cyan circles, respectively. Non-SPCMs are in black. Bright and faint X-ray SPCMs are separated by the red line. (b) Cumulative distributions of X-ray median energy are compared among the bright X-ray SPCMs (solid green), faint X-ray SPCMs (dashed green), non-SPCMs (black), and simulated faint X-ray sources with spurious 2MASS matches produced by the random position shifting via 100 draws (orange). The figure legends provide numbers of plotted source samples. \label{fig_fx_vs_me}}
\end{figure*}

\begin{figure*}
\centering
\includegraphics[angle=0.,width=6.5in]{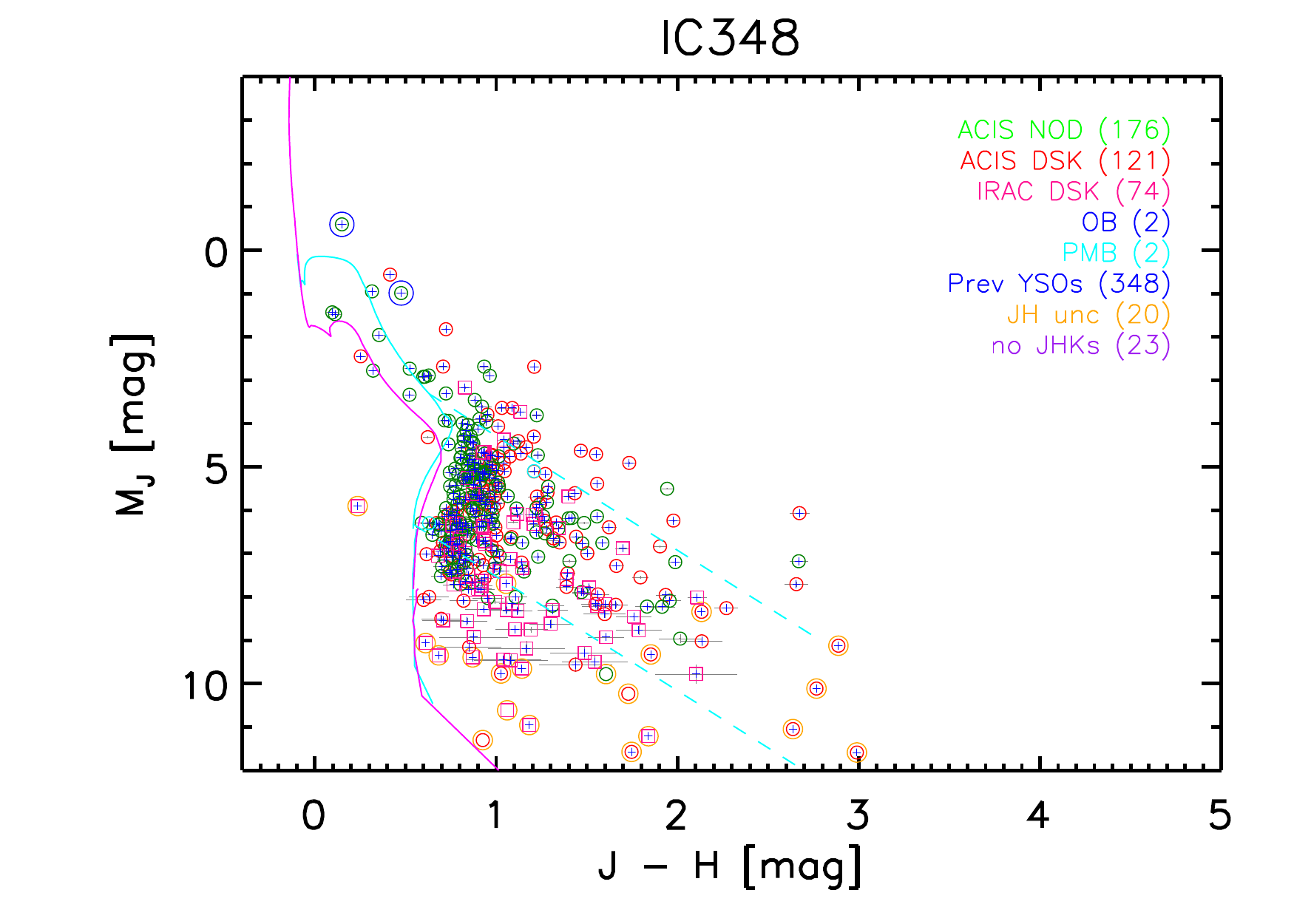}
\caption{2MASS color-magnitude diagrams in $J$- and $H$-bands for SPCMs stratified by YSO class. An example is given for IC~348 SFR. The figure set presenting other SFiNCs SFRs is available in the on-line journal. To assist with the review process a single pdf file comprising the entire figure set is also provided (f10\_figureset\_merged\_j\_vs\_jh.pdf). SPCM sources are color-coded according to their YSO class: IRAC disky (pink squares), ACIS disky (red circles), ACIS diskless (green circles), PMBs (cyan circles), and OB-type stars (blue circles). Sources that have uncertain NIR magnitudes are additionally marked by large orange circles (one or more NIR magnitudes are often upper limits). SPCMs that have been listed in previous YSO catalogs are further marked by small blue crosses. The unabsorbed 3~Myr (cyan) and 10~Myr (magenta) PMS isochrones are from \citet{Bressan2012} and \citet{Baraffe2015} for $M>0.1$~M$_{\odot}$ and $M<0.1$~M$_{\odot}$, respectively. In the case of 3~Myr isochrones, their $A_V = 20$~mag reddening vectors (using the extinction law from \citet{Cardelli1989}) for 0.1 and 1~M$_{\odot}$ are shown as dashed cyan lines. Uncertainties on the $J$ and $H$ magnitudes are shown as gray error bars.  The legends indicate the numbers of plotted sources; the number of SPCMs without 2MASS counterparts is also indicated, in purple color.  The values on y-axis are absolute magnitudes (corrected for distance). \label{fig_j_vs_jh}}
\end{figure*}

\begin{figure*}
\centering
\includegraphics[angle=0.,width=6.5in]{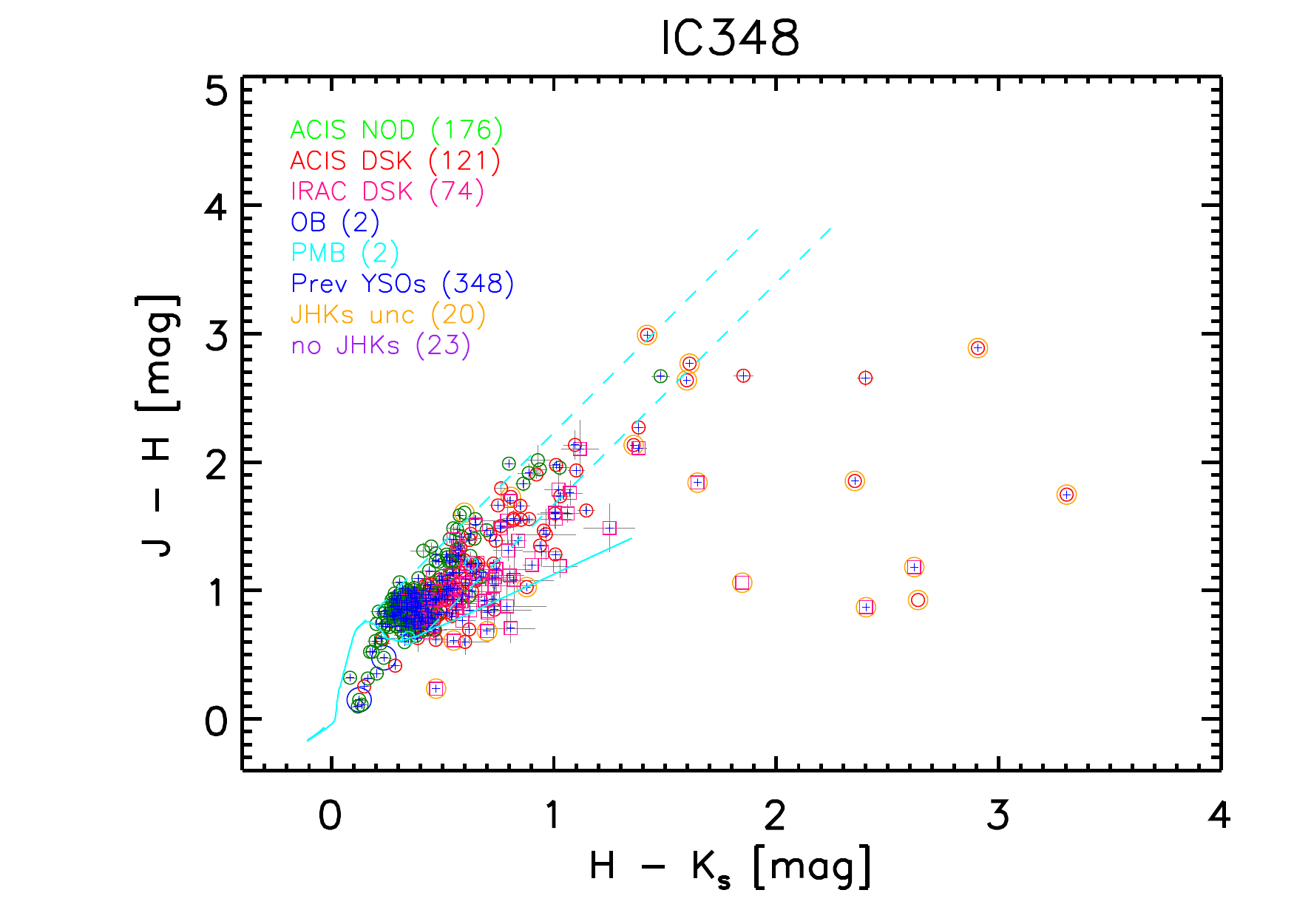}
\caption{2MASS color-color diagrams in $JHK_s$ bands for SPCMs stratified by YSO class. An example is given for IC~348 SFR. The figure set presenting other SFiNCs SFRs is available in the on-line journal. To assist with the review process a single pdf file comprising the entire figure set is also provided (f11\_figureset\_merged\_jh\_vs\_hk.pdf). SPCM sources are color-coded according to their YSO class: IRAC disky (pink squares), ACIS disky (red circles), and ACIS diskless (green circles), PMBs (cyan circles), and OB-type stars (blue circles). Sources that have uncertain NIR magnitudes are additionally marked by large orange circles (one or more NIR magnitudes are often upper limits). SPCMs that have been listed in previous YSO catalogs are further marked by small blue crosses.  The solid cyan curve shows the locus of unabsorbed 3~Myr stars from \citet{Bressan2012} joined with the baseline for de-reddening disky YSOs \citet{Getman2014a}. The $A_V = 30$~mag reddening vectors (using the extinction law from \citet{Cardelli1989}) for 0.1 and 0.8~M$_\odot$ are shown as dashed cyan lines. Uncertainties on the individual photometric colors are shown as gray error bars. The legends indicate the numbers of plotted sources; the number of SPCMs without 2MASS counterparts is also indicated, in purple color. \label{fig_jh_vs_hk}}
\end{figure*}

\begin{figure*}
\centering
\includegraphics[angle=0.,width=6.5in]{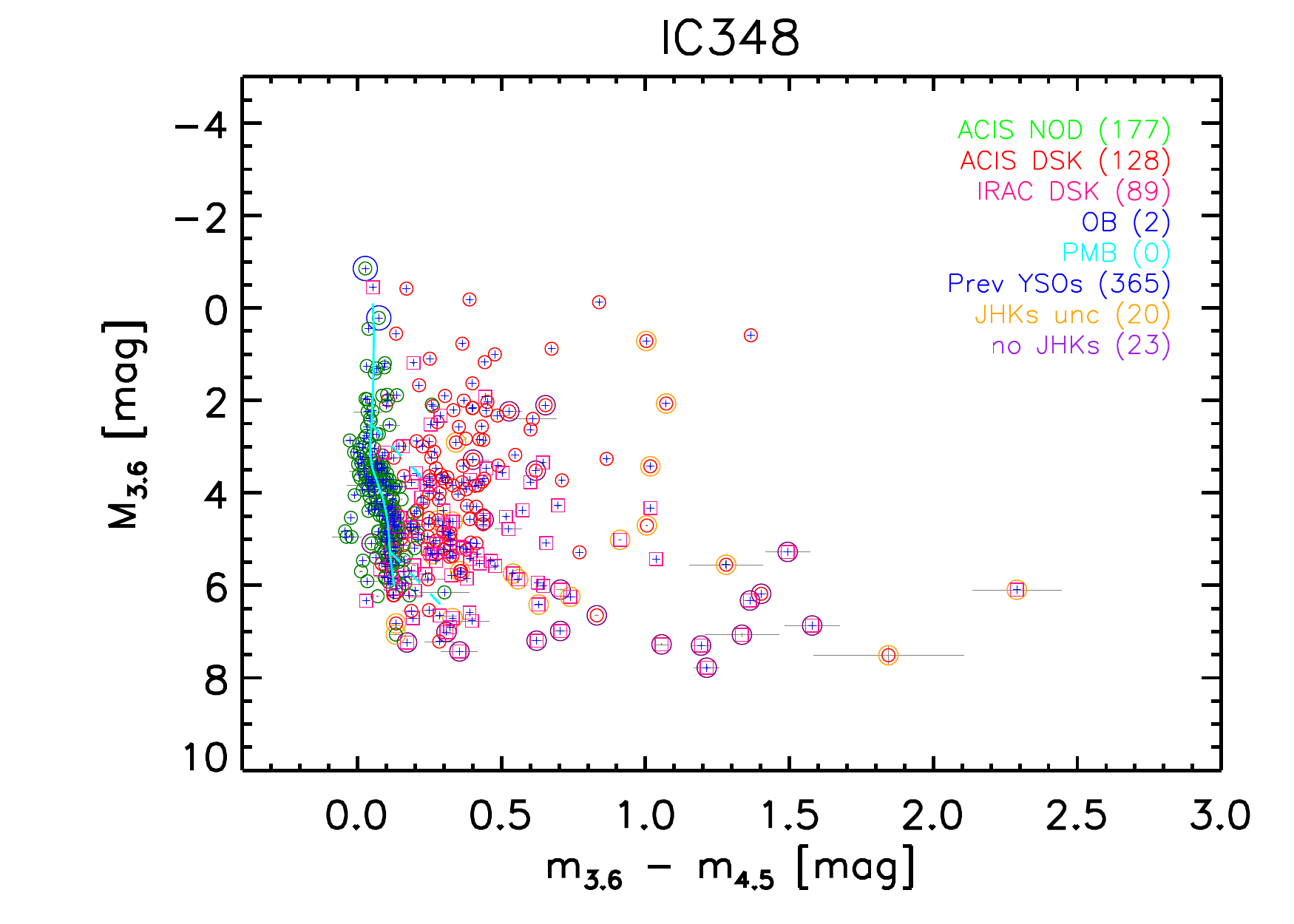}
\caption{IRAC MIR color-magnitude diagrams in [3.6] and [4.5] for SPCMs stratified by YSO class. An example is given for IC~348 SFR. The figure set presenting other SFiNCs SFRs is available in the on-line journal. To assist with the review process a single pdf file comprising the entire figure set is also provided (f12\_figureset\_merged\_ch1\_vs\_ch12.pdf). SPCM sources are color-coded according to their YSO class: IRAC disky (pink squares), ACIS disky (red circles), ACIS diskless (green circles), PMBs (cyan circles), and OB-type stars (blue circles). Sources that have uncertain NIR magnitudes are additionally marked by large orange circles (one or more NIR magnitudes are often upper limits); and sources without 2MASS counterparts are marked by large purple circles. For every SFiNCs SFR, the solid cyan curve represents the smoothed version of the locus of the IC~348 diskless SPCMs. The reddening vectors of $A_K = 2$~mag (using the extinction law from \citet{Flaherty2007}), originating from the IC~348 locus at $\sim 1$ and $0.1$~M$_\odot$ (according to the 3~Myr PMS models of \citet{Baraffe2015}), are shown as dashed cyan lines. Uncertainties on the [3.6] and [4.5] magnitudes are shown as gray error bars. SPCMs that have been listed in previous YSO catalogs are further marked by small blue crosses. The legends indicate the numbers of plotted sources. The values on y-axis are absolute magnitudes (corrected for distance). \label{fig_ch1_vs_ch12}}
\end{figure*}

\begin{figure*}
\centering
\includegraphics[angle=0.,width=6.5in]{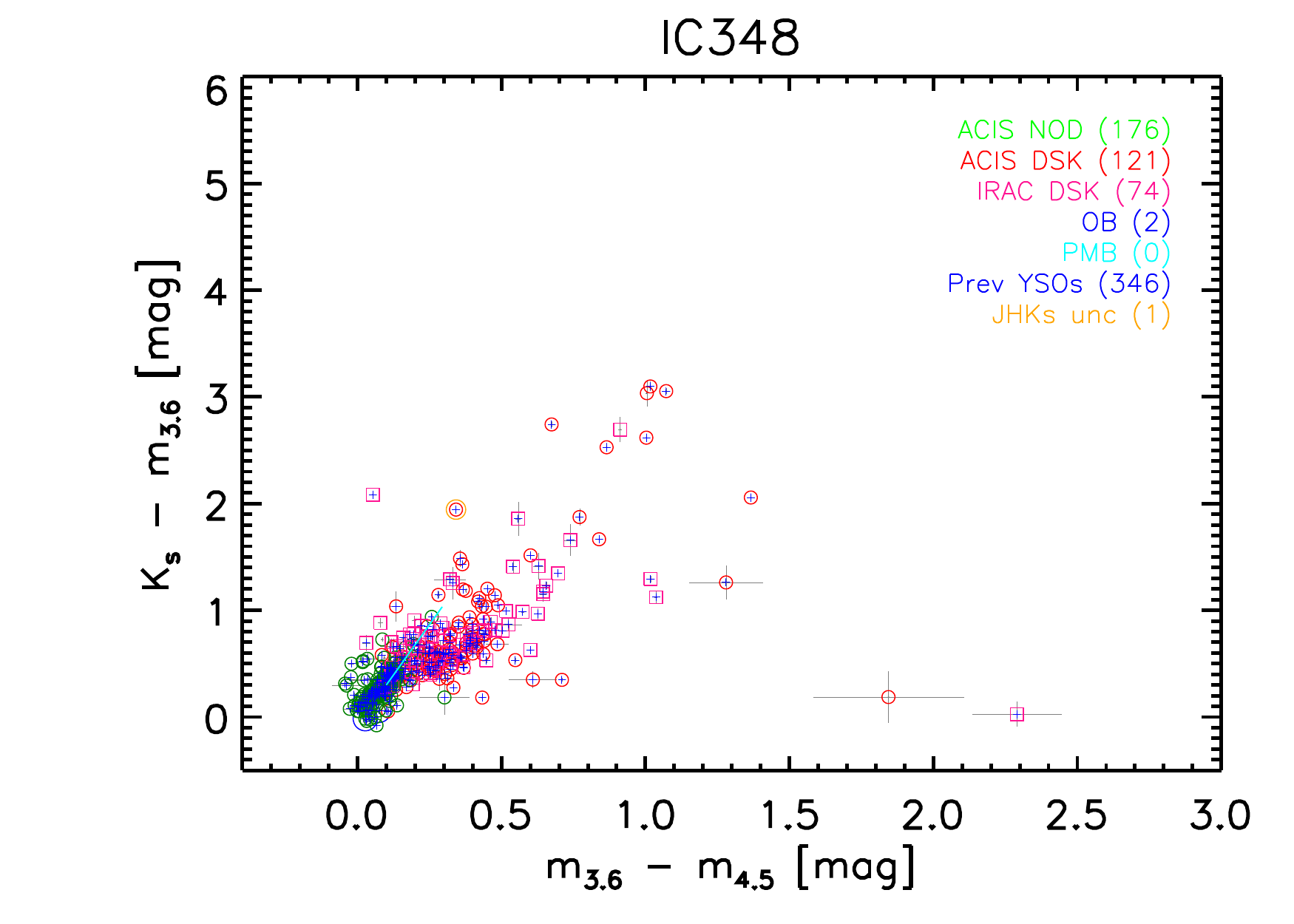}
\caption{IRAC MIR color-color diagrams in $K_s$, [3.6], and [4.5] bands for SPCMs stratified by YSO class. An example is given for IC~348 SFR. The figure set presenting other SFiNCs SFRs is available in the on-line journal. To assist with the review process a single pdf file comprising the entire figure set is also provided (f13\_figureset\_merged\_ksch1\_vs\_ch12.pdf). SPCM sources are color-coded according to their YSO class: IRAC disky (pink squares), ACIS disky (red circles), ACIS diskless (green circles), PMBs (cyan circles), and OB-type stars (blue circles). Sources that have uncertain NIR magnitudes are additionally marked by large orange circles (one or more NIR magnitudes are often upper limits). The reddening vector of $A_K = 2$~mag (using the extinction law from \citet{Flaherty2007}) is shown by the cyan line.  Uncertainties on the individual photometric colors are shown as gray error bars. SPCMs that have been listed in previous YSO catalogs are further marked by small blue crosses. The legends indicate the numbers of plotted sources. \label{fig_ksch1_vs_ch12}}
\end{figure*}

\begin{figure*}
\centering
\includegraphics[angle=0.,width=6.5in]{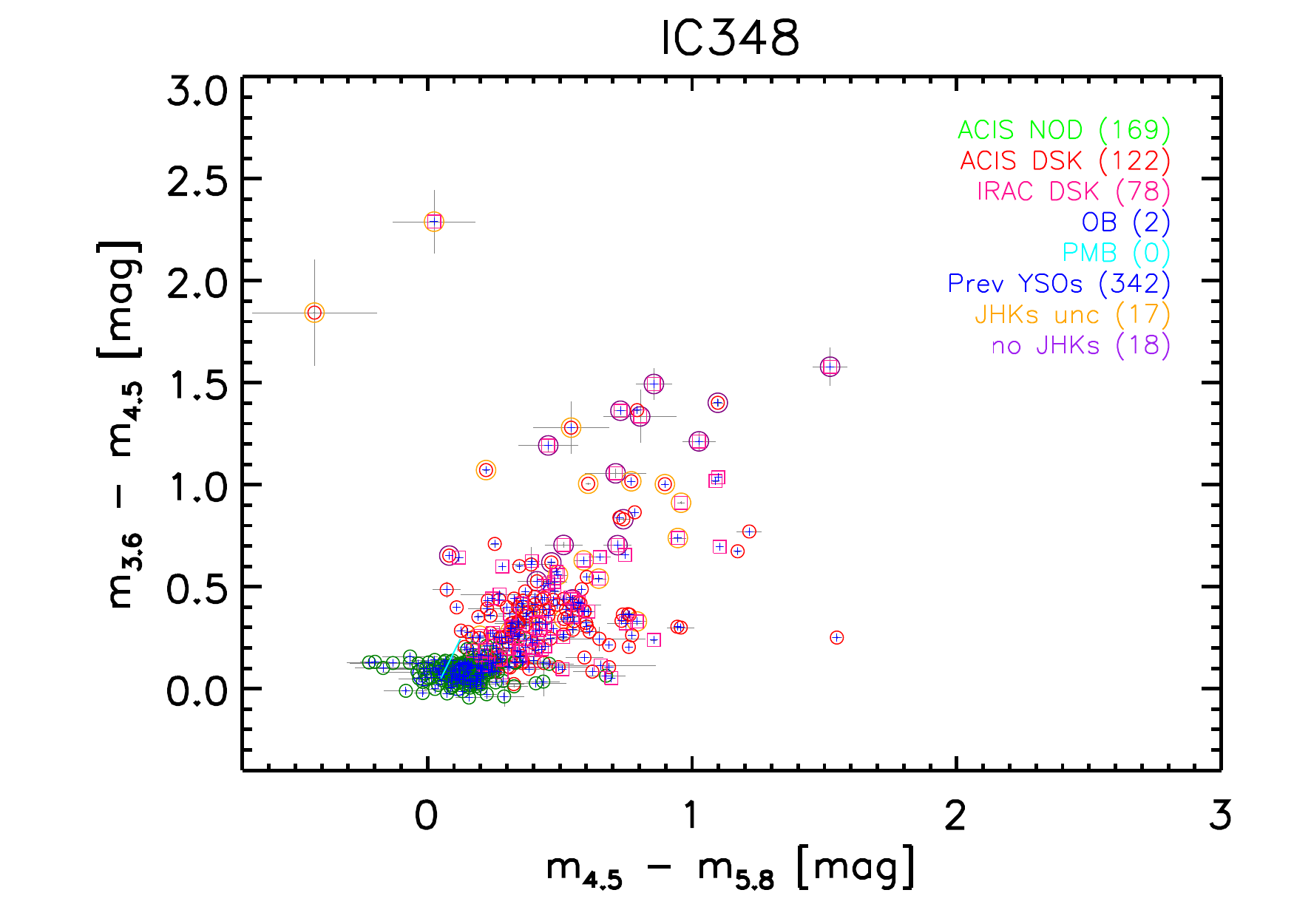}
\caption{IRAC MIR color-color diagrams in [3.6], [4.5], and [5.8] bands for SPCMs stratified by YSO class. An example is given for IC~348 SFR. The figure set presenting other SFiNCs SFRs is available in the on-line journal. To assist with the review process a single pdf file comprising the entire figure set is also provided (f14\_figureset\_merged\_ch12\_vs\_ch23.pdf). SPCM sources are color-coded according to their YSO class: IRAC disky (pink squares), ACIS disky (red circles), ACIS diskless (green circles), PMBs (cyan circles), and OB-type stars (blue circles). Sources that have uncertain NIR magnitudes are additionally marked by large orange circles (one or more NIR magnitudes are often upper limits); and sources without 2MASS counterparts are marked by large purple circles. The reddening vector of $A_K = 2$~mag (using the extinction law from \citet{Flaherty2007}) is shown by the cyan line.  Uncertainties on the individual photometric colors are shown as gray error bars. SPCMs that have been listed in previous YSO catalogs are further marked by small blue crosses. The legends indicate the numbers of plotted sources. \label{fig_ch12_vs_ch23}}
\end{figure*}

\begin{figure*}
\centering
\includegraphics[angle=0.,width=6.5in]{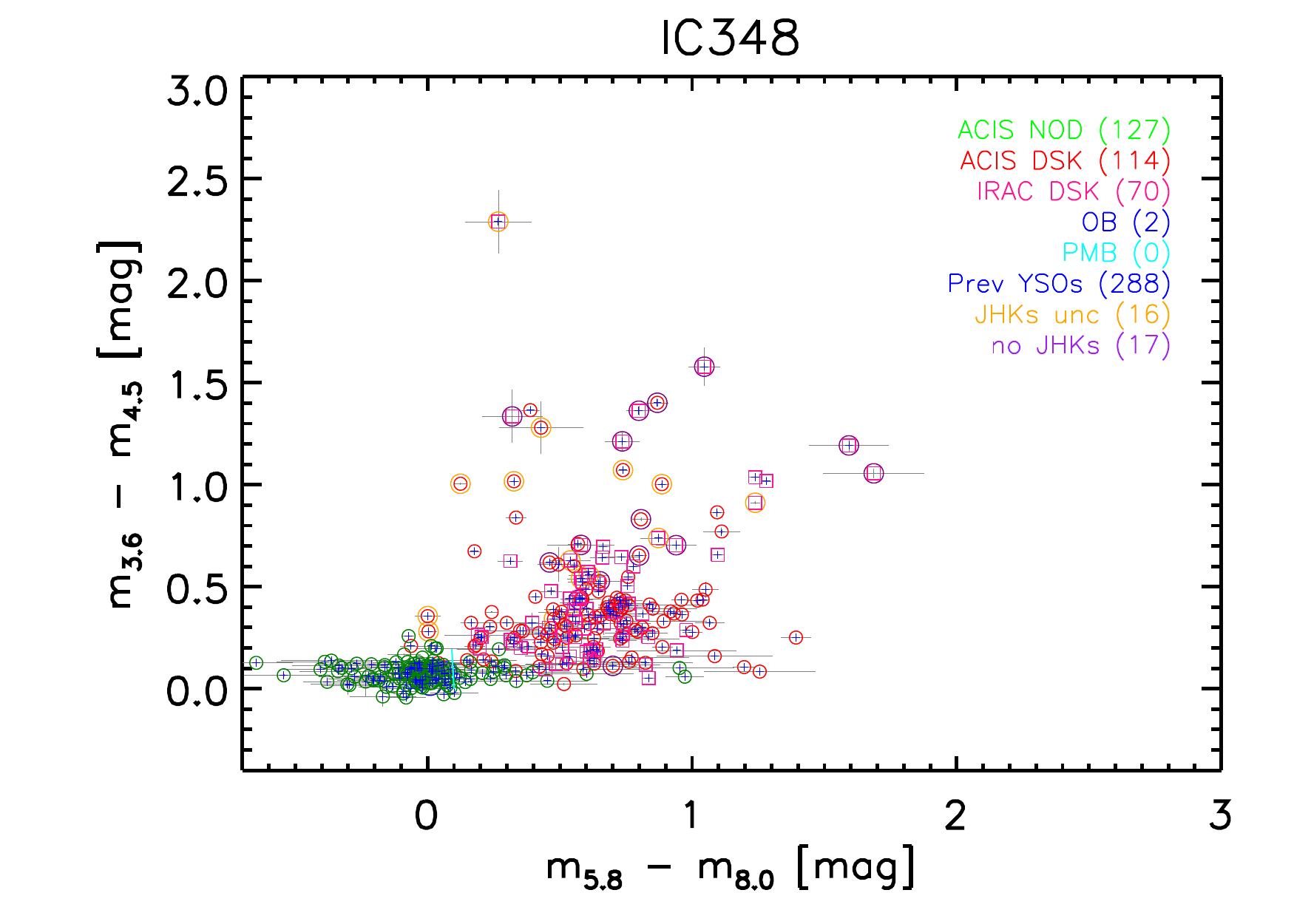}
\caption{IRAC MIR color-color diagrams in [3.6], [4.5], [5.8], and [8.0] bands for SPCMs stratified by YSO class. An example is given for IC~348 SFR. The figure set presenting other SFiNCs SFRs is available in the on-line journal. To assist with the review process a single pdf file comprising the entire figure set is also provided (f15\_figureset\_merged\_ch12\_vs\_ch34.pdf). SPCM sources are color-coded according to their YSO class: IRAC disky (pink squares), ACIS disky (red circles), and ACIS diskless (green circles). Sources that have uncertain NIR magnitudes are additionally marked by large orange circles (one or more NIR magnitudes are often upper limits); and sources without 2MASS counterparts are marked by large purple circles. The reddening vector of $A_K = 2$~mag (using the extinction law from \citet{Flaherty2007}) is shown by the cyan line.  Uncertainties on the individual photometric colors are shown as gray error bars. SPCMs that have been listed in previous YSO catalogs are further marked by small blue crosses. The legends indicate the numbers of plotted sources. \label{fig_ch12_vs_ch34}}
\end{figure*}

\begin{figure*}
\centering
\includegraphics[angle=0.,width=6.5in]{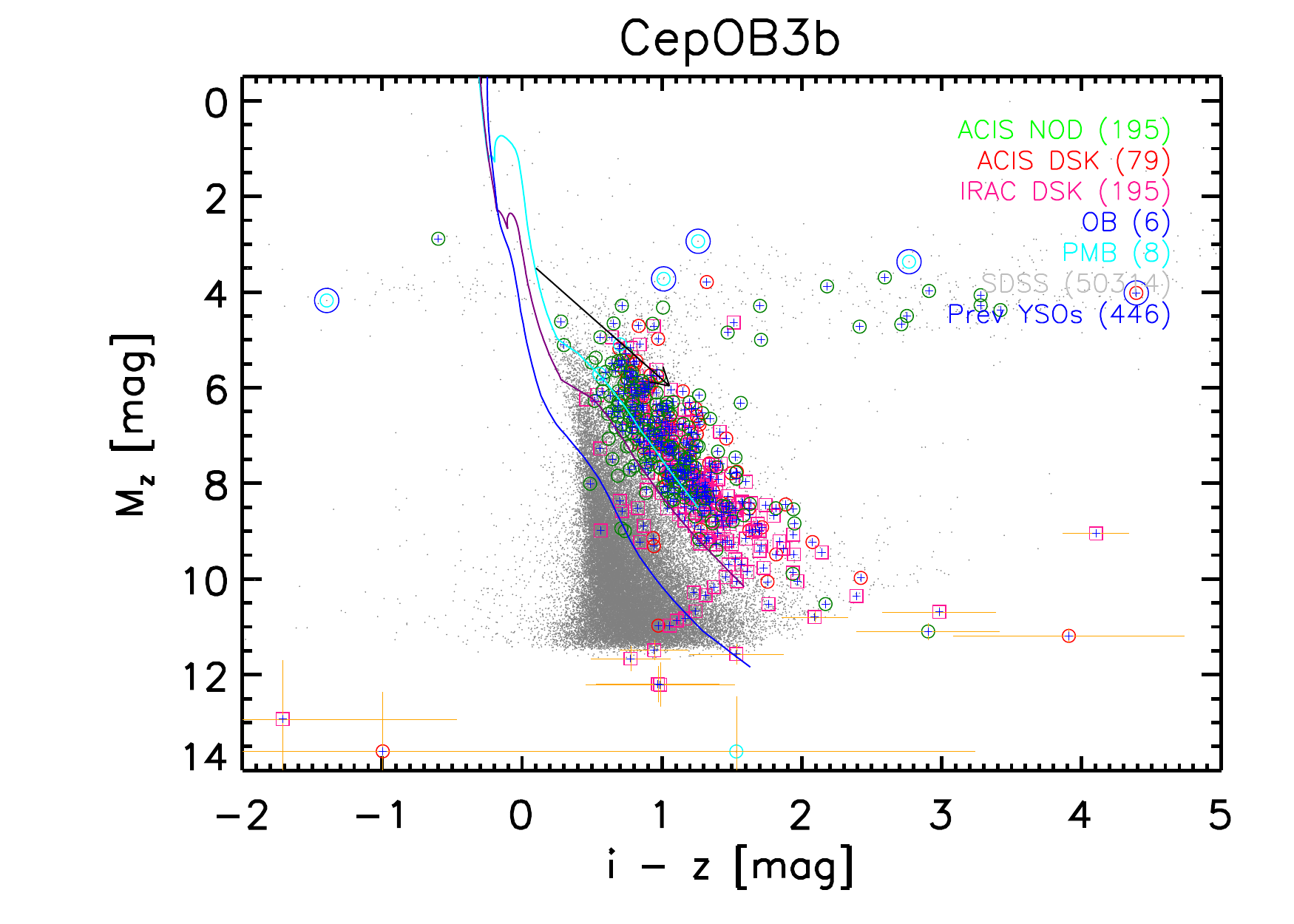}
\caption{SDSS optical color-magnitude diagrams in the $I$ and $Z$ bands for SPCMs stratified by YSO class. The SDSS DR~12 data \citep{Alam2015} are available for the Orion and Cepheus as well as IC~5146 and Serpens Main SFiNCs regions. For any of those SFRs, the SDSS coverage is only partial due to the presence of diffuse nebular background. An example is given here for the Cep~OB3b SFR. The figure set presenting other SFiNCs SFRs with available SDSS data can be found in the on-line journal. To assist with the review process a single pdf file comprising the entire figure set is also provided (f16\_figureset\_merged\_z\_vs\_iz.pdf).  All SPCMs with available SDSS photometry are plotted as colored circles. All SDSS sources in the field that have photometric errors below 0.2~mag are plotted as gray dots.  The SPCM sources are color-coded according to their YSO class: IRAC disky (pink squares), ACIS disky (red circles), ACIS diskless (green circles), PMBs (cyan circles), and OB-type stars (blue circles). Error bars (orange) are shown for SPCMs with photometric errors above 0.2~mag. The unabsorbed 3~Myr (cyan), 10~Myr (magenta), and 100~Myr (blue) PMS isochrones are from \citet{Bressan2012}. Reddening vector $A_V = 5$~mag (using the extinction law from \citet{Cardelli1989}) is shown as the black arrow. SPCMs that have been listed in previous YSO catalogs are further marked by small blue crosses. The legends indicate the numbers of plotted sources. The values on y-axis are absolute magnitudes (corrected for distance). \label{fig_z_vs_iz}}
\end{figure*}

\begin{figure*}
\centering
\includegraphics[angle=0.,width=7.0in]{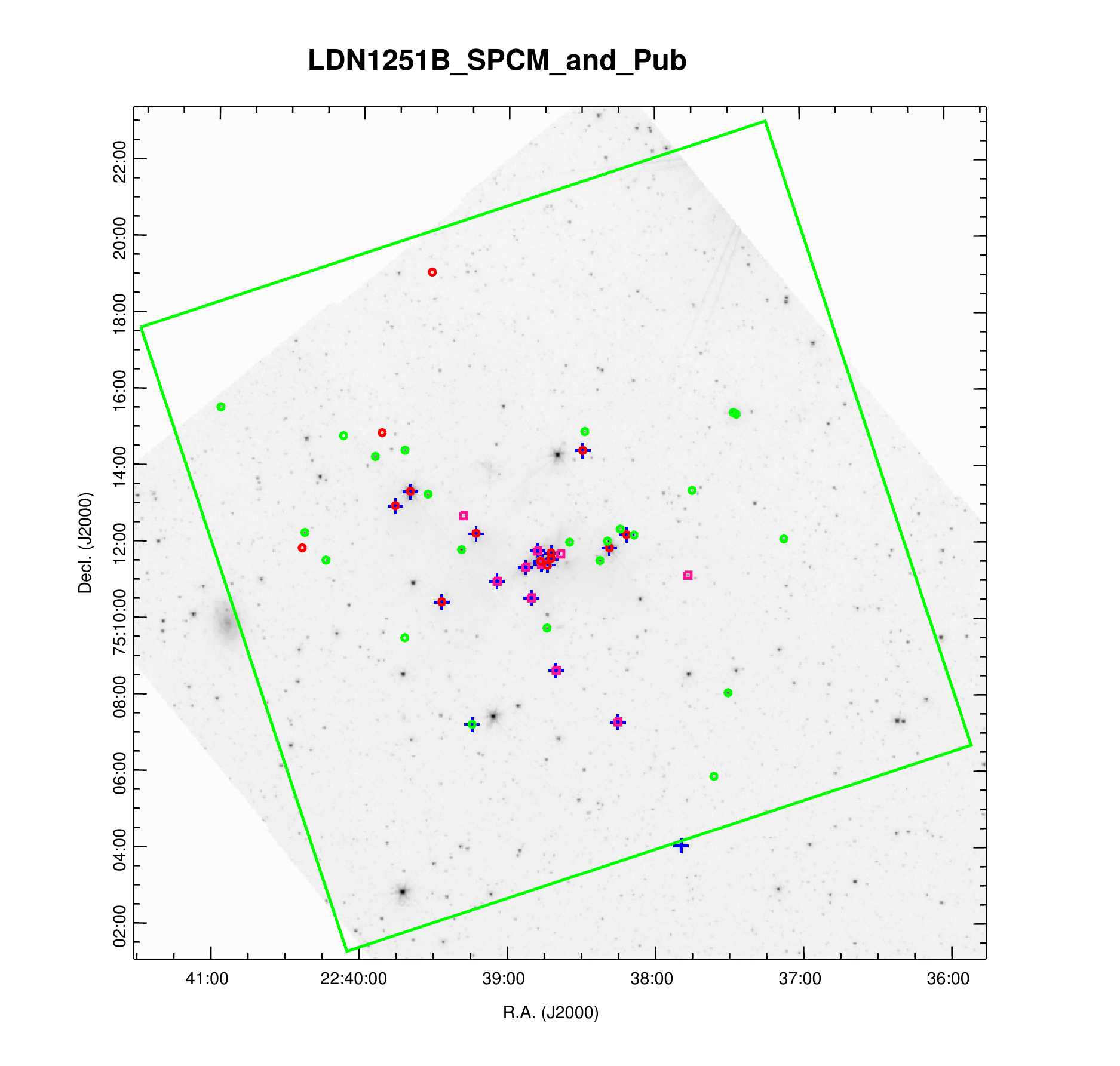}
\caption{SPCMs (green, red, pink, cyan, and blue circles) and YSOs from previously published catalogs (blue $+$) superimposed on the low-resolution IRAC-[3.6] images. SPCM's colors and numbers are similar to Figure \ref{fig_spcm_maps}. An example is given for the LDN~1251B SFR. The figure set presenting other SFiNCs SFRs is available in the on-line journal; this omits Be~59 and RCW~120 due to the absence of previous IR/X-ray YSO catalogs. To assist with the review process a single pdf file comprising the entire figure set is also provided (f17\_figureset\_merged\_spcm\_pub\_maps.pdf). The images are shown in inverted colors with logarithmic scales. {\it Chandra}-ACIS-I field of view is outlined by the green polygons.  \label{fig_spcm_pub_maps}}
\end{figure*}

\begin{figure*}
\centering
\includegraphics[angle=0.,width=6.5in]{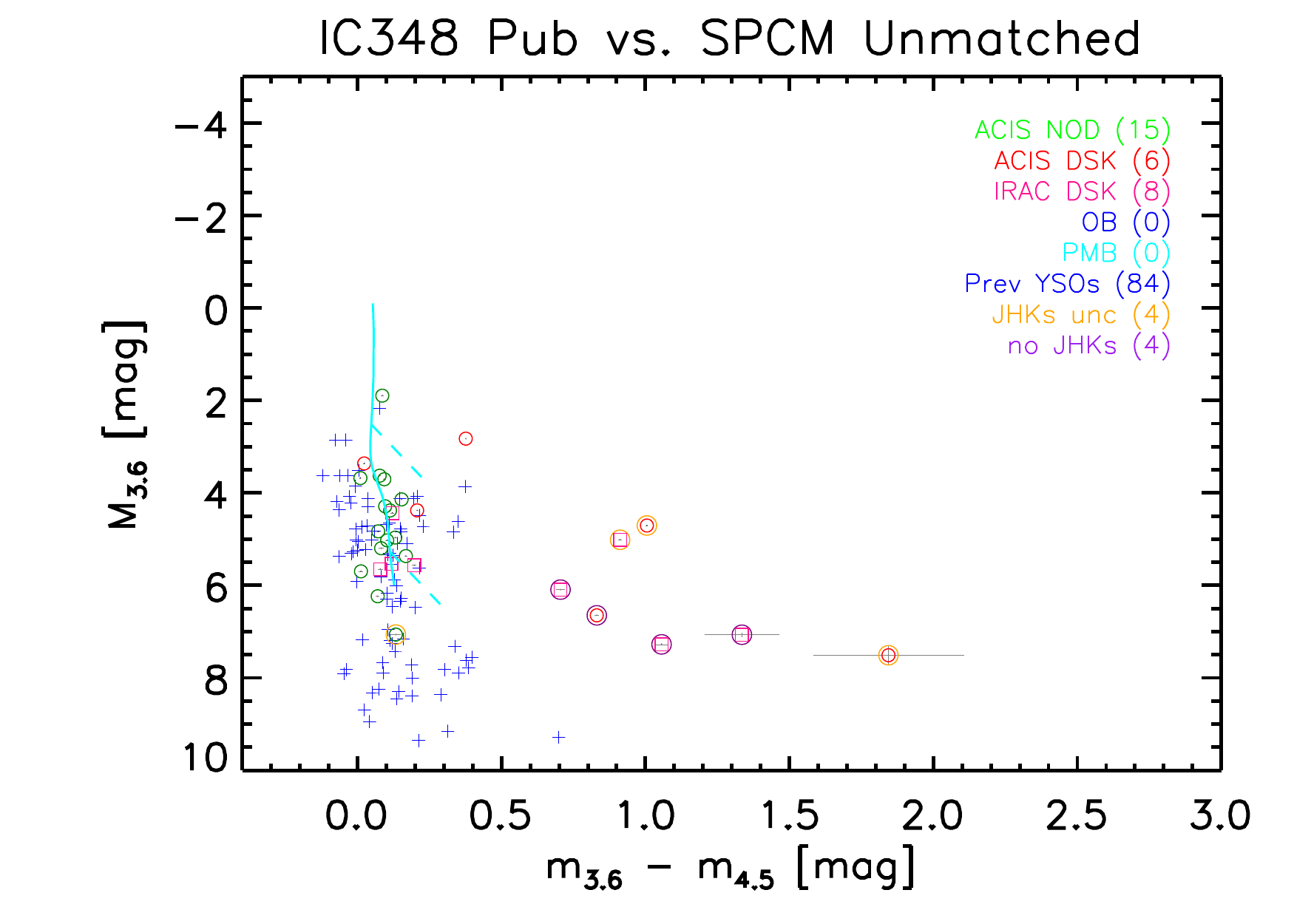}
\caption{IRAC MIR color-magnitude diagrams in [3.6] and [4.5] for the YSOs that are uncommon between the SPCM and previously published catalogs (Table~\ref{tbl_spcm_vs_previous}). An example is given for the IC~348 SFR. The figure set presenting other SFiNCs SFRs is available in the on-line journal; this omits Be~59 and RCW~120 due to the absence of previous IR/X-ray YSO catalogs. To assist with the review process a single pdf file comprising the entire figure set is also provided (f18\_figureset\_merged\_ch1\_vs\_ch12\_nomatch.pdf). YSOs from the previous YSO catalogs are marked by blue crosses. SPCM sources are color-coded according to their YSO class: IRAC disky (pink squares), ACIS disky (red circles), ACIS diskless (green circles), PMBs (cyan circles), and OB-type stars (blue circles). Sources that have uncertain NIR magnitudes are additionally marked by large orange circles (one or more NIR magnitudes are often upper limits); and sources without 2MASS counterparts are marked by large purple circles. For every SFiNCs SFR, the solid cyan curve represents the smoothed version of the locus of the IC~348 diskless SPCMs. The reddening vectors of $A_K = 2$~mag (using the extinction law from \citet{Flaherty2007}), originating from the IC~348 locus at $\sim 1$ and $0.1$~M$_\odot$ (according to the 3~Myr PMS models of \citet{Baraffe2015}), are shown as dashed cyan lines. For SPCMs, the uncertainties on the [3.6] and [4.5] magnitudes are shown as gray error bars. The legends indicate the numbers of plotted sources. The values on y-axis are absolute magnitudes (corrected for distance). \label{fig_ch1_vs_ch12_nomatch}}
\end{figure*}


\floattable 
\begin{deluxetable*}{ccccccccccc}
\centering \tabletypesize{\scriptsize} \rotate \tablewidth{0pt} \tablecolumns{10}

\tablecaption{SFiNCs Star Forming Regions \label{tbl_sfincs_sfrs}}

\tablehead{\colhead{Name} &  \colhead{R.A.(J2000)} & \colhead{Decl.(J2000)} & \colhead{{\it l}} & \colhead{{\it b}} &\colhead{$D$} & \colhead{Dist. Ref.} & \colhead{OB} & \colhead{{\it Spitzer} Ref.} & \colhead{{\it Chandra} Ref.}\\
\colhead{} &  \colhead{$^{h}:^{m}:^{s}$} & \colhead{$\arcdeg$$:$$\arcmin$$:$$\arcsec$} & \colhead{{deg}} & \colhead{{deg}} &\colhead{pc} & \colhead{} & \colhead{} & \colhead{} & \colhead{}\\
\colhead{(1)} & \colhead{(2)} & \colhead{(3)} & \colhead{(4)} & \colhead{(5)} & \colhead{(6)}  & \colhead{(7)}  & \colhead{(8)} & \colhead{(9)} & \colhead{(10)}}

\startdata
NGC 7822 (Be 59) & 00:02:15.27 & +67:25:59.7 & 118.2262 & 5.0120 & 900 & Ma08;Pa08 & O5V & \nodata & \nodata\\
IRAS 00013+6817 (SFO 2) & 00:04:03.85 & +68:33:33.7 & 118.6050 & 6.0865 & 900 & Ma08;Pa08 & \nodata & Gu09 & \nodata\\
NGC 1333 & 03:29:04.76 & +31:20:10.7 & 158.3063 & -20.4982 & 235 & Hi08 & B5V & Gu09 & Wi10\\
IC 348 & 03:44:30.30 & +32:07:44.8 & 160.5025 & -17.8368 & 300 & He08 & B5V & Lu16 & St12\\
LkH$\alpha$ 101 & 04:30:09.86 & +35:16:09.2 & 165.3491 & -9.0203 & 510 & Wo10 & B4 & Gu09 & Wo10\\
NGC 2068-2071 & 05:46:43.54 & +00:07:34.2 & 205.2813 & -14.2981 & 414 & Me07 & B2/3 & Me12 & Gr04;Sk09\\
ONC Flank S & 05:35:05.79 & -05:39:07.9 & 209.2379 & -19.5421 & 414 & Me07 & B9V & Me12 & Ra04\\
ONC Flank N & 05:35:14.38 & -04:49:49.5 & 208.4786 & -19.1380 & 414 & Me07 & B1V & Me12 & Ra04\\
OMC 2-3 & 05:35:23.44 & -05:07:03.1 & 208.7676 & -19.2349 & 414 & Me07 & B0.5V & Me12 & Ts02\\
Mon R2 & 06:07:46.15 & -06:22:53.6 & 213.7009 & -12.6029 & 830 & Ra68 & B0 & Gu09 & Na03\\
GGD 12-15 & 06:10:49.51 & -06:11:45.7 & 213.8772 & -11.8404 & 830 & Ra68 & B8/A0 & Gu09 & \nodata\\
RCW 120 & 17:12:23.25 & -38:28:45.4 & 348.2381 & 0.4639 & 1350 & Za07 & O9: & \nodata & \nodata\\
Serpens Main & 18:29:56.30 & +01:13:11.5 & 31.5647 & 5.3394 & 415 & Dz10 & \nodata & Gu09 & Gi07;Wi07\\
Serpens South & 18:30:02.56 & -02:03:42.2 & 28.6501 & 3.8133 & 415 & Dz10 & \nodata & Po13 & \nodata\\
IRAS 20050$+$2720 & 20:07:06.56 & +27:30:02.3 & 65.7962 & -2.6014 & 700 & Wi89 & B3 & Gu09 & Gu12\\
Sh 2-106 & 20:27:24.36 & +37:22:54.9 & 76.3802 & -0.6137 & 1400 & No05;Sc07 & B0V & Gu09 & Gi04\\
IC 5146 & 21:53:29.82 & +47:15:22.4 & 94.3896 & -5.5133 & 800 & Ga14 & B2V & Gu09 & \nodata\\
NGC 7160 & 21:53:49.98 & +62:35:10.2 & 104.0159 & 6.4318 & 870 & Co02 & B0.5V & Si06 & \nodata\\
LDN 1251B & 22:38:48.05 & +75:11:50.4 & 114.6324 & 14.5068 & 300 & Ku08 & \nodata & Ev03 & Si09\\
Cep OB3b & 22:55:39.32 & +62:38:11.2 & 110.0643 & 2.6936 & 700 & Mo09;Dz11 & O7V & Al12 & Ge06;Al12\\
Cep A & 22:56:14.13 & +62:02:17.8 & 109.8676 & 2.1246 & 700 & Mo09;Dz11 & O9V & Gu09 & Pr09\\
Cep C & 23:05:49.42 & +62:30:27.2 & 111.0778 & 2.0919 & 700 & Mo09;Dz11 & B9: & Gu09 & \nodata\\
\enddata

\tablecomments{Column 1: SFR name. Columns 2-5: Coordinates: right ascension, declination, Galactic longitude, and Galactic latitude. Column 6: Distance from the Sun, in parsecs. Column 7: Literature reference to the distance value. Column 8: The earliest OB star within the {\it Chandra} field from the catalogs of \citet{Skiff2009} and SIMBAD. Columns 9-10: Literature references to the previous {\it Spitzer} and {\it Chandra} studies of the region. Reference code in Columns 7, 9, and 10: Al12 \citep{Allen2012},  Co02 \citep{Contreras2002}, Dz10 \citep{Dzib2010}, Dz11 \citep{Dzib2011}, Ev03 \citep{Evans2003}, Ga14 \citep{GarciaRojas2014}, Ge06 \citep{Getman2006}, Gi04 \citep{Giardino2004}, Gi07 \citep{Giardino2007}, Gr04 \citep{Grosso2004}, Gu09 \citep{Gutermuth2009}, Gu12 \citep{Gunther2012}, He08 \citep{Herbst2008}, Hi08 \citep{Hirota2008},  Ku08 \citep{Kun2008}, Lu16 \citep{Luhman2016}, Ma08 \citep{Majaess2008}, Me07 \citep{Menten2007}, Me12 \citep{Megeath2012}, Mo09 \citep{Moscadelli2009}, Na03 \citep{Nakajima2003}, No05 \citep{Noel2005}, Pa08 \citep{Pandey2008},  Po13 \citep{Povich2013}, Pr09 \citep{Pravdo2009}, Ra68 \citep{Racine1968}, Ra04 \citep{Ramirez2004}, Sc07 \citep{Schneider2007},  Si06 \citep{Sicilia-Aguilar2006}, Si09 \citep{Simon2009}, Sk09 \citep{Skinner2009}, St12 \citep{Stelzer2012}, Ts02 \citep{Tsuboi2002}, Wi89 \citep{Wilking1989}, Wi07 \citep{Winston2007}, Wi10 \citep{Winston2010}, Wo10 \citep{Wolk2010}, Za07 \citep{Zavagno2007}.}

\end{deluxetable*}


\floattable 
\begin{deluxetable*}{crrcrccrcl}
\centering \tabletypesize{\tiny} \tablewidth{0pt} \tablecolumns{10}

\tablecaption{Log of SFiNCs {\em Chandra}-ACIS Observations \label{tbl_chandra_obslog}}

\tablehead{
\colhead{Region} & 
\colhead{ObsID} & 
\colhead{Sequence} & 
\colhead{Start Time} & 
\colhead{Exposure} & 
\colhead{R.A.} &
\colhead{Decl.} &
\colhead{Roll} & 
\colhead{ACIS Mode} &  
\colhead{PI} \\
\colhead{} & 
\colhead{} & 
\colhead{} & 
\colhead{(UT)} & 
\colhead{(s)} & 
\colhead{$\alpha_{\rm J2000}$} & 
\colhead{$\delta_{\rm J2000}$} & 
\colhead{(\arcdeg)} & 
\colhead{} &
\colhead{} \\
\colhead{(1)} & \colhead{(2)} & \colhead{(3)} & \colhead{(4)} & \colhead{(5)} & \colhead{(6)}  & \colhead{(7)}  & \colhead{(8)} & \colhead{(9)} & \colhead{(10)}}

\startdata
NGC 7822 (Be 59) &14536 &  200837 & 2014-01-14T17:52 &   45488 & 00:02:16.99 & +67:25:09.0 &  301 &    VFaint & K. Getman \\ %
IRAS 00013+6817 (SFO 2) & 16344 &  200966 & 2015-03-17T03:46 &    1841 & 00:04:04.69 & +68:33:12.9 &  355 &    VFaint & G. Garmire \\ %
IRAS 00013+6817 (SFO 2) & 17643 &  200966 & 2015-04-07T06:26 &   21486 & 00:04:04.69 & +68:33:12.9 &   16 &    VFaint & G. Garmire \\ %
NGC 1333 & 642 &  200067 & 2000-07-12T23:15 &   37611 & 03:29:05.59 & +31:19:18.9 &   95 &     Faint & E. Feigelson \\ %
NGC 1333 & 6436 &  200410 & 2006-07-05T15:03 &   36484 & 03:29:01.99 & +31:20:53.9 &   94 &    VFaint & S. Wolk \\ %
NGC 1333 & 6437 &  200411 & 2006-07-11T10:25 &   39616 & 03:29:01.99 & +31:20:53.9 &   94 &    VFaint & S. Wolk \\ %
IC 348 & 13425 &  200747 & 2011-10-17T06:12 &    9912 & 03:44:31.50 & +32:08:33.6 &  118 &    VFaint & K. Flaherty \\ %
IC 348 & 13426 &  200748 & 2011-10-19T22:06 &    9912 & 03:44:31.50 & +32:08:33.6 &  119 &    VFaint & K. Flaherty \\ %
IC 348 & 13427 &  200749 & 2011-10-22T13:32 &    9912 & 03:44:31.50 & +32:08:33.6 &  121 &    VFaint & K. Flaherty\\ %
IC 348 & 13428 &  200750 & 2011-11-17T05:35 &    9224 & 03:44:31.50 & +32:08:33.6 &  163 &    VFaint & K. Flaherty \\ %
IC 348 & 13429 &  200751 & 2011-10-28T09:53 &   10408 & 03:44:31.50 & +32:08:33.6 &  125 &    VFaint & K. Flaherty \\ %
IC 348 & 13430 &  200752 & 2011-10-31T06:55 &    9912 & 03:44:31.50 & +32:08:33.6 &  128 &    VFaint & K. Flaherty \\ %
IC 348 & 13431 &  200753 & 2011-11-03T09:46 &    9915 & 03:44:31.50 & +32:08:33.6 &  132 &    VFaint & K. Flaherty \\ %
IC 348 & 13432 &  200754 & 2011-11-06T18:23 &    9912 & 03:44:31.50 & +32:08:33.6 &  136 &    VFaint & K. Flaherty \\ %
IC 348 & 13433 &  200755 & 2011-11-10T01:36 &   10790 & 03:44:31.50 & +32:08:33.6 &  143 &    VFaint & K. Flaherty \\ %
IC 348 & 13434 &  200756 & 2011-11-13T06:49 &    9911 & 03:44:31.50 & +32:08:33.6 &  150 &    VFaint & K. Flaherty \\ %
IC 348 & 606 &  200031 & 2000-09-21T19:58 &   52285 & 03:44:30.00 & +32:07:59.9 &  109 &    VFaint & T. Preibisch  \\ %
IC 348 & 8584 &  200471 & 2008-03-15T09:02 &   49509 & 03:44:13.19 & +32:06:00.0 &  288 &     Faint & N. Calvet  \\ %
IC 348 & 8933 &  200514 & 2008-03-18T17:35 &   39632 & 03:43:59.89 & +31:58:21.6 &  289 &    VFaint & S. Wolk  \\ %
IC 348 & 8944 &  200525 & 2008-03-13T17:53 &   39142 & 03:43:59.89 & +31:58:21.6 &  288 &    VFaint & S. Wolk  \\ %
LkH$\alpha$ 101 & 5428 &  200361 & 2005-03-08T17:25 &   39645 & 04:30:14.40 & +35:16:22.1 &  280 &    VFaint & S. Wolk  \\ %
LkH$\alpha$ 101 & 5429 &  200362 & 2005-03-06T16:50 &   39651 & 04:30:14.40 & +35:16:22.1 &  280 &    VFaint & S. Wolk  \\ %
NGC 2068-2071 & 10763 &  200472 & 2008-11-27T22:50 &   19696 & 05:46:02.40 & -00:09:00.0 &   43 &     Faint & N. Calvet  \\ %
NGC 2068-2071 & 1872 &  200100 & 2000-10-18T05:14 &   93969 & 05:46:43.50 & +00:03:29.9 &   79 &     Faint & N. Grosso  \\ %
NGC 2068-2071 & 5382 &  200317 & 2005-04-11T01:29 &   18208 & 05:46:13.09 & -00:06:05.0 &  261 &     Faint & J. Kastner  \\ %
NGC 2068-2071 & 5383 &  200318 & 2005-08-27T14:31 &   19879 & 05:46:13.09 & -00:06:05.0 &  100 &     Faint & J. Kastner  \\ %
NGC 2068-2071 & 5384 &  200319 & 2005-12-09T14:48 &   19702 & 05:46:13.09 & -00:06:05.0 &   22 &     Faint & J. Kastner \\ %
NGC 2068-2071 & 6413 &  200388 & 2005-12-14T15:45 &   18100 & 05:46:13.09 & -00:06:05.0 &   10 &     Faint & J. Kastner \\ %
NGC 2068-2071 & 6414 &  200389 & 2006-05-01T03:45 &   21648 & 05:46:13.09 & -00:06:05.0 &  250 &     Faint & J. Kastner \\ %
NGC 2068-2071 & 6415 &  200390 & 2006-08-07T18:23 &   20454 & 05:46:13.09 & -00:06:05.0 &  110 &     Faint & J. Kastner  \\ %
NGC 2068-2071 & 7417 &  200430 & 2007-11-06T19:40 &   67178 & 05:47:04.79 & +00:21:42.8 &   68 &     Faint & S. Skinner  \\ %
NGC 2068-2071 & 8585 &  200472 & 2008-11-28T12:15 &   28470 & 05:46:02.40 & -00:09:00.0 &   43 &     Faint & N. Calvet  \\ %
NGC 2068-2071 & 9915 &  200531 & 2008-09-18T04:03 &   19895 & 05:46:13.09 & -00:06:04.7 &   91 &     Faint & D. Weintraub  \\ %
NGC 2068-2071 & 9916 &  200532 & 2009-01-23T03:58 &   18405 & 05:46:13.09 & -00:06:04.7 &  299 &     Faint & D. Weintraub  \\ %
NGC 2068-2071 & 9917 &  200533 & 2009-04-21T16:18 &   29784 & 05:46:13.09 & -00:06:04.7 &  255 &     Faint & D. Weintraub  \\ %
ONC Flank S & 2548 &  200156 & 2002-09-06T12:57 &   46759 & 05:35:05.59 & -05:41:04.7 &  101 &     Faint & J. Stauffer  \\ %
ONC Flank N & 2549 &  200157 & 2002-08-26T13:49 &   48804 & 05:35:19.09 & -04:48:31.3 &  102 &     Faint & J. Stauffer  \\ %
OMC 2-3 & 634 &  200059 & 2000-01-01T13:05 &   79647 & 05:35:19.97 & -05:05:29.8 &  329 &     Faint & K. Koyama  \\ %
Mon R2 & 1882 &  200110 & 2000-12-02T23:14 &   96364 & 06:07:49.49 & -06:22:54.6 &   37 &     Faint & K. Koyama  \\ %
GGD 12-15 & 12392 &  200726 & 2010-12-15T14:05 &   67317 & 06:10:49.99 & -06:12:00.0 &   18 &    VFaint & J. Forbrich  \\ %
RCW 120 & 13276 &  200746 & 2013-02-11T07:39 &   29688 & 17:12:20.80 & -38:29:30.5 &   93 &    VFaint & G. Garmire  \\ %
RCW 120 & 13621 &  200775 & 2012-06-30T12:28 &   49117 & 17:12:20.80 & -38:29:30.5 &  303 &    VFaint & K. Getman  \\ %
Serpens Main & 4479 &  200248 & 2004-06-19T21:42 &   88449 & 18:29:49.99 & +01:15:29.9 &  161 &     Faint & F. Favata  \\ %
Serpens South & 11013 &  200642 & 2010-06-07T23:39 &   97485 & 18:30:02.99 & -02:01:58.1 &  136 &     Faint & E. Winston  \\ %
IRAS 20050+2720 & 6438 &  200412 & 2006-12-10T02:35 &   22669 & 20:07:13.60 & +27:28:48.7 &  314 &    VFaint & S. Wolk  \\ %
IRAS 20050+2720 & 7254 &  200412 & 2006-01-07T19:48 &   20852 & 20:07:13.60 & +27:28:48.7 &  344 &    VFaint & S. Wolk \\ %
IRAS 20050+2720 & 8492 &  200412 & 2007-01-29T03:53 &   50481 & 20:07:13.60 & +27:28:48.7 &   12 &    VFaint & S. Wolk  \\ %
Sh 2-106 & 1893 &  200121 & 2001-11-03T00:08 &   44384 & 20:27:25.49 & +37:22:48.6 &  283 &     Faint & Y. Maeda  \\ %
IC 5146 & 15723 &  200936 & 2015-02-24T18:06 &   37468 & 21:53:30.30 & +47:16:03.6 &   10 &    VFaint & M. Kuhn \\ %
IC 5146 & 6401 &  200378 & 2006-02-22T08:12 &   26571 & 21:52:34.09 & +47:13:43.6 &    8 &     Faint & B. Stelzer \\ %
NGC 7160 & 10818 &  200519 & 2008-11-21T13:00 &   20685 & 21:53:47.99 & +62:36:00.0 &  278 &     Faint & J. Miller  \\ %
NGC 7160 & 10819 &  200519 & 2008-11-22T12:26 &   14761 & 21:53:47.99 & +62:36:00.0 &  278 &     Faint & J. Miller  \\ %
NGC 7160 & 10820 &  200519 & 2008-11-23T03:59 &   15746 & 21:53:47.99 & +62:36:00.0 &  278 &     Faint & J. Miller  \\ %
NGC 7160 & 8938 &  200519 & 2008-11-18T22:42 &   18021 & 21:53:47.99 & +62:36:00.0 &  278 &     Faint & J. Miller  \\ %
LDN 1251B & 7415 &  200428 & 2007-08-11T13:05 &   29664 & 22:38:46.99 & +75:11:30.0 &  161 &    VFaint & T. Simon  \\ %
LDN 1251B & 8588 &  200428 & 2007-08-10T10:46 &   27982 & 22:38:46.99 & +75:11:30.0 &  161 &    VFaint & T. Simon  \\ %
Cep OB3b & 10809 &  200536 & 2009-04-04T12:49 &   21323 & 22:55:47.50 & +62:38:10.2 &   34 &    VFaint & T. Allen  \\ %
Cep OB3b & 10810 &  200536 & 2009-05-07T20:34 &   22852 & 22:55:47.50 & +62:38:10.2 &   71 &    VFaint & T. Allen  \\ %
Cep OB3b & 10811 &  200535 & 2009-04-28T07:05 &   24405 & 22:53:31.69 & +62:35:33.3 &   63 &    VFaint & T. Allen  \\ %
Cep OB3b & 10812 &  200535 & 2009-05-03T01:48 &   24814 & 22:53:31.69 & +62:35:33.3 &   63 &    VFaint & T. Allen  \\ %
Cep OB3b & 3502 &  200197 & 2003-03-11T12:21 &   30090 & 22:56:46.99 & +62:40:00.0 &    7 &     Faint & G. Garmire  \\ %
Cep OB3b & 9919 &  200535 & 2009-05-08T03:29 &   22465 & 22:53:31.69 & +62:35:33.3 &   68 &    VFaint & T. Allen  \\ %
Cep OB3b & 9920 &  200536 & 2009-04-16T10:45 &   27682 & 22:55:47.50 & +62:38:10.2 &   47 &    VFaint & T. Allen  \\ %
Cep A & 8898 &  200479 & 2008-04-08T12:13 &   77953 & 22:56:19.84 & +62:01:46.9 &   39 &     Faint & S. Pravdo  \\ %
Cep C & 10934&  200570 & 2010-09-21T10:42 &   43989 & 23:05:50.99 & +62:30:55.0 &  193 &     Faint & K. Covey  \\ %
\enddata

\tablecomments{Column 1: SFR name, sorted by R.A. Columns 2-3: {\it Chandra} observation id and sequence number. Column 4: Start time of a {\it Chandra} observation. Column 5: Exposure time is the net usable time after various filtering steps are applied in the data reduction process. Columns 6-7: The aimpoint of a {\it Chandra} observation is obtained from the satellite aspect solution before astrometric correction is applied. Units of right ascension are hours, minutes, and seconds; units of declination are degrees, arcminutes, and arcseconds. Column 8. Roll angle of a {\it Chandra} observation. Column 9. {\it Chandra}-ACIS observing mode. The ACIS modes are described in \S6.12 of {\it Chandra} Proposer' Observatory Guide, \url{http://asc.harvard.edu/proposer/POG/}. Column 10. The principal investigator of a {\it Chandra} observation.}

\end{deluxetable*}


\floattable 
\begin{deluxetable*}{ccccccccccccccccc}
\centering \rotate \tabletypesize{\tiny} \tablewidth{0pt} \tablecolumns{17}

\tablecaption{SFiNCs X-ray Sources and Basic Properties \label{tbl_acis_src_properties}}

\tablehead{
\multicolumn{3}{c}{Source} &
\multicolumn{4}{c}{Position} &
\multicolumn{5}{c}{Extraction} &
\multicolumn{5}{c}{Characteristics} \\[0pt]
\multicolumn{3}{c}{\hrulefill} &  
\multicolumn{4}{c}{\hrulefill} &
\multicolumn{5}{c}{\hrulefill} &
\multicolumn{5}{c}{\hrulefill} \\
\colhead{SFR} & \colhead{Seq. No.} & \colhead{CXOU J} &
\colhead{$\alpha$ (J2000)} & \colhead{$\delta$ (J2000)} & \colhead{Error} & \colhead{$\theta$} &
\colhead{$C_{t,net}$} & \colhead{$\sigma_{t,net}$} & \colhead{$B_{t}$} & \colhead{$C_{h,net}$} & \colhead{PSF Frac.} &   
\colhead{SNR} & \colhead{$\log P_B$} & \colhead{Anom.} & \colhead{Var.}  & \colhead{$E_{median}$}  \\
\colhead{} & \colhead{} & \colhead{} &
\colhead{(\arcdeg)} & \colhead{(\arcdeg)} & \colhead{(\arcsec)} & \colhead{(\arcmin)} &
\colhead{(cts)} & \colhead{(cts)} & \colhead{(cts)} & \colhead{(cts)} & \colhead{} &
\colhead{} & \colhead{} & \colhead{} & \colhead{} &  \colhead{(keV)}
 \\
\colhead{(1)} & \colhead{(2)} & \colhead{(3)} & \colhead{(4)} & \colhead{(5)} & \colhead{(6)}  & \colhead{(7)}  & \colhead{(8)} & \colhead{(9)} & \colhead{(10)} & \colhead{(11)} & \colhead{(12)} & \colhead{(13)} & \colhead{(14)} & \colhead{(15)} & \colhead{(16)}  & \colhead{(17)}}

\startdata
NGC 7822 & 1 & 000033.87$+$672446.2 & 0.141150 & 67.412846 & 0.6 & 10.3 & 30.5 & 6.1 & 3.5 & 21.9 & 0.54 & 5.2 & $<$-5 & .... & a & 2.6 \\ 
NGC 7822 & 2 & 000033.92$+$672452.8 & 0.141362 & 67.414691 & 0.8 & 10.3 & 7.5 & 3.5 & 4.5 & 3.7 & 0.39 & 2.0 & -2.9 & .... & a & 1.6 \\ 
NGC 7822 & 3 & 000034.62$+$672537.7 & 0.144257 & 67.427159 & 0.9 & 10.2 & 30.4 & 6.5 & 10.6 & 18.2 & 0.91 & 4.5 & $<$-5 & .... & a & 2.6 \\ 
NGC 7822 & 4 & 000036.43$+$672658.5 & 0.151798 & 67.449596 & 0.6 & 10.2 & 71.9 & 9.0 & 8.1 & 32.9 & 0.91 & 7.9 & $<$-5 & .... & a & 1.9 \\ 
NGC 7822 & 5 & 000041.15$+$672526.6 & 0.171487 & 67.424076 & 1.5 & 9.6 & 8.5 & 4.1 & 7.5 & 8.7 & 0.91 & 1.9 & -2.4 & .... & a & 5.0 \\ 
NGC 7822 & 6 & 000044.02$+$672844.3 & 0.183441 & 67.478987 & 1.4 & 10.1 & 11.9 & 4.6 & 8.1 & 10.7 & 0.91 & 2.4 & -3.3 & .... & a & 4.0 \\ 
NGC 7822 & 7 & 000044.33$+$672306.9 & 0.184745 & 67.385269 & 1.6 & 9.4 & 5.1 & 3.3 & 5.9 & 5.5 & 0.89 & 1.4 & -2.2 & .... & b & 4.1 \\ 
NGC 7822 & 8 & 000045.20$+$672805.8 & 0.188345 & 67.468297 & 1.1 & 9.7 & 19.2 & 5.2 & 6.8 & 2.0 & 0.91 & 3.5 & $<$-5 & .... & a & 1.7 \\ 
NGC 7822 & 9 & 000046.19$+$672358.2 & 0.192477 & 67.399503 & 0.8 & 9.1 & 30.2 & 6.1 & 5.8 & 8.0 & 0.91 & 4.7 & $<$-5 & .... & a & 1.6 \\ 
NGC 7822 & 10 & 000048.35$+$672648.8 & 0.201499 & 67.446893 & 0.7 & 9.1 & 40.3 & 6.8 & 5.7 & 28.1 & 0.90 & 5.7 & $<$-5 & .... & a & 3.0 \\ 
\enddata

\tablecomments{This table is available in its entirety (15364 SFiNCs X-ray sources) in the machine-readable form in the on-line journal. A portion is shown here for guidance regarding its form and content. The format of this table is similar to that of Table~1 in \citet{Broos2010}.  Column 1: Star Forming Region. Column 2: X-ray catalog sequence number, sorted by R.A. Column 3: IAU designation. Columns 4-5: Right ascension and declination (in decimal degrees) for epoch J2000.0. Column 6:  Estimated standard deviation of the random component of the position error, $\sqrt{\sigma_x^2 + \sigma_y^2}$.  The single-axis position errors, $\sigma_x$ and $\sigma_y$, are estimated from the single-axis standard deviations of the PSF inside the extraction region and the number of counts extracted. Column 7: Off-axis angle. Columns 8 and 9: Net counts extracted in the total energy band ($0.5-8$~keV); average of the upper and lower $1\sigma$ errors on Column 9. Column 10: Background counts expected in the source extraction region (total band). Column 11: Net counts extracted in the hard energy band ($2-8$~keV). Column 12: Fraction of the PSF (at 1.497 keV) enclosed within the extraction region. A reduced PSF fraction (significantly below 90\%) may indicate that the source is in a crowded region. Column 13: Photometric significance computed as net counts divided by the upper error on net counts. Column 14: Logarithmic probability that extracted counts (total band) are solely from background.  Some sources have $P_B$ values above the 1\% threshold that defines the catalog because local background estimates can rise during the final extraction iteration after sources are removed from the catalog. Column 15: Source anomalies: (g) fractional time that source was on a detector (FRACEXPO from {\em mkarf}) is $<$0.9; (e) source on field edge; (p) X-ray properties may be biased due to photon pile-up; (s) source on readout streak; (a) photometry and spectrum may contain $>10$\% afterglow events. Column 16: Variability characterization based on K-S statistic (total band) from the single ObsId showing the most variability: (a) no evidence for variability ($0.05<P_{KS}$); (b) possibly variable ($0.005<P_{KS}<0.05$); (c) definitely variable ($P_{KS}<0.005$).  No value is reported for sources with fewer than four counts or for sources in chip gaps or on field edges. Column 17:  Background-corrected median photon energy (total band).}
\end{deluxetable*}


\floattable
\begin{deluxetable*}{cccccccccccccccccc}
\centering \rotate \tabletypesize{\tiny} \tablewidth{0pt} \tablecolumns{18}

\tablecaption{SFiNCs X-ray Fluxes\label{tbl_acis_src_fluxes}}

\tablehead{
\multicolumn{3}{c}{Source} &
\multicolumn{2}{c}{{\it AE} Fluxes} &
\multicolumn{13}{c}{Spectral Properties from XPHOT}\\                        
\multicolumn{3}{c}{\hrulefill} &  
\multicolumn{2}{c}{\hrulefill} &
\multicolumn{13}{c}{\hrulefill} \\
\colhead{} & \colhead{} & \colhead{} &
\colhead{$\log$} & \colhead{$\log$} & \colhead{$\log$} & \colhead{} &
\colhead{$\log$} & \colhead{} & \colhead{$\log$} & \colhead{} & \colhead{} &   
\colhead{$\log$} & \colhead{} & \colhead{} & \colhead{$\log$}  & \colhead{} & \colhead{}\\
\colhead{SFR} & \colhead{Seq. No.} & \colhead{CXOU J} &
\colhead{$PF_h$} & \colhead{$PF_t$} & \colhead{$F_h$} & \colhead{$\sigma$} &
\colhead{$F_t$} & \colhead{$\sigma$} & \colhead{$N_H$} & \colhead{$\sigma_{stat}$} & \colhead{$\sigma_{sys}$} &   
\colhead{$F_{hc}$} & \colhead{$\sigma_{stat}$} & \colhead{$\sigma_{sys}$} & \colhead{$F_{tc}$}  & \colhead{$\sigma_{stat}$} & \colhead{$\sigma_{sys}$}\\
\multicolumn{3}{c}{} &
\multicolumn{2}{c}{[photon cm$^{-2}$ s$^{-1}$]} &
\multicolumn{4}{c}{[erg cm$^{-2}$ s$^{-1}$]} &
\multicolumn{3}{c}{[cm$^{-2}$]} &
\multicolumn{6}{c}{[erg cm$^{-2}$ s$^{-1}$]}\\
\multicolumn{3}{c}{\hrulefill} &  
\multicolumn{2}{c}{\hrulefill} &
\multicolumn{4}{c}{\hrulefill} &  
\multicolumn{3}{c}{\hrulefill} &
\multicolumn{6}{c}{\hrulefill} \\
\colhead{(1)} & \colhead{(2)} & \colhead{(3)} & \colhead{(4)} & \colhead{(5)} & \colhead{(6)}  & \colhead{(7)}  & \colhead{(8)} & \colhead{(9)} & \colhead{(10)} & \colhead{(11)} & \colhead{(12)} & \colhead{(13)} & \colhead{(14)} & \colhead{(15)} & \colhead{(16)}  & \colhead{(17)} & \colhead{(18)}}

\startdata
NGC 7822 & 1  &  000033.87$+$672446.2  &   -5.293 &   -5.161 &  -13.661 &    0.122 &  -13.527 &    0.105 &  22.38 &   0.11 &   0.08 &  -13.536 &    0.125 &    0.043 &  -13.048 &    0.114 &    0.172\\
NGC 7822 & 2  &  000033.92$+$672452.8  &   -5.924 &   -5.632 &  -14.361 &    0.391 &  -14.233 &    0.262 &  21.95 &   0.65 &   0.12 &  -14.304 &    0.394 &    0.034 &  -13.844 &    0.328 &    0.251\\
NGC 7822 & 3  &  000034.62$+$672537.7  &   -5.640 &   -5.432 &  -13.900 &    0.142 &  -13.808 &    0.109 &  22.40 &   0.10 &   0.11 &  -13.764 &    0.145 &    0.065 &  -13.283 &    0.118 &    0.298\\
NGC 7822 & 4  &  000036.43$+$672658.5  &   -5.383 &   -5.061 &  -13.626 &    0.095 &  -13.560 &    0.066 &  22.08 &   0.12 &   0.08 &  -13.560 &    0.096 &    0.020 &  -13.196 &    0.078 &    0.126\\
NGC 7822 & 5  &  000041.15$+$672526.6  &   -5.969 &   -6.001 &  -13.999 &    0.220 &  -13.972 &    0.239 &  23.48 &   0.25 & \nodata &  -13.266 &    0.326 &    0.087 &  -12.838 &    0.339 &    0.227\\
NGC 7822 & 6  &  000044.02$+$672844.3  &   -5.872 &   -5.834 &  -14.008 &    0.198 &  -13.967 &    0.193 &  23.00 &   0.27 &   0.08 &  -13.625 &    0.263 &    0.114 &  -13.184 &    0.263 &    0.263\\
NGC 7822 & 7  &  000044.33$+$672306.9  &   -6.154 &   -6.192 &  -14.324 &    0.302 &  -14.310 &    0.339 &  23.11 &   0.36 &   0.07 &  -13.848 &    0.417 &    0.153 &  -13.395 &    0.447 &    0.428\\
NGC 7822 & 8  &  000045.20$+$672805.8  &   -6.598 &   -5.643 & \nodata & \nodata & \nodata & \nodata & \nodata & \nodata & \nodata & \nodata & \nodata & \nodata & \nodata & \nodata & \nodata\\
NGC 7822 & 9  &  000046.19$+$672358.2  &   -6.008 &   -5.448 &  -14.370 &    0.226 &  -14.073 &    0.101 &  21.84 &   0.15 &   0.16 &  -14.326 &    0.226 &    0.031 &  -13.732 &    0.119 &    0.248\\
NGC 7822 & 10  &  000048.35$+$672648.8  &   -5.463 &   -5.331 &  -13.729 &    0.105 &  -13.633 &    0.089 &  22.53 &   0.11 &   0.07 &  -13.559 &    0.112 &    0.051 &  -13.090 &    0.102 &    0.183\\
\enddata

\tablecomments{This table is available in its entirety (15364 SFiNCs X-ray sources) in the machine-readable form in the on-line journal. A portion is shown here for guidance regarding its form and content. Fluxes given in Columns 4-5 are produced by $\it AE$ \citep{Broos2010}. Fluxes and column densities given in Columns 6-18 are produced by {\it XPHOT} \citep{Getman2010}. {\it XPHOT} assumes X-ray spectral shapes of young, low-mass stars. Intrinsic {\it XPHOT} quantities (Columns 10-18) will be unreliable for high-mass stellar members of the SFiNCs SFRs as well as for non-members, such as Galactic field stars and extragalactic objects. The fluxes and the column densities are given in a log scale. Column 1: Star Forming Region. Column 2: X-ray catalog sequence number, sorted by R.A. Column 3: IAU designation. Columns 4-5: Incident X-ray photon fluxes in the hard ($2-8$)~keV and total ($0.5-8$)~keV bands, respectively. Columns 6-9: Apparent X-ray fluxes in the hard and total bands, and their 1$\sigma$ statistical uncertainties. Columns 10-12: X-ray column density and its 1$\sigma$ statistical and systematic uncertainties. Columns 13-18: Corrected for absorption, X-ray fluxes in the hard and total bands, and their 1$\sigma$ statistical and systematic uncertainties.}
\end{deluxetable*}


\floattable
\begin{deluxetable*}{cccccccrl}
\centering \rotate \tabletypesize{\tiny} \tablewidth{0pt} \tablecolumns{9}

\tablecaption{Log of SFiNCs {\it Spitzer}-IRAC Observations \label{tbl_irac_obslog}}

\tablehead{
\colhead{Region} & 
\colhead{AOR} & 
\colhead{PID} & 
\colhead{Start Time} & 
\colhead{Stop Time} & 
\multicolumn{2}{c}{Center} & 
\colhead{IRAC Mode} &
\colhead{PI} \\
\colhead{} & 
\colhead{} & 
\colhead{} & 
\colhead{(UT)} & 
\colhead{(UT)} & 
\colhead{$\alpha_{\rm J2000}$} & 
\colhead{$\delta_{\rm J2000}$} &
\colhead{} &
\colhead{} \\
\colhead{(1)} & \colhead{(2)} & \colhead{(3)} & \colhead{(4)} & \colhead{(5)} & \colhead{(6)}  & \colhead{(7)}  & \colhead{(8)} & \colhead{(9)}}

\startdata
NGC 7822 & 48001280 & 90179 & 2013-04-12 10:27:38  & 2013-04-12 16:34:37  & 0.570833 & 67.419194 & IRAC Map PC  & Getman, Konstantin V \\
IRAS 00013+6817 & 3658240 & 6 & 2003-12-23 13:33:58  & 2003-12-23 14:06:02  & 0.994833 & 68.594583 & IRAC Map  & Fazio, Giovanni \\
NGC 1333 & 16034304 & 178 & 2005-09-16 09:56:12  & 2005-09-16 10:59:36  & 52.437500 & 30.913889 & IRAC Map  & Evans, Neal \\
NGC 1333 & 3652864 & 6 & 2004-02-10 08:29:23  & 2004-02-10 09:23:31  & 52.252458 & 31.311917 & IRAC Map  & Fazio, Giovanni \\
NGC 1333 & 5793280 & 178 & 2004-09-08 17:03:35  & 2004-09-08 17:37:59  & 52.252458 & 31.311917 & IRAC Map  & Evans, Neal \\
NGC 1333 & 18323968 & 30516 & 2007-02-15 21:50:58  & 2007-02-15 22:34:19  & 52.293333 & 31.225000 & IRAC Map  & Looney, Leslie W \\
NGC 1333 & 18325760 & 30516 & 2007-02-15 21:07:00  & 2007-02-15 21:50:22  & 52.185417 & 31.094167 & IRAC Map  & Looney, Leslie W \\
IC 348 & 16034048 & 178 & 2005-09-16 09:10:33  & 2005-09-16 09:54:45  & 56.150000 & 31.927222 & IRAC Map  & Evans, Neal \\
IC 348 & 34977024 & 60160 & 2009-10-02 21:25:30  & 2009-10-02 21:57:37  & 56.083667 & 32.050278 & IRAC Map PC  & Muzerolle, James \\
IC 348 & 34977280 & 60160 & 2009-10-04 02:44:04  & 2009-10-04 03:26:21  & 56.083667 & 32.050278 & IRAC Map PC  & Muzerolle, James \\
\enddata

\tablecomments{This table is available in its entirety (423 {\it Spitzer}-IRAC AORs for the 22 SFiNCs SFRs) in the machine-readable form in the on-line journal. A portion is shown here for guidance regarding its form and content. Column 1: Star Forming Region. Column 2: Astronomical Object Request number. Column 3: {\it Spitzer} program identification number. Columns 4-5: Start and stop times of the observation, in UT. Columns 6-7: Approximate center of the observation; right ascension and declination for epoch (J2000.0).  Column 8: IRAC mode. The PC (Post-Cryo) mode has been introduced during the warm mission of the {\it Spitzer} observatory with only two shortest-wavelength IRAC modules in operation. Column 9: The principal investigator of the observation.}
\end{deluxetable*}


\floattable
\begin{deluxetable*}{ccccccccccccc}
\centering \rotate \tabletypesize{\tiny} \tablewidth{0pt} \tablecolumns{13}

\tablecaption{SFiNCs IRAC Sources and Photometry \label{tbl_irac_sources}}

\tablehead{
\colhead{Region} & 
\colhead{Source} & 
\colhead{R.A.} &
\colhead{Decl.} &
\colhead{[3.6]} &
\colhead{$\sigma$\_[3.6]} &
\colhead{[4.5]} &
\colhead{$\sigma$\_[4.5]} &
\colhead{[5.8]} &
\colhead{$\sigma$\_[5.8]} &
\colhead{[8.0]} &
\colhead{$\sigma$\_[8.0]} &
\colhead{ApertureFl}\\
\colhead{} & 
\colhead{} & 
\colhead{(deg)} & 
\colhead{(deg)} & 
\colhead{(mag)} & 
\colhead{(mag)} & 
\colhead{(mag)} & 
\colhead{(mag)} & 
\colhead{(mag)} & 
\colhead{(mag)} & 
\colhead{(mag)} & 
\colhead{(mag)} & 
\colhead{} \\
\colhead{(1)} & \colhead{(2)} & \colhead{(3)} & \colhead{(4)} & \colhead{(5)} & \colhead{(6)}  & \colhead{(7)}  & \colhead{(8)} & \colhead{(9)} & \colhead{(10)}  & \colhead{(11)}  & \colhead{(12)} & \colhead{(13)}}

\startdata
NGC 7822 & G118.0586+05.2740 &     0.0017167 &    67.6576472 & 15.190 &  0.022 & 15.134 &  0.027 & \nodata & \nodata & \nodata & \nodata &  4\\
NGC 7822 & G117.9462+04.7218 &     0.0018208 &    67.0942639 & 14.366 &  0.027 & 14.438 &  0.019 & \nodata & \nodata & \nodata & \nodata &  4\\
NGC 7822 & G117.9954+04.9637 &     0.0018333 &    67.3410528 & 14.171 &  0.075 & 14.056 &  0.053 & \nodata & \nodata & \nodata & \nodata &  4\\
NGC 7822 & G117.9429+04.7056 &     0.0018708 &    67.0777194 & 14.289 &  0.017 & 14.247 &  0.014 & \nodata & \nodata & \nodata & \nodata &  4\\
NGC 7822 & G118.0391+05.1771 &     0.0023458 &    67.5588750 & 13.123 &  0.022 & 13.023 &  0.019 & \nodata & \nodata & \nodata & \nodata &  2\\
NGC 7822 & G117.9819+04.8961 &     0.0023958 &    67.2720972 & 15.710 &  0.108 & 15.415 &  0.106 & \nodata & \nodata & \nodata & \nodata &  2\\
NGC 7822 & G118.0296+05.1298 &     0.0025708 &    67.5106333 & 15.856 &  0.092 & 15.759 &  0.091 & \nodata & \nodata & \nodata & \nodata &  2\\
NGC 7822 & G117.9481+04.7292 &     0.0027542 &    67.1018444 & 14.157 &  0.033 & 14.158 &  0.037 & \nodata & \nodata & \nodata & \nodata &  4\\
NGC 7822 & G118.0293+05.1283 &     0.0027625 &    67.5090611 & 13.802 &  0.033 & 13.602 &  0.032 & \nodata & \nodata & \nodata & \nodata &  4\\
NGC 7822 & G118.0301+05.1319 &     0.0027750 &    67.5127861 & 15.815 &  0.053 & 15.960 &  0.078 & \nodata & \nodata & \nodata & \nodata &  3\\
\enddata

\tablecomments{This table is available in its entirety (1638654 {\it Spitzer}-IRAC sources in/around the 22 SFiNCs SFRs) in the machine-readable form in the on-line journal. A portion is shown here for guidance regarding its form and content. The format of the table is similar to that of MYStIX Table~3 in \citet{Kuhn2013b}. Column 1: SFiNCs SFR name. Column 2: Source name in the GLLL.llll$+$BB.bbbb format. Columns 3-4: Right ascension and declination in decimal degrees (J2000.0). Columns 5-12: IRAC magnitudes and their 1$\sigma$ errors for the 3.6, 4.5, 5.8, and 8.0~$\mu$m bands. Column 13: Aperture size flag: 2, 3, and 4 --- 2, 3, and 4-pixel aperture with contaminating flux $<10$\% of source flux; -1 --- 2-pixel aperture with contaminating flux $>10$\% of source flux.}
\end{deluxetable*}


\floattable
\begin{deluxetable*}{ccccclccl}
\centering \rotate \tabletypesize{\tiny} \tablewidth{0pt} \tablecolumns{9}

\tablecaption{SFiNCs Probable Cluster Members: IR Photometry \label{tbl_spcm_irprops}}

\tablehead{
\colhead{Region} & 
\colhead{Source} & 
\colhead{R.A.} &
\colhead{Decl.} &
\colhead{2MASS} &
\colhead{$JHK_s$} &
\colhead{NIR\_Fl} &
\colhead{IRAC} &
\colhead{IRAC mags}\\
\colhead{} & 
\colhead{} & 
\colhead{(deg)} & 
\colhead{(deg)} &
\colhead{} &
\colhead{(mag)} & 
\colhead{} &
\colhead{} &
\colhead{(mag)}\\
\colhead{(1)} & \colhead{(2)} & \colhead{(3)} & \colhead{(4)} & \colhead{(5)} & \colhead{(6)}  & \colhead{(7)}  & \colhead{(8)} & \colhead{(9)}}

\startdata
NGC 7822 & 000033.87+672446.2 &     0.141150 &    67.412846 & 00003378+6724462 & 15.19$\pm$0.05 13.58$\pm$0.04 12.89$\pm$0.03 & AAA000 & G118.0624+05.0235 & 12.38$\pm$0.01 12.26$\pm$0.01 \nodata \nodata\\
NGC 7822 & 000036.43+672658.5 &     0.151798 &    67.449596 & 00003633+6726582 & 13.68$\pm$0.03 12.64$\pm$0.03 12.17$\pm$0.03 & AAA000 & G118.0737+05.0586 & 11.78$\pm$0.00 11.73$\pm$0.00 \nodata \nodata\\
NGC 7822 & 000045.20+672805.8 &     0.188345 &    67.468297 & 00004532+6728055 & 13.90$\pm$0.03 12.85$\pm$0.03 12.45$\pm$0.03 & AAA000 & G118.0915+05.0742 & 12.06$\pm$0.01 11.98$\pm$0.01 \nodata \nodata\\
NGC 7822 & 000046.19+672358.2 &     0.192477 &    67.399503 & 00004605+6723575 & 14.08$\pm$\nodata 13.01$\pm$0.04 12.60$\pm$0.03 & UAA0cc & G118.0791+05.0063 & 12.19$\pm$0.01 12.16$\pm$0.03 \nodata \nodata\\
NGC 7822 & 000050.10+672721.4 &     0.208781 &    67.455954 & 00005018+6727204 & 15.09$\pm$0.05 13.89$\pm$0.04 13.42$\pm$0.04 & AAA000 & G118.0967+05.0603 & 13.01$\pm$0.01 12.92$\pm$0.01 \nodata \nodata\\
NGC 7822 & 000051.39+672648.8 &     0.214161 &    67.446914 & 00005151+6726487 & 14.66$\pm$0.04 13.33$\pm$0.04 12.89$\pm$0.03 & AAA000 & G118.0970+05.0513 & 12.55$\pm$0.01 12.40$\pm$0.01 \nodata \nodata\\
NGC 7822 & 000053.45+672615.0 &     0.222725 &    67.437501 & 00005355+6726148 & 15.43$\pm$0.06 13.99$\pm$0.05 13.34$\pm$0.04 & AAA000 & G118.0984+05.0415 & 12.79$\pm$0.02 12.72$\pm$0.01 \nodata \nodata\\
NGC 7822 & 000054.01+672119.8 &     0.225079 &    67.355504 & 00005405+6721190 & 15.41$\pm$0.05 14.03$\pm$0.04 13.56$\pm$0.05 & AAA000 & G118.0830+04.9607 & 13.00$\pm$0.01 12.87$\pm$0.01 \nodata \nodata\\
NGC 7822 & 000055.58+672647.8 &     0.231621 &    67.446638 & 00005552+6726466 & 15.30$\pm$0.05 13.96$\pm$0.04 13.56$\pm$0.05 & AAA000 & G118.1032+05.0494 & 13.05$\pm$0.02 12.96$\pm$0.01 \nodata \nodata\\
NGC 7822 & 000056.24+672835.1 &     0.234343 &    67.476426 & 00005640+6728347 & 16.19$\pm$0.09 15.00$\pm$0.08 14.30$\pm$0.06 & AAAccc & G118.1104+05.0786 & 13.71$\pm$0.02 13.65$\pm$0.02 \nodata \nodata\\
\enddata

\tablecomments{This table is available in its entirety (8492 SFiNCs probable cluster members across the 22 SFiNCs SFRs) in the machine-readable form in the on-line journal. A portion is shown here for guidance regarding its form and content. Column 1: SFiNCs SFR name. Column 2: Source' IAU designation. Columns 3-4: Right ascension and declination in decimal degrees (J2000.0). Column 5: 2MASS source lable. Column 6: 2MASS photometry in the $J$, $H$, and $K_s$-bands, respectively. Column 7: 2MASS photometry quality and confusion/contamination flag. Columns 8-9: IRAC source label and IRAC photometry in the [3.6], [4.5], [5.8], and [8.0] bands, respectively. These quanties were excerpted from Table \ref{tbl_irac_sources}.}

\end{deluxetable*}


\floattable
\begin{deluxetable*}{ccccccccccccccc}
\centering \rotate \tabletypesize{\tiny} \tablewidth{0pt} \tablecolumns{15}

\tablecaption{SFiNCs Probable Cluster Members: Main X-ray and Other Properties \label{tbl_spcm_otherprops}}

\tablehead{
\colhead{Region} & 
\colhead{Source} & 
\colhead{ACIS Label} &
\colhead{XFOV} &
\colhead{$C_{t,net}$} &
\colhead{$E_{median}$} &
\colhead{$\log(PF_t)$} &
\colhead{$\log(N_H)$} &
\colhead{$\log(L_{tc})$} &
\colhead{${\alpha}_{IRAC}$} &
\colhead{Class} &
\colhead{$A_V$} &
\colhead{$AgeJX$} &
\colhead{$OB$\tablenotemark{a}} &
\colhead{$Id$\tablenotemark{b}}\\
\colhead{} & 
\colhead{} & 
\colhead{} & 
\colhead{flag} &
\colhead{(cts)} &
\colhead{(keV)} & 
\colhead{[ph cm$^{-2}$ s$^{-1}$]} &
\colhead{[cm$^{-2}$]} &
\colhead{[erg s$^{-1}$]} &
\colhead{} &
\colhead{} &
\colhead{(mag)} &
\colhead{(Myr)} &
\colhead{flag} &
\colhead{flag}\\
\colhead{(1)} & \colhead{(2)} & \colhead{(3)} & \colhead{(4)} & \colhead{(5)} & \colhead{(6)}  & \colhead{(7)}  & \colhead{(8)} & \colhead{(9)} & \colhead{(10)}  & \colhead{(11)}  & \colhead{(12)} & \colhead{(13)} & \colhead{(14)} & \colhead{(15)}}

\startdata
NGC 7822 & 000033.87+672446.2 & c2 & 1 & 30.5 &  2.62 & -5.16 & 22.4 & 30.94 &  -2.05$\pm$0.04 & NOD    &  8.7 & \nodata & \nodata & \nodata\\
NGC 7822 & 000036.43+672658.5 & c11 & 1 & 71.9 &  1.88 & -5.06 & 22.1 & 30.79 &  -2.30$\pm$0.02 & NOD    &  3.8 & \nodata & \nodata & \nodata\\
NGC 7822 & 000045.20+672805.8 & c24 & 1 & 19.2 &  1.66 & -5.64 &  \nodata & \nodata &  -2.21$\pm$0.04 & NOD    &  3.6 & \nodata & \nodata & \nodata\\
NGC 7822 & 000046.19+672358.2 & c26 & 1 & 30.2 &  1.58 & -5.45 & 21.8 & 30.25 &  -2.38$\pm$0.11 & NOD    &  \nodata & \nodata & \nodata & \nodata\\
NGC 7822 & 000050.10+672721.4 & c31 & 1 & 9.1 &  1.77 & -5.98 & 22.1 & 29.90 &  -2.19$\pm$0.05 & NOD    &  4.9 & 2.5 & \nodata & \nodata\\
NGC 7822 & 000051.39+672648.8 & c32 & 1 & 38.3 &  2.12 & -5.36 & 22.2 & 30.63 &  -1.92$\pm$0.04 & NOD    &  5.7 & \nodata & \nodata & \nodata\\
NGC 7822 & 000053.45+672615.0 & c36 & 1 & 56.5 &  2.37 & -5.20 & 22.3 & 30.82 &  -2.23$\pm$0.10 & NOD    &  7.4 & \nodata & \nodata & \nodata\\
NGC 7822 & 000054.01+672119.8 & c39 & 1 & 45.5 &  2.37 & -5.21 & 22.3 & 30.81 &  -2.01$\pm$0.04 & NOD    &  6.2 & \nodata & \nodata & \nodata\\
NGC 7822 & 000055.58+672647.8 & c41 & 1 & 12.9 &  2.08 & -5.84 & 22.3 & 30.21 &  -2.15$\pm$0.09 & NOD    &  5.6 & 4.3 & \nodata & \nodata\\
NGC 7822 & 000056.24+672835.1 & c42 & 1 & 40.6 &  3.00 & -4.93 & 22.5 & 31.25 &  -2.29$\pm$0.13 & NOD    &  \nodata & \nodata & \nodata & Y\\
\enddata

\tablecomments{This table is available in its entirety (8492 SFiNCs probable cluster members across the 22 SFiNCs SFRs) in the machine-readable form in the on-line journal. A portion is shown here for guidance regarding its form and content. Column 1: SFiNCs SFR name. Column 2: Source' IAU designation. Column 3: ACIS source' label. Column 4: A flag indicating whether the source is located inside (XFOV$=1$) or outside (XFOV=$0$) the {\it Chandra} ACIS-I field of view. Columns 5-9: The main X-ray properties of the SFiNCs YSO excerpted from Tables \ref{tbl_acis_src_properties} and \ref{tbl_acis_src_fluxes}. These include net counts, median energy, incident X-ray photon flux, column density, and intrinsic luminosity. All quantities were computed in the total $(0.5-8)$~keV band. Column 10. Apparent SED slope measured in the IRAC wavelength range from 3.6 to 8.0~$\mu$m. Column 11. YSO class: diskless (``NOD''), disky (``DSK''), possible member without a disk class (``PMB''). The suffix ``FRG'' indicates that the source could be a foreground star or a YSO member of the region. Columns 12-13: Source extinction in the $V$-band and stellar age estimated using the methods of \citet{Getman2014a}. Column 14: OB flag indicating whether the source is associated with a known OB-type star from the catalogs of \citet{Skiff2009} and/or SIMBAD. If the flag$=$`Y', the related information on the OB star' name, spectral type, and primary catalog of origin is placed in the note section of this table. Column 15: Flag `Y' indicates that the X-ray-IR source match was rejected by the tool $match\_xy$ but was reinstated as a legitimate match upon the visual inspection. Such cases are often associated with source' binarity and multiplicity. The related comments are given in the note section of this table.}
\tablenotetext{a}{Notes on individual OB-type stars, including OB name, spectral type, and primary catalog of origin.\\
NGC 7822 000146.83+673025.8: BD+66 1673; O5V((f))n from SKIFF\\
NGC 7822 000200.07+672511.5: 2MASS J00020012+6725109; B3V from SKIFF\\
NGC 7822 000210.16+672545.5: BD+66 1674; B0IIIn from SKIFF\\
NGC 7822 000210.24+672432.3: BD+66 1675; O7V from SKIFF\\
NGC 7822 000219.06+672538.5: NGC 7822 x; O9 from SKIFF\\
NGC 7822 000219.68+673424.3: LS I +67 10; B1: from SKIFF\\
NGC 7822 000229.79+672543.7: NGC 7822 y; B3? from SKIFF\\
NGC 7822 000210.62+672408.6: 2MASS J00021063+6724087; B8III from SKIFF\\
NGC 7822 000213.58+672503.6: LS I +67 9; B0.5Vn from SKIFF\\
NGC 1333 032857.19+311419.1: BD+30 547; BV from SIMBAD\\
NGC 1333 032910.39+312159.2: [SVS76] NGC 1333 3; B5:V + F2: from SKIFF\\
NGC 1333 032919.81+312457.4: BD+30 549; B8/9V from SKIFF\\
NGC 1333 032909.63+312256.4: [SVS76] NGC 1333 7; B:III from SKIFF\\
\nodata}
\tablenotetext{b}{Notes on specific X-ray-IR source matches, based on the visual inspection of the 2MASS, IRAC, and ACIS images. The ``SERs highly overlap'' comment means that the IR and X-ray sources look like single sources and their source extraction regions (SERs) highly overlap. The last two digits given at the end of the individual notes indicate whether the source' X-ray-IR properties are consistent with the trends of the X-ray incident photon flux vs. the $J$-band magnitude and the X-ray median energy vs. the $J-H$ color seen for the majority of the X-ray-IR pairs: ``0'' --- inconsistent; ``1'' --- consistent.\\
NGC 7822 000056.24+672835.1: SERs highly overlap; 00\\
NGC 7822 000115.84+672813.1: X-ray is a v. double with c79; 2MASS is a single; SERs highly overlap; 11\\
NGC 7822 000127.37+672218.9: SERs highly overlap; 10\\
NGC 7822 000134.82+672152.3: SERs highly overlap; 10\\
NGC 7822 000136.77+672521.9: SERs highly overlap; 11\\
NGC 7822 000138.66+672800.6: SERs highly overlap; 01\\
NGC 7822 000141.10+672405.8: SERs highly overlap; 10\\
NGC 7822 000146.83+673025.8: SERs highly overlap; 01\\
NGC 7822 000151.85+673153.5: X-ray is a v. triple; 2MASS is a single; SERs highly overlap; 01\\
NGC 7822 000152.06+673156.1: X-ray is a v. triple with c279 and c274\\
\nodata}

\end{deluxetable*}

\floattable
\begin{deluxetable*}{cccccccccc}
\centering \rotate \tabletypesize{\tiny} \tablewidth{0pt} \tablecolumns{8}
\tablecaption{Comparison between SPCM and previously published catalogs \label{tbl_spcm_vs_previous}}
\tablehead{
\colhead{Region} & 
\colhead{SPCM} &
\colhead{Pub} & 
\colhead{Pub} &
\colhead{IR\_Xray} &
\colhead{SPCM-Pub} &
\colhead{FOV} &
\colhead{Census}\\
\colhead{} & \colhead{\#} & \colhead{Ref.} & \colhead{\#} & \colhead{Flag} & \colhead{\#} & \colhead{Flag} & \colhead{Increase}\\
\colhead{} & \colhead{sources} & \colhead{} & \colhead{sources} & \colhead{} & \colhead{sources} & \colhead{} & \colhead{\%}\\
\colhead{(1)} & \colhead{(2)} & \colhead{(3)} & \colhead{(4)} & \colhead{(5)} & \colhead{(6)} & \colhead{(7)} & \colhead{(8)}}
\startdata
NGC 7822 (Be 59)           & 626 & \nodata   & \nodata &  \nodata & \nodata  & \nodata & \nodata \\
IRAS 00013+6817 (SFO 2)    & 71  & Gu09      &  48      & 10       & 44      & C & 56 \\
NGC 1333                   & 181 & Lu16      &  203     & 11       & 155     & S & 13\\
IC 348                     & 396 & Lu16      &  478     & 11       & 367     & C & 6\\
LkH$\alpha$ 101            & 250 & Wo10      &  211     & 11       & 200     & C & 24\\
NGC 2068-2071              & 387 & Me12      &  273     & 10       & 254     & C & 49\\
ONC Flank S                & 386 & Me12,Ra04 &  281,190,(384) & 11       & 256,186,(354) & C & 8\\
ONC Flank N                & 327 & Me12,Ra04 &  217,166,(311) & 11       & 203,160,(295) & C & 10\\
OMC 2-3                    & 530 & Me12,Ts02 &  394,108,(443) & 11       & 355,101,(389) & C & 32\\
Mon R2                     & 652 & Gu09,Na03 &  235,290,(426) & 11       & 229,281,(411) & L & 57\\
GGD 12-15                  & 222 & Gu09      &  119     & 10       & 119     & C & 86\\
RCW 120                    & 420 & \nodata   & \nodata  & \nodata  & \nodata & \nodata & \nodata\\
Serpens Main               & 159 & Wi07      & 137      & 11       & 110     & C & 36\\
Serpens South              & 645 & Po13      & 666      & 10       & 542     & C & 15\\
IRAS 20050$+$2720          & 380 & Gu12      & 330      & 11       & 286     & C & 28\\
Sh 2-106                   & 264 & Gu09,Gi04 & 79,93,(158)   & 11       & 76,85,(146)   & C & 75\\
IC 5146                    & 256 & Gu09      & 149      & 10       & 148     & C & 72\\
NGC 7160                   & 143 & Si06      & 132      & 10       & 21      & S & 92\\
LDN 1251B                  & 49  & Ev03      & 21       & 10       & 19      & C & 143\\
Cep OB3b                   & 1636& Al12      & 2575     & 11       & 1487    & S & 6\\
Cep A                      & 335 & Gu09,Pr09 & 96,29,(113)    & 11       & 92,28,(101)   & L & 207\\
Cep C                      & 177 & Gu09      & 114      & 10       & 109     & C & 60\\
\enddata
\tablecomments{Column 1: SFR. Column 2: Total number of the SPCM sources. Column 3: Literature references to the previous YSO catalogs. Column 4: Total number of YSOs from the previous catalog(s). In the case of two separate previous catalogs, the numbers age given for both catalogs, as well as for the merged catalog removing duplicate sources. For the merged catalog, the number is in parenthesis.   Column 5: Flag indicating the type of the previous YSO catalog(s): ``10'' - IR without X-ray, ``11'' - IR and X-ray. Column 6: Number of source matches between SPCM and the previous catalog(s). As for Column 4, the numbers for the merged catalogs are given in parenthesis. Column 7: Flag comparing the fields of view: ``C'' - both SPCM and the previous catalog(s) have fields of view of a comparable size, ``S'' - the SPCM field of view is smaller, ``L'' - the SPCM field of view is larger. Column 8: The increase in the census of YSOs by SFiNCs relative to the previous studies (in \%): $(Col.2 - Col.6)/Col.4$. Reference code in Column 3:, Al12 \citep{Allen2012}, Ev03 \cite{Evans2003}, Gi04 \citep{Giardino2004}, Gu09 \citep{Gutermuth2009}, Gu12 \citep{Gunther2012}, Lu16 \citep{Luhman2016}, Me12 \citep{Megeath2012}, Na03 \citep{Nakajima2003}, Po13 \citep{Povich2013}, Pr09 \citep{Pravdo2009}, Ra04 \citep{Ramirez2004}, Si06 \citep{Sicilia-Aguilar2006}, Ts02 \citep{Tsuboi2002}, Wi07 \citep{Winston2007}, Wo10 \citep{Wolk2010}.}
\end{deluxetable*}

\clearpage
\newpage

\appendix

\section{Comparison of the X-ray and MIR fluxes between SFiNCs and previously published catalogs} \label{sec_appendix_flux_comparison}

\begin{figure*}
\centering
\includegraphics[angle=0.,width=6.5in]{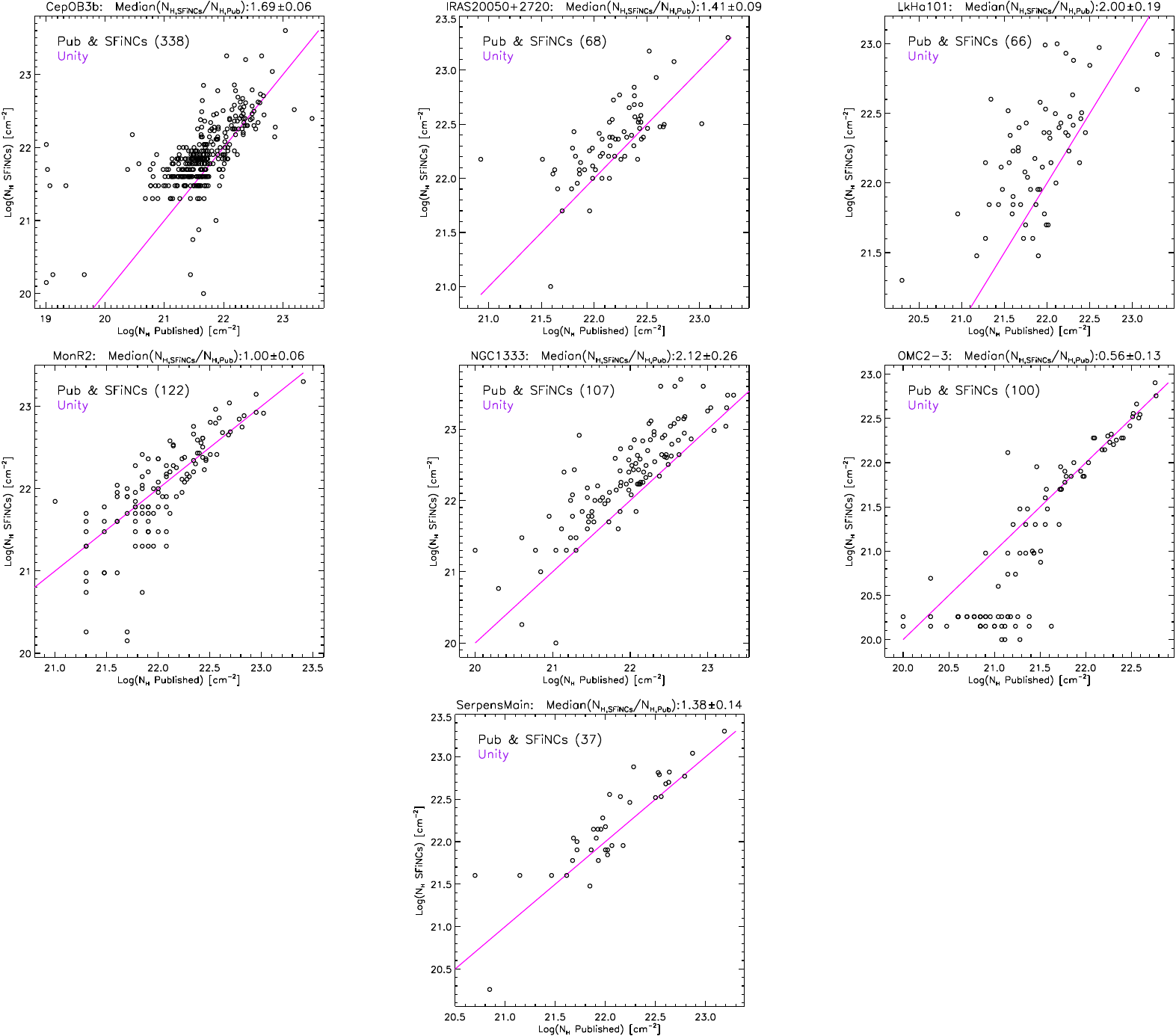}
\caption{Comparison of the X-ray column densities ($N_H$) between SFiNCs and previously published catalogs. $N_H$s were previously reported for 7 SFiNCs SFRs. Sources common between SFiNCs and Pub are shown in black. A handful of Pub sources with the reported $N_H$ values of 0 cm$^{-2}$ were omitted from the analysis. Unity lines are shown in magenta. The median of the $N_H$ ratio between the two catalogs is given in the figure title, as well as in Table \ref{tbl_new_sfincs_vs_pub}. The figure legends give the numbers of plotted sources. \label{fig_comparison_of_nh}}
\end{figure*}

\begin{figure*}
\centering
\includegraphics[angle=0.,width=6.5in]{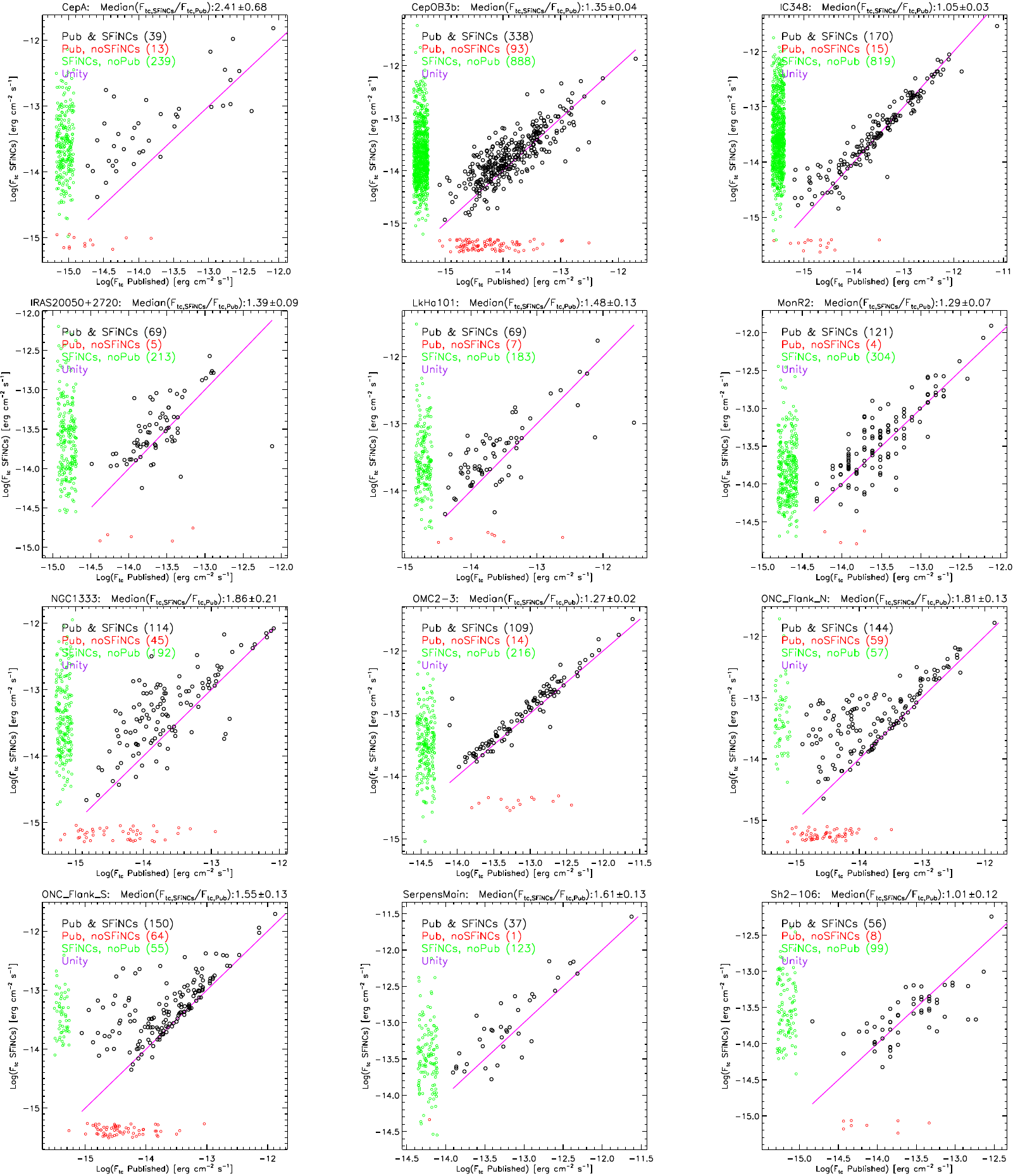}
\caption{Comparison of the intrinsic X-ray fluxes between SFiNCs and previously published catalogs. Intrinsic X-ray fluxes were previously reported for 12 SFiNCs SFRs. Sources that are common between the two catalogs are shown in black. The unique SFiNCs sources with available SFiNCs flux estimates are in green, and unique previously published sources with available Pub flux estimates are in red. Unity lines are shown in magenta. The median of the flux ratio between the two catalogs is given in the figure title, as well as in Table \ref{tbl_new_sfincs_vs_pub}. The figure legends give the numbers of plotted sources. \label{fig_comparison_of_ftc}}
\end{figure*}

The previously published catalogs are abbreviated here as Pub. The measurements of intrinsic (corrected for absorption) X-ray fluxes and X-ray column densities were previously reported for 12 and 7 SFiNCs SFRs, respectively. The comparison of the column densities and fluxes between SFiNCs and Pub is presented in Figures \ref{fig_comparison_of_nh} and \ref{fig_comparison_of_ftc} and Table \ref{tbl_new_sfincs_vs_pub}. Two important results are evident from this analysis.

First, both Figures \ref{fig_comparison_of_nh} and \ref{fig_comparison_of_ftc} show clear column density ($N_H$) and flux ($F_{tc}$) biases. The SFiNCs column densities are generally higher ($> 40$\%) than the Pub densities. The SFiNCs fluxes are systematically ($> 30$\%) higher than the Pub fluxes.  For the vast majority of the Pub sources their $N_H$ and $F_{tc}$ estimates were derived by fitting the data with a one-temperature thermal plasma model using the {\it XSPEC} \citep{Arnaud96} or {\it Sherpa} \citep{Freeman2001} packages. Meanwhile the SFiNCs $N_H$ and $F_{tc}$ measurements were obtained with {\it XPHOT} using more realistic two-temperature plasma models \citep{Getman2010}. A situation analogous to the SFiNCs-Pub $N_H$ bias can be found in \citet{Maggio2007}, where the authors were improving the spectral fits of $\sim 150$ X-ray bright PMS stars in Orion Nebula Cluster by substituting the one- and two-tempearture (1T/2T) plasma model fits of \citet{Getman2005} with more realistic two- and three-temperature (2T/3T) model fits. The column densities from the 2T/3T fits were systematically larger by 0.1 dex \citep[Figure 4 in][]{Maggio2007}. Another example of a systematic increase in flux and column density with the choice of more reasonable X-ray PMS models can be found in \citet[][their Figures 7 and 8]{Gudel2007}, where for over 100 Taurus PMS stars the authors compare the X-ray spectral results between the traditional 1- and 2-T model fits and the fits with more realistic distributions of the differential emission measure.

Further, it is interesting to note that the choice of the thermal plasma emission model (for instance, MEKAL vs. APEC) for the same number of model components appears to have a negligible effect on the $N_H$ outcome \citep[Figure 7a in][]{Hasenberger2016}. The choice of the adopted Solar abundances would affect $N_H$ estimates  \citep[Figure 7b in][]{Hasenberger2016}, but all Pub studies use the same solar abundance table \citep{Anders1989} that {\it XPHOT} is calibrated to; and the value of coronal metal abundance is generally similar across the Pub studies (0.3 of solar photospheric abundances) and is consistent with that of {\it XPHOT}. Based on these lines of evidence, we believe that the observed $N_H$ and $F_{tc}$ biases between SFiNS and Pub are generally a consequence of inability of one-temperature plasma models to fully recover the soft (often unseen) component of the PMS X-ray emission \citep{Getman2010}. In the case of IC~348, where \citet{Stelzer2012} uses the {\it XPHOT} package to derive fluxes for the majority of their Pub sources (but not for their faintest ones), the SFiNCs-Pub bias is small. 

It is also interesting to note here that the linear quantile regression analysis\footnote{Description of the quantile regression analysis can be found on-line at \url{https://en.wikipedia.org/wiki/Quantile_regression}.} applied to the disky and diskless SPCMs across all 22 SFiNCs SFRs (Table \ref{tbl_spcm_otherprops}), suggests that the median $N_H/A_V$ ratio, as a function of $A_V$, ranges between $1.6 \times 10^{21}$~cm$^{-2}$~mag$^{-1}$ \citep{Vuong2003} and $2.2 \times 10^{21}$~cm$^{-2}$~mag$^{-1}$ \citep{Watson2011} and a wider $N_H/A_V$ spread is present for the disky SPCMs (graph is not shown). Detailed analyses of the $N_H/A_V$ ratios for the individual SFiNCs SFRs will be presented in a future SFiNCs paper.

Second, since SFiNCs produces most sensitive X-ray source catalogs and {\it XPHOT} allows derivation of fluxes for many faint sources, the number of unique SFiNCs sources with available flux estimates (green) is generally significantly higher than the number of unique Pub sources with available fluxes (red). However, we caution that these intrinsic {\it XPHOT} flux estimates are valid only in cases where the SFiNCs sources are found to be YSO members of the SFiNCs SFRs.

\begin{figure*}
\centering
\includegraphics[angle=0.,width=6.5in]{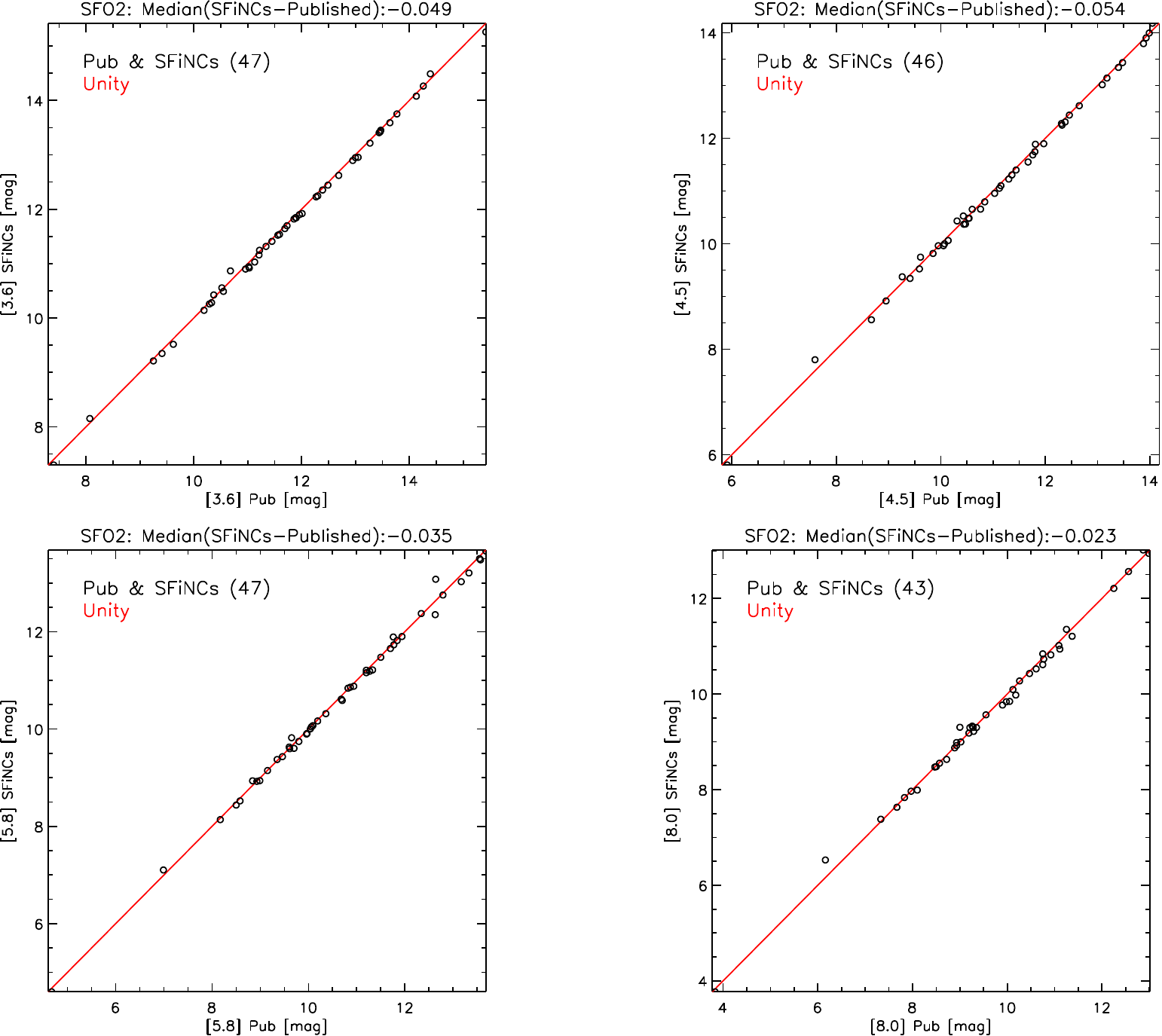}
\caption{Comparison of the IRAC magnitudes between SFiNCs and previously published catalogs. An example is given for the SFO~2 SFR. The figure set presenting other SFiNCs SFRs with previously published IRAC photometry is available in the on-line journal. To assist with the review process a single pdf file comprising the entire figure set is also provided (fA3\_figureset\_merged\_irac\_phot.pdf). Unity lines are shown in red. Figure legends give the numbers of plotted sources. The median of the magnitude difference between the two catalogs is given in the figure title, whereas the mean and standard deviation of the difference are listed in Table \ref{tbl_new_sfincs_vs_pub_ir}. \label{fig_comparison_of_mags}}
\end{figure*}

The comparison of the MIR IRAC magnitudes between SFiNCs and Pub is given in Figure \ref{fig_comparison_of_mags} and Table \ref{tbl_new_sfincs_vs_pub_ir}. The IRAC photometry was previously published for 21 SFiNCs SFRs (omitting only one SFR, Be~59). Except for the RCW~120 and LDN 1251B SFRs, the previously published source catalogs are limited to YSO samples; the vast majority of these sources are relatively bright MIR sources with $[3.6] > 14$~mag.  For the majority of the SFiNCs regions their Pub photometry was derived using the aperture photometry tool PhotVis \citep[][and references therein]{Gutermuth2009}. In the case of the GLIMPSE data of RCW~120, the Pub photometry was obtained through point response function fitting \citep{Benjamin2003}. The SFiNCs-Pub magnitude differences generally have small biases and dispersions (Table \ref{tbl_new_sfincs_vs_pub_ir}). Typical biases are of $0.04$, $0.03$, $0.02$, and $0.07$~mag in the [3.6],[4.5],[5.8], and [8.0] bands, respectively. Typical dispersions are $0.15$, $0.14$, $0.16$, and $0.19$~mag in the [3.6],[4.5],[5.8], and [8.0] bands, respectively.

\floattable
\begin{deluxetable*}{ccccccc}
\centering \rotate \tabletypesize{\tiny} \tablewidth{0pt} \tablecolumns{7}
\tablecaption{X-ray Flux Comparison Between SFiNCs And Earlier Published Catalogs \label{tbl_new_sfincs_vs_pub}}
\tablehead{
\colhead{Region} & 
\colhead{SFiNCs} & 
\colhead{Pub} &
\colhead{Pub Ref} &
\colhead{SFiNCs \& Pub} &
\colhead{$N$;$F_{tc,SFiNCs}/F_{tc,Pub}$} &
\colhead{$N$;$N_{H,SFiNCs}/N_{H,Pub}$}\\
\colhead{} & \colhead{(All/ACIS-I)} & \colhead{(Total)} & \colhead{} & \colhead{(Matches)} & \colhead{} & \colhead{}\\
\colhead{} & \colhead{(sources)} & \colhead{(sources)} & \colhead{} & \colhead{(sources)} & \colhead{} & \colhead{}\\
\colhead{(1)} & \colhead{(2)} & \colhead{(3)} & \colhead{(4)} & \colhead{(5)} & \colhead{(6)} & \colhead{(7)}}
\startdata
NGC 1333   &   683/671   & 180 & Wi10 & 139 & 114;$1.86\pm0.21$ & 107;$2.12\pm0.26$\\
IC 348  &       1554/1534 & 290 & St12 & 280 & 170;$1.05\pm0.03$ & \nodata\\
LkH$\alpha$ 101   &    581/574  & 213 & Wo10 & 193 & 69;$1.48\pm0.14$ & 66;$2.00\pm0.19$\\
ONC Flank S &   408/387  & 214 &  Ra04 & 204  & 150;$1.55\pm0.12$ & \nodata\\
ONC Flank N &   446/421  & 203  & Ra04 & 195 & 144;$1.81\pm0.14$ & \nodata\\
OMC 2-3    &   569/557  & 398  & Ts02 & 304 & 109;$1.27\pm0.02$ & 100;$0.56\pm0.13$\\
Mon R2     &   780/764   & 368 & Na03 & 359 & 121;$1.29\pm0.07$ & 122;$1.00\pm0.06$\\
Serpens Main &  351/341 & 85  & Wi07 & 79  & 37;$1.61\pm0.13$   & 37;$1.38\pm0.14$\\
IRAS 20050 &    790/781  & 348 & Gu12 & 197 & 69;$1.39\pm0.09$  & 68;$1.41\pm0.09$\\
Sh 2-106   &   337/331  & 93  & Gi04 & 93   & 56;$1.01\pm0.11$ & \nodata\\
Cep OB3b &        2196/2148 & 431 & Ge06 & 408 & 338;$1.35\pm0.04$ & 338;$1.69\pm0.06$\\
Cep A &         530/521 &  52 & Pr09 &  46 &  39;$2.41\pm0.68$ & \nodata\\
LDN 1251B    &   340/334  & 43  & Si09 & 43  & \nodata & \nodata\\
Be 59    &  736/736  &  \nodata  &  \nodata &  \nodata  &  \nodata & \nodata\\
SFO 2    &  148/148  &  \nodata  &  \nodata &  \nodata  &  \nodata & \nodata\\
NGC 2068-2071    &   2080/2025  &  \nodata  &  \nodata &  \nodata  &  \nodata & \nodata\\
GGD 12-15    &   365/359  &  \nodata  &  \nodata &  \nodata  &  \nodata & \nodata\\
RCW 120    &   678/678  &  \nodata  &  \nodata &  \nodata  &  \nodata & \nodata\\
Serpens South    &   357/347  &  \nodata  &  \nodata &  \nodata  &  \nodata & \nodata\\
IC 5146    &   432/408  &  \nodata  &  \nodata &  \nodata  &  \nodata & \nodata\\
NGC 7160    &  729/715  &  \nodata  &  \nodata &  \nodata  &  \nodata & \nodata\\
Cep C    &   274/270  &  \nodata  &  \nodata &  \nodata  &  \nodata & \nodata\\
\enddata

\tablecomments{Previously published catalogs are abbreviated here as ``Pub''. Column 1: SFR. Column 2: Total number of SFiNCs X-ray sources. This includes both all X-ray and ACIS-I sources only. In the remaining columns, the numbers are given for the ACIS-I sources only. Column 3: Total number of Pub X-ray sources. Column 4. Literature reference to the Pub catalog. Column 5. Number of source matches between SFiNCs and Pub. Column 6. Characterization of the SFiNCs-Pub X-ray flux bias. The column gives the number of SFiNCs-Pub sources with available flux measurements, as well as the median and its bootstrap error for the ratio of these fluxes, $F_{tc,SFiNCs}/F_{tc,Pub}$. Column 7. Characterization of the SFiNCs-Pub X-ray column density bias. The column gives the number of SFiNCs-Pub sources with available $N_H$ measurements, as well as the median and its bootstrap error for the $N_{H,SFiNCs}/N_{H,Pub}$ ratio. Reference code in Column 4: Ge06 \citep{Getman2006}, Gi04 \citep{Giardino2004}, Gu12 \citep{Gunther2012}, Na03 \citep{Nakajima2003}, Pr09 \citep{Pravdo2009}, Ra04 \citep{Ramirez2004}, Si09 \citep{Simon2009}, St12 \citep{Stelzer2012}, Ts02 \citep{Tsuboi2002}, Wi07 \citep{Winston2007}, Wi10 \citep{Winston2010}, Wo10 \citep{Wolk2010}.}
\end{deluxetable*}


\floattable
\begin{deluxetable*}{ccccccccc}
\centering \rotate \tabletypesize{\tiny} \tablewidth{0pt} \tablecolumns{9}

\tablecaption{MIR Flux Comparison Between SFiNCs And Earlier Published Catalogs \label{tbl_new_sfincs_vs_pub_ir}}

\tablehead{
\colhead{Region} & 
\colhead{SFiNCs} & 
\colhead{Pub} &
\colhead{Pub Ref} &
\colhead{SFiNCs \& Pub} &
\colhead{[3.6]} &
\colhead{[4.5]} &
\colhead{[5.8]} &
\colhead{[8.0]}\\
\colhead{} & \colhead{Total} & \colhead{Total} & \colhead{} & \colhead{Matches} & \colhead{$N$;$\Delta m$} & \colhead{$N$;$\Delta m$} & \colhead{$N$;$\Delta m$}  & \colhead{$N$;$\Delta m$}\\
\colhead{} & \colhead{(sources)} & \colhead{(sources)} & \colhead{} & \colhead{(sources)} & \colhead{(sources;mag)} & \colhead{(sources;mag)}  & \colhead{(sources;mag)}  & \colhead{(sources;mag)}\\
\colhead{(1)} & \colhead{(2)} & \colhead{(3)} & \colhead{(4)} & \colhead{(5)} & \colhead{(6)}  & \colhead{(7)} & \colhead{(8)} & \colhead{(9)}}

\startdata
SFO 2 & 4626 & 48 & Gu09 & 47 & 47;-0.039$\pm$0.057 & 46;-0.029$\pm$0.075 & 47;-0.023$\pm$0.099 & 43;-0.019$\pm$0.112\\
NGC 1333 & 10270 & 133 & Gu09 & 129 & 128;-0.047$\pm$0.151 & 129;-0.039$\pm$0.096 & 127;-0.020$\pm$0.099 & 121;-0.064$\pm$0.117\\
IC 348 & 3300 & 307 & La06 & 298 & 294;0.011$\pm$0.107 & 295;0.014$\pm$0.099 & 265;-0.078$\pm$0.163 & 213;-0.057$\pm$0.282\\
LkH$\alpha$ 101 & 3149 & 103 & Gu09 & 103 & 103;-0.039$\pm$0.088 & 102;-0.032$\pm$0.096 & 77;0.002$\pm$0.140 & 50;-0.074$\pm$0.219\\
NGC 2068-2071 & 7686 & 273 & Me12 & 269 & 266;-0.014$\pm$0.147 & 267;-0.010$\pm$0.127 & 240;-0.009$\pm$0.115 & 224;0.036$\pm$0.156\\
ONC Flank S & 4810 & 297 & Me12 & 267 & 262;0.006$\pm$0.222 & 260;-0.014$\pm$0.137 & 247;-0.003$\pm$0.172 & 217;-0.134$\pm$0.228\\
ONC Flank N & 4113 & 228 & Me12 & 214 & 214;-0.031$\pm$0.109 & 214;-0.033$\pm$0.117 & 195;-0.017$\pm$0.140 & 161;-0.107$\pm$0.160\\
OMC 2-3 & 5654 & 425 & Me12 & 369 & 364;-0.015$\pm$0.172 & 367;-0.016$\pm$0.111 & 303;0.001$\pm$0.183 & 230;-0.099$\pm$0.182\\
Mon R2 & 14482 & 235 & Gu09 & 234 & 229;0.065$\pm$0.244 & 233;-0.016$\pm$0.204 & 195;-0.037$\pm$0.181 & 171;-0.076$\pm$0.247\\
GGD 12-15 & 11416 & 119 & Gu09 & 119 & 119;-0.037$\pm$0.153 & 119;-0.045$\pm$0.150 & 117;-0.041$\pm$0.158 & 112;-0.080$\pm$0.140\\
RCW 120 & 29061 & 20596 & Be03 & 17480 & 17460;0.031$\pm$0.189 & 17450;0.035$\pm$0.222 & 6153;0.067$\pm$0.204 & 3584;0.037$\pm$0.240\\
Serpens Main & 40691 & 97 & Gu09 & 96 & 96;-0.041$\pm$0.219 & 96;-0.052$\pm$0.189 & 94;-0.043$\pm$0.241 & 94;-0.121$\pm$0.217\\
Serpens South & 49340 & 666 & Po13 & 650 & 650;0.014$\pm$0.160 & 650;-0.010$\pm$0.140 & 633;-0.001$\pm$0.117 & 575;-0.020$\pm$0.170\\
IRAS 20050+2720 & 22971 & 177 & Gu09 & 176 & 175;0.128$\pm$0.299 & 176;0.025$\pm$0.295 & 172;-0.010$\pm$0.205 & 169;-0.075$\pm$0.144\\
Sh 2-106 & 43543 & 79 & Gu09 & 78 & 77;0.031$\pm$0.161 & 78;0.010$\pm$0.133 & 55;-0.002$\pm$0.188 & 36;0.031$\pm$0.222\\
IC 5146 & 21413 & 149 & Gu09 & 149 & 149;-0.061$\pm$0.114 & 149;-0.057$\pm$0.118 & 130;-0.006$\pm$0.150 & 101;-0.084$\pm$0.179\\
NGC 7160 & 24763 & 132 & Si06 & 107 & 107;0.073$\pm$0.135 & 107;0.043$\pm$0.121 & 107;-0.024$\pm$0.135 & 107;0.015$\pm$0.185\\
LDN 1251B & 4392 & 5043 & Ev03 & 2664 & 2558;0.068$\pm$0.159 & 2528;0.008$\pm$0.204 & 625;-0.036$\pm$0.254 & 403;-0.179$\pm$0.415\\
Cep OB3b & 57065 & 2575 & Al12 & 2555 & 2555;-0.004$\pm$0.085 & 2555;-0.008$\pm$0.084 & 1582;-0.031$\pm$0.088 & 1225;-0.017$\pm$0.132\\
Cep A & 35824 & 96 & Gu09 & 95 & 94;0.013$\pm$0.140 & 94;-0.023$\pm$0.153 & 92;-0.029$\pm$0.122 & 80;-0.114$\pm$0.175\\
Cep C & 7175 & 114 & Gu09 & 113 & 113;-0.043$\pm$0.117 & 113;-0.034$\pm$0.102 & 113;0.006$\pm$0.116 & 108;-0.082$\pm$0.103\\
\enddata

\tablecomments{Previously published catalogs are abbreviated here as ``Pub''. Column 1: SFR. Column 2: Total number of sources in the SFiNCs IRAC cut-out catalogs. Column 3: Total number of MIR sources in Pub. Column 4. Literature reference to the Pub catalog. Column 5. Number of source matches between SFiNCs and Pub. Columns 6-9. Comparison of IRAC magnitudes. The column gives the number of SFiNCs-Pub sources with available magnitude estimates, as well as the mean and standard deviation of the magnitude difference, $m_{SFiNCs} - m_{Pub}$. Reference code in Column 4: Al12 \citep{Allen2012}, Be03 \citep{Benjamin2003}, Ev03 \citep{Evans2003}, Gu09 \citep{Gutermuth2009}, La06 \citep{Lada2006}, Me12 \citep{Megeath2012}, Po13 \citep{Povich2013}, Si06 \citep{Sicilia-Aguilar2006}.}
\end{deluxetable*}

\clearpage

\section{SPCM source atlas} \label{sec_appendix_source_atlas}

\begin{figure*}
\centering
\includegraphics[angle=0.,width=6.5in]{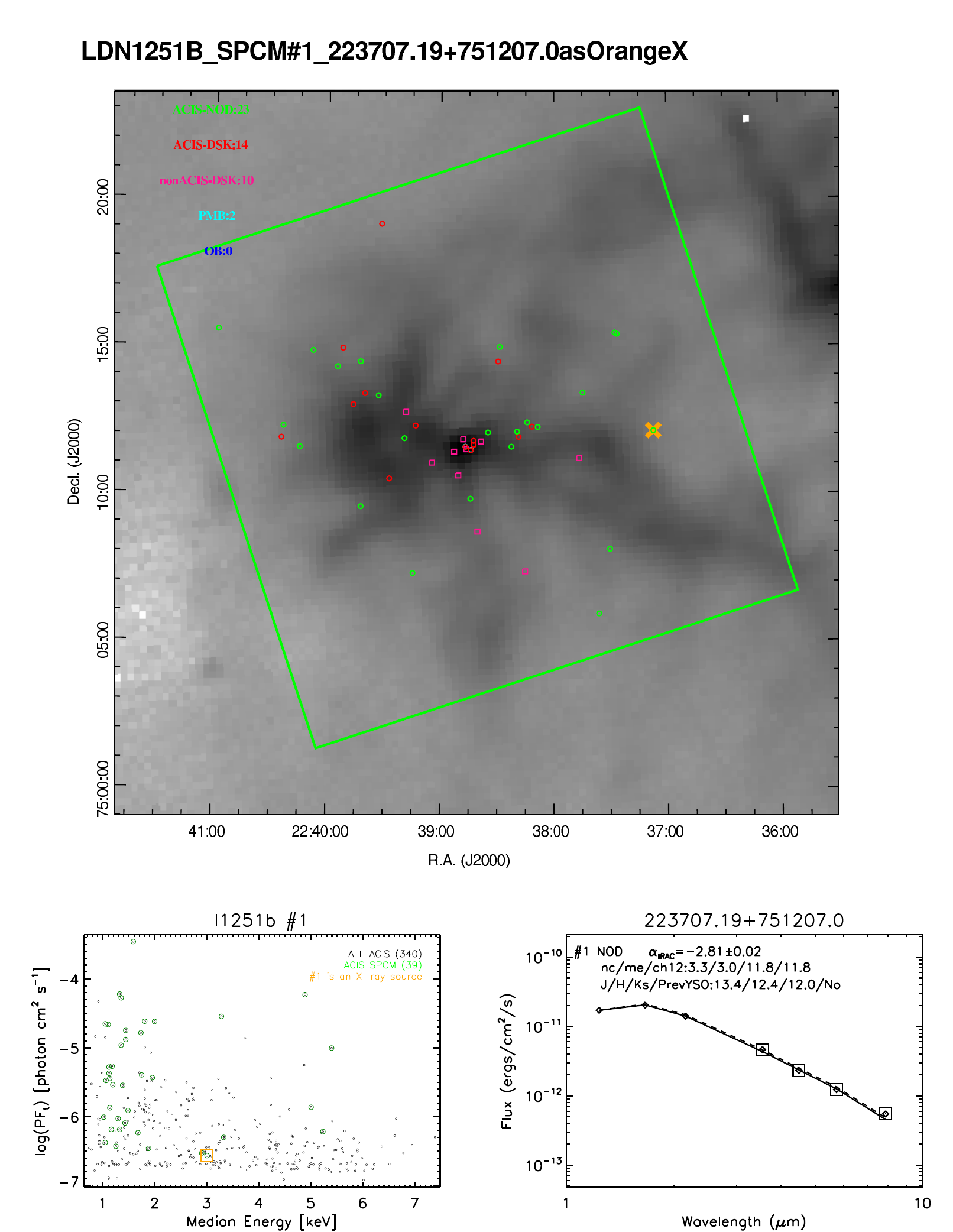}
\caption{SPCM source atlas. An example is given here for the SPCM source \#1 in LDN~1251B SFR. The figure set presenting the entire atlas for all 8492 SPCMs is available in the on-line journal. To assist with the review process the 16984 pages of the SPCM atlas (2 pages per source) are available in PDF format at \url{https://drive.google.com/drive/folders/0B4lwbriAuoXfSXJtbzB6V2tyNTA?usp=sharing}. The description of the atlas is given in Appendix \S \ref{sec_appendix_source_atlas}. \label{fig_source_atlas}}
\end{figure*}

We produce a source atlas in which some tabulated and graphical information is collected onto two pages per source. Figure set \ref{fig_source_atlas} shows a sample page for the SPCM source \#1 in the LDN 1251B SFR. The graphs present various projections of the multi-dimensional SPCM data set, including the SCPM's spatial positions and X-ray/IR photometric properties.

On the first page of the atlas, the upper panel shows the map of the spatial distributions of SPCMs, similar to that of Figure \ref{fig_spcm_maps}, but with the SPCM source of interest additionally marked by orange X. This provides information on the location of the source with respect to both other YSO members and molecular cloud in the region. The bottom left panel is similar to the X-ray color-magnitude diagram from Figure \ref{fig_fx_vs_me}, but with the SPCM source of interest additionally marked by orange square. This gives information on the X-ray photometric flux and median energy of the source as well as the comparison with the X-ray photometry of other SPCM and non-SPCM sources in the region. The bottom right panel presents the IR SED of the SPCM source of interest (along with the tabulated values of its apparent SED slope and disk class), similar to that of Figure \ref{fig_seds}. In this paper, IR SED is used as the primary source for distinguishing between disky and diskless YSOs (\S \ref{yso_selection_section}). The figure legends further give various useful tabulated quantities such as X-ray net counts, X-ray median energy, IR magnitudes in the $J$, $H$, $K_{s}$, [3.6], and [4.5] bands, and the flag indicating the presence/absence of a counterpart in the previously published YSO catalogs that are listed in Table \ref{tbl_spcm_vs_previous}.

On the second page of the atlas, the six panels present the IR color-magnitude and color-color diagrams from Figures \ref{fig_j_vs_jh} -- \ref{fig_ch12_vs_ch34}, where the SPCM source of interest is additionally marked by a large black circle. On these diagrams, the information on the location of the source of interest with respect to other SPCMs and to the expected theoretical loci of YSOs helps us to: elucidate the nature of the source (typically, a foreground star versus a YSO member), reaffirm its disk classification derived from the SED analysis, infer its approximate mass and absorption values, and compare the IR and X-ray (as the X-ray median energy from the previous page) absorption estimates.

For all 8492 SPCMs, the 16984 pages of the SPCM atlas are available in PDF format at \url{https://drive.google.com/drive/folders/0B4lwbriAuoXfSXJtbzB6V2tyNTA?usp=sharing}.

\begin{figure*}
\centering
\includegraphics[angle=0.,width=5.5in]{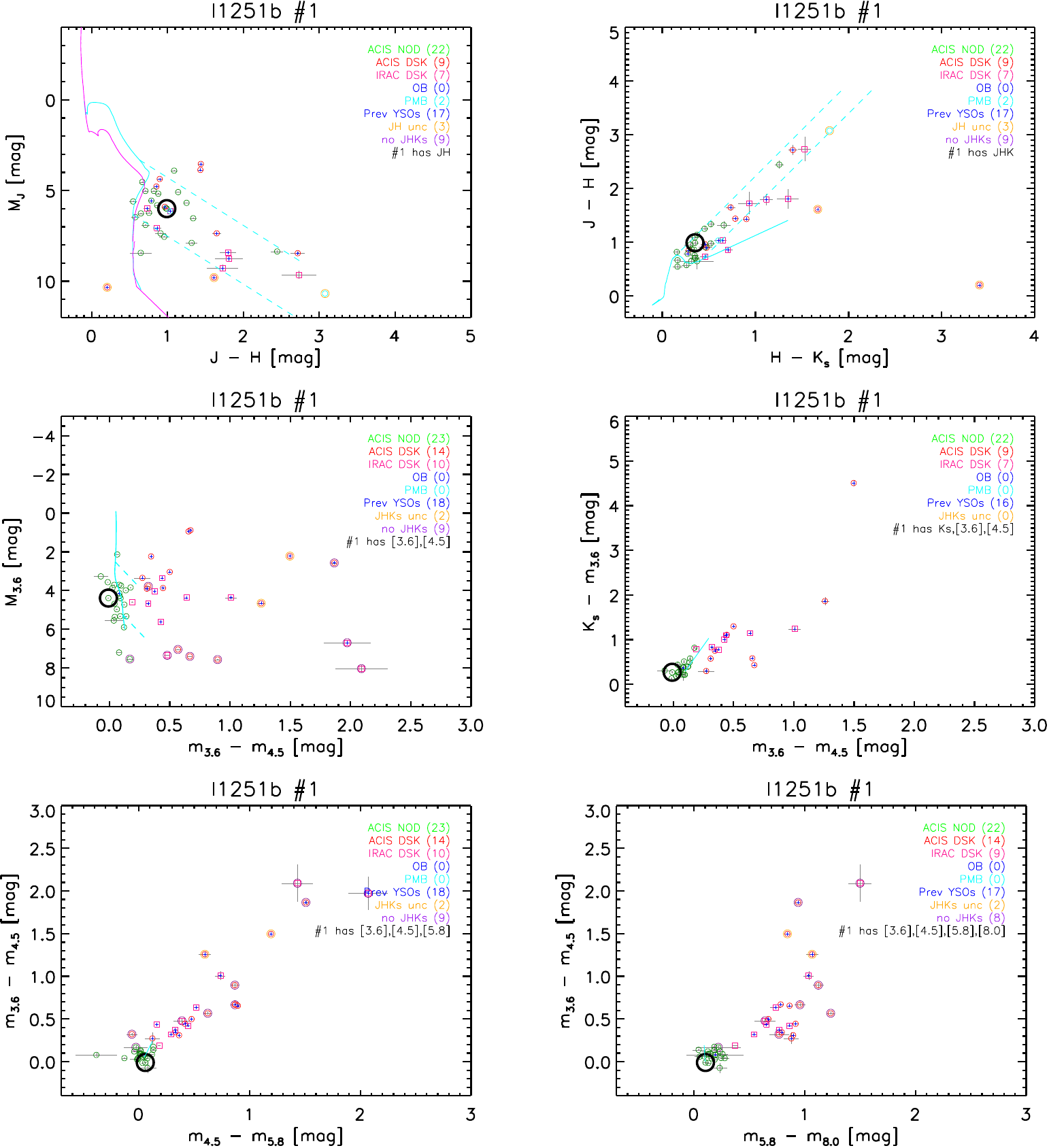}
\end{figure*}

\end{document}